\documentclass[10pt,a4paper]{article}
\usepackage[dvips]{color}
\usepackage{epsfig}
\usepackage{amsmath}
\usepackage{graphicx}

\begin{document}
\title{\bf Interacting Ghost Dark Energy Models with Variable $G$ and $\Lambda$}
\author{{J. Sadeghi$^{a}$ \thanks{Email: pouriya@ipm.ir},\hspace{1mm} M. Khurshudyan$^{b, c, d}$ \thanks{Email: martiros.khurshudyan@nano.cnr.it},
\hspace{1mm} A. Movsisyan$^{e}$
\thanks{Email: artmovsissyan@yandex.ru}\hspace{1mm} and H. Farahani$^{a}$ \thanks{Email:
h.farahani@umz.ac.ir}}\\
$^{a}${\small {\em Department of Physics, Mazandaran University, Babolsar, Iran}}\\
{\small {\em P .O .Box 47416-95447, Babolsar, Iran}}\\
$^{b}${\small {\em CNR NANO Research Center S3, Via Campi 213a, 41125 Modena MO}}\\
$^{c}${\small {\em Dipartimento di Scienze Fisiche, Informatiche e Matematiche,}}\\
{\small {\em Universita degli Studi di Modena e Reggio Emilia, Modena, Italy}}\\
$^{d}${\small {\em Department of Theoretical Physics, Yerevan State
University, 1 Alex Manookian, 0025, Yerevan, Armenia}}\\
$^{e}${\small {\em R\&D Center of Semiconductor Devices \& Nanotechnologies, Yerevan State
University} } }  \maketitle

\begin{abstract}
In this paper we consider several phenomenological models of
variable $\Lambda$. Model of a flat Universe with variable $\Lambda$
and $G$ is accepted. It is well known, that varying $G$ and
$\Lambda$ gives rise to modified field equations and modified
conservation laws, which gives rise to many different manipulations
and assumptions in literature. We will consider two component fluid,
which parameters will enter to $\Lambda$. Interaction between fluids
with energy densities $\rho_{1}$ and $\rho_{2}$ assumed as
$Q=3Hb(\rho_{1}+\rho_{2})$. We have numerical analyze of important
cosmological parameters like EoS parameter of the composed fluid and
deceleration parameter $q$ of the model.
\end{abstract}
\section*{\large{Introduction}}
In modern cosmology, despite to hard work and interesting ideas,
several crucial question still are open, which makes authors to
propose different models and different approaches to find keys for
the problems. Modern era in theoretical cosmology starts, when
observations of high redshift type SNIa supernovae [1-3] reveal the
speeding up expansion of our Universe. Then, other series of
observations like to investigation of surveys of clusters of
galaxies show that the density of matter is very much less than
critical density [4], observations of Cosmic Microwave Background
(CMB) anisotropy indicate that the Universe is flat and the total
energy density is very close to the critical $\Omega_{\small{tot}}
\simeq1$ [5]. Faced with these results we started to find realistic
models to explain experimental data concerning to the nature of the
accelerated expansion of the Universe and a huge number of
hypothesis were proposed. For instance, in general relativity
framework, the desirable result could be achieved by so-called dark
energy: an exotic and mysterious component of the Universe, with
negative pressure (we thought that the energy density is always
positive) and with negative EoS parameter $\omega<0$. Dark energy
occupies about 73$\% $ of the energy of our Universe, other
component (dark matter) about 23$\%$, and usual baryonic matter
occupy about 4$\%$. The simplest model for a dark energy is a
cosmological constant $\omega_{\Lambda}=-1$ introduced by Einstein.
This model has two famous problems: fine-tuning problem and
cosmological coincidence problem. Absence of a fundamental mechanism
which sets the cosmological constant zero or very small value makes
researchers to go deeper and deeper in theories to understand the
solution of the problem, because in the framework of quantum field
theory, the expectation value of vacuum energy is 123 order of
magnitude larger than the observed value [6] of cosmological
constant. The second problem asks why are we living in an epoch in
which the densities of dark energy and matter are comparable. To
alleviate these problems alternative models of dark energy suggest a
dynamical form of dark energy, which at least in an effective level,
can originate from a variable cosmological constant [7, 8], or from
various fields, such as a canonical scalar field [9-14]
(quintessence), a phantom field, that is a scalar field with a
negative sign of the kinetic term [15-23] or the combination of
quintessence and phantom in a unified model named quintom [24-37].
By using some basic of quantum gravitational principles we can
formulate several other models for dark energy, and in literature
they are known as holographic dark energy paradigm
[38-49] and agegraphic dark energy models [50-52].\\
Interaction between components is proved to be other way which can solve coincidence problem. From observations no piece of evidence has been
so far presented against interactions between dark energy and dark matter.
From theoretical side we have not any known symmetry which prevents or suppresses a non-minimal coupling between dark energy and dark matter.\\
Research in theoretical cosmology proposes two possible ways to
explain later time accelerated expansion of the Universe. Remember
that field equations make connection between geometry and matter
content of Universe in a simple way. Therefore, there is two
possibilities either we should modify matter content which is coded
in energy-stress tensor or we should modify geometrical part
including different functions of Ricci scalar etc. Different type of
couplings between geometry and matter could give desirable effects
as well. Recently, huge number of articles appears, where we are
trying to make connection between scalar field and other models of
dark energy. In literature, often used idea of fluid despite to
other ideas, because over the years we learned that modifications of
geometrical part of field equations can be codded in fluid
expression. From this point of view an important component becomes
to be  Equation of State (EoS), which makes connection between
energy density and pressure. Well studied examples are barotropic
fluid $P=\omega \rho$ with its modifications like $P=\omega(t)
\rho^{n}$. In this contexts other interesting parametrization is a
barotropic fluid with more general form (see for instance in [53]
and references therein) or Chaplygin gas models [54-59],
\begin{equation}\label{eq:mcg}
P=\mu \rho -\frac{B}{\rho^{\alpha}},
\end{equation}
where $\mu$ is a positive constant. This model is more appropriate
choice to have constant negative pressure at low energy density and
high pressure at high energy density. The special case of
$\mu=\frac{1}{3}$ is the best fitted value to describe evolution of
the universe from radiation regime to the $\Lambda$CDM regime.  In
[60] one of the authors motivated by a series of works [61-64]
proposed a model of varying generalized Chaplygin gas and considered
its sign-changeable interaction of the form $Q=q(3Hb\rho+\gamma
\dot{\rho} )$ with Tachyonic Fluid. We also can consider fluids with
more general form of EoS given as,
\begin{equation}\label{eq:gform}
f(\rho,P)=0.
\end{equation}
Today, we do not feel lack of models for DE, which could be seen from the number of references given above.
However, all of them are phenomenological models and wait to be
proved by observational data. The same can be said for DM, which
thought to operate on large scales and be responsible for structure
formation, evolution etc.\\
Apart attempts of fluid modifications in modern cosmology
modification of geometrical part also very popular subject of
discussions. Examples are $F(R)$, $F(T)$, $F(G)$ etc just to mention
a few. Models of this origin however contain some future
singularities which can be solved in principle. But, there are
models also, that can explain accelerated expansion without any DE,
for instance, Cardassian Universe [65-70]. In this model, one need
to modify Friedmann equations and therefore having usual matter is enough.\\
It is well known that Einstein equations of general relativity do
not permit any variations in the gravitational constant $G$ and
cosmological constant $\Lambda$ because of the fact that the
Einstein tensor has zero divergence and energy conservation law is
also zero. So, some modifications of Einstein equations are
necessary. This is because, if we simply allow $G$ and $\Lambda$ to
be a variable in Einstein equations, then energy conservation law is
violated. Therefore, the study of the varying $G$ and $\Lambda$ can
be done only through modified field equations and modified
conservation laws. It was Dirac who proposed possibility of
variation in $G$, which open a door for a lot of works and
manipulations. This period in cosmology can be called era of Dirac's
Large Number Hypothesis. For instance, observation of spinning-down
rate of pulsar $PSR J2019+2425$ provides the result,
\begin{equation}\label{eq:gvar1}
\left|\frac{\dot{G}}{G} \right|\leq (1.4-3.2) \times 10^{-11} yr^{-1}.
\end{equation}
Depending on the observations of pulsating white dwarf star $G 117-B
15A$, Benvenuto et al. [71] have set up the astroseismological bound
as,
\begin{equation}\label{eq:gvar2}
 -2.50 \times 10^{-10} \leq \left|\frac{\dot{G}}{G} \right|\leq 4 \times 10^{-10} yr^{-1}.
\end{equation}
For a review to "Large Number Hypothesis" (LNH) we refer our readers
to
[72] and references therein.\\
In this paper we would like to propose different phenomenological
models cosmological constant:
\begin{enumerate}
\item $\Lambda(t)=\rho_{1}+\rho_{2}e^{-tH}$,
\item $\Lambda(t)=H^{2}+(\rho_{1}+\rho_{2})e^{-tH}$,
\item $\Lambda(t)=t^{-2}+(\rho_{1}-\rho_{2})e^{-tH}$.
\end{enumerate}
We will consider two component composed fluid for our Universe and
$\rho_{1}$ and $\rho_{2}$ referred to the energy densities of the
fluid components. We use an interaction between fluids of the
$Q=3Hb(\rho_{1}+\rho_{2})$ form and analyze important cosmological
parameters like EoS parameter of a fluid and deceleration parameter
$q$ of the model. This article is organized in following way.
Section introduction is devoted to introduce basic ideas and gives
some general information related to the research field and our
motivation. Next section review FRW Universe with variable $G$ and
$\Lambda$. In section "Interacting Fluids and Model Setup" we recall
the basics of origin of the fluids and general settings how problem
can be solved. In other sections we consider various models of
variable $\Lambda$ and then we give conclusions.

\section*{\large{FRW Universe with Variable $G$ and $\Lambda$}}
Flat FRW Universe described by the following metric,
\begin{equation}\label{s5}
ds^2=-dt^2+a(t)^2\left(dr^{2}+r^{2}d\Omega^{2}\right),
\end{equation}
where $d\Omega^{2}=d\theta^{2}+\sin^{2}\theta d\phi^{2}$, and $a(t)$
represents the scale factor. Also, field equations that govern our
model with variable $G(t)$ and $\Lambda(t)$ (see for instance [73])
are,
\begin{equation}\label{s6}
R^{ij}-\frac{1}{2}Rg^{ij}=-8 \pi G(t) \left[ T^{ij} -
\frac{\Lambda(t)}{8 \pi G(t)}g^{ij} \right],
\end{equation}
where $G(t)$ and $\Lambda(t)$ are function of time. These leads to
the following Friedmann equations,
\begin{equation}\label{s7}
H^{2}=\frac{\dot{a}^{2}}{a^{2}}=\frac{8\pi G(t)\rho}{3}+\frac{\Lambda(t)}{3},
\end{equation}
and,
\begin{equation}\label{s8}
\frac{\ddot{a}}{a}=-\frac{4\pi
G(t)}{3}(\rho+3P)+\frac{\Lambda(t)}{3}.
\end{equation}
Energy conservation $T^{;j}_{ij}=0$ reads as,
\begin{equation}\label{s9}
\dot{\rho}+3H(\rho+P)=0.
\end{equation}
Combination of the equations (7), (8) and (9) gives the relationship
between $\dot{G}(t)$ and $\dot{\Lambda}(t)$ as the following,
\begin{equation}\label{eq:glambda}
\dot{G}=-\frac{\dot{\Lambda}}{8\pi\rho}.
\end{equation}
Subject of our interest is to consider composed fluids. Basic
components thought to be a barotropic fluid with EoS
$P_{b}=\omega(t)\rho_{b}$, where varying EoS parameter given by
$\omega=\omega_{0}+\omega_{1}\frac{t\dot{H}}{H}$, and ghost dark
energy. Among various models of dark energy, a new model of dark
energy called Veneziano ghost dark (GD) energy, which supposed to
exist to solve the $U(1)_{A}$ problem in low-energy effective theory
of QCD, and has attracted a lot of interests in recent years
[74-86]. Indeed, the contribution of the ghosts field to the vacuum
energy in curved space or time-dependent background can be regarded
as a possible candidate for the dark energy. It is completely
decoupled from the physics sector. Veneziano ghost is unphysical in
the QFT formulation in Minkowski space-time, but exhibits important
non trivial physical effects in the expanding Universe and these
effects give rise to a vacuum energy density $\rho_{D}\sim
\Lambda^{3}_{QCD}H\sim (10^{-3}eV)^{4}$. With $H\sim 10^{-33}eV$ and
$\Lambda_{QCD}\sim 100 eV$ we have the right value for the force of
accelerating the Universe today. It is hard to accept such linear
behavior and it is thought that there should be some exponentially
small corrections. However, it can be argued that the form of this
behavior can be result of the fact of the very complicated
topological structure of strongly coupled QCD. This model has
advantage compared to other models of dark energy, which can be
explained by standard model and general relativity. Comparison with
experimental data, reveal that the current data does not favorite
compared to the $\Lambda$CDM model, which is not conclusive and
future study of the problem is needed. Energy density of ghost dark
energy may reads as,
\begin{equation}\label{eq:GDE}
\rho_{\small{GD}}=\theta H,
\end{equation}
where $H$ is Hubble parameter $H=\dot{a}/a$ and $\theta$ is constant
parameter of the model, which should be determined. The relation
(11) generalized by the Ref. [87] as the following,
\begin{equation}\label{eq:GDEgen}
\rho_{\small{GD}}=\theta H+\vartheta H^{2},
\end{equation}
where $\theta$ and $\vartheta$ are constant parameters of the model.
Such kind of fluids could be named as a geometrical fluids, because
it is clear that it contains information about geometry of the
space-time and metric. Recently a model of varying ghost dark energy
were proposed in the Ref. [88]. We will assume that components are
interacting on Universe with variable $G$ and $\Lambda$. This model
is a phenomenological and we are interested by the evolution of the
Universe with this setup. As there is an interaction between
components, there is not energy conservation for the components
separately, but for the whole mixture the energy conservation is
hold. The forms of interaction term considered in literature very
often are of the following forms: $Q=3Hb\rho_{dm}$,
$Q=3Hb\rho_{\small{de}}$, $Q=3Hb\rho_{\small{tot}}$, where $b$ is a
coupling constant and positive $b$ means that dark energy decays
into dark matter, while negative $b$ means dark matter decays into
dark energy. Other forms for interaction term considered in
literature are $Q=\gamma\dot{\rho}_{\small{dm}}$,
$Q=\gamma\dot{\rho}_{\small{de}}$,
$Q=\gamma\dot{\rho}_{\small{tot}}$, and
$Q=3Hb\gamma\rho_{i}+\gamma\dot{\rho_{i}}$, where $i=\{dm,de,tot\}$.
These types of interactions are either positive or negative and can
not change sign. However, Cai and Su found that the sign of
interaction $Q$ in the dark sector changed in the redshift range of
$0.45 \leq z \leq 0.9$ and a sign-changeable interaction in the
Refs. [64] and [89] introduced as,
\begin{equation}\label{13}
Q=q(\gamma\dot{\rho}+3b H\rho),
\end{equation}
where $\gamma$ and $b$ are dimensionless constants, the energy
density $\rho$ could be $\rho_{m}$, $\rho_{\small{de}}$, and
$\rho_{tot}$. $q$ is the deceleration parameter given by,
\begin{equation}\label{eq:decparameter}
q=-\frac{1}{H^{2}} \frac{\ddot{a}}{a}=-1-\frac{\dot{H}}{H^{2}}.
\end{equation}
For $\Lambda$ were considered over years different forms based on
phenomenological approach, some of examples are, for instance,
$\Lambda \propto (\dot{a}/a)$, $\Lambda \propto \ddot{a}/a$ or
$\Lambda \propto \rho$ to mention a few. As we are interested by toy
models we pay our attention to the problem from a numerical
investigation point of view and we believe that after some effort we
also can provide exact solutions for the problem, which will be done
in other forthcoming articles.\\
In the next section we consider two components fluid Universe with
the sign-changeable interaction (13).

\section*{\large{Interacting Fluid and Model Setups}}
Two-component fluid, in our case, will be described by total energy
density $\rho=\rho_{b}+\rho_{GD}$ and pressure $P=P_{b}+P_{GD}$,
where $b$ stands for barotropic fluid. It is well know that in case
of an interaction $Q$ between fluid components we should consider
following conservation equations,
\begin{equation}\label{s1}
\dot{\rho}_{GD}+3H(\rho_{GD}+{P_{GD}})=-Q,
\end{equation}
and,
\begin{equation}\label{s2}
\dot{\rho}_{b}+3H(\rho_{b}+{P_{b}})=Q.
\end{equation}
We will use interaction term of the form,
\begin{equation}\label{s17}
Q=3Hb(\rho_{b}+\rho_{GD})
\end{equation}
which introduced in the introduction. Taking into account the
equation (16), form of interaction term (17) and GD density (11) we
can write,
\begin{equation}\label{s3}
 \dot{\rho}_{b}+3H\rho_{b}(1-b+\omega(t))-3\theta b H^{2}=0.
\end{equation}
From the equation (15) we will have pressure for GD energy as the
following,
\begin{equation}\label{s4}
P_{GD}=-b\rho_{b}-\theta (b+1)H-\frac{\theta}{3}\frac{\dot{H}}{H}.
\end{equation}
One of the cosmological parameters, which we are interested, is EoS
parameter of the composed fluid which reads as,
\begin{equation}\label{eq:Eospar}
\omega_{tot}=\frac{P_{b}+P_{GD}}{\rho_{b}+\rho_{GD}},
\end{equation}
which reduced to the following expression,
\begin{equation}\label{eq:Eosparf}
\omega_{tot}=\frac{(\omega(t)-b)\rho_{b}-\theta (1+b)
H-\frac{\theta\dot{H}}{3H}}{\rho_{b}+\theta H},
\end{equation}
where we used equations (11) and (18). EoS of GD energy also can be
expressed as a function of $\rho_{b}$ and other parameters of the
models as the following,
\begin{equation}\label{eq:EosGD}
\omega_{GD}=-(1+b)-b\frac{\rho_{b}} {\theta H} -\frac{\dot{H} }{ 3 H^{2} }.
\end{equation}
In order to obtain cosmological parameters we will use the following
models of cosmological constant.
\subsection*{\large{Model 1}}
In the first model we consider,
\begin{equation}\label{23}
\Lambda(t)=\rho_{b}+\rho_{GD}e^{-tH}.
\end{equation}
Using this relation in Friedmann equation together with the results
of the previous section we can obtain,
\begin{equation}\label{s5}
A\dot{H}+\frac{3}{2}H^{2}+BH+C=0,
\end{equation}
where $A$,$B$ and $C$ define as the following,
\begin{equation}\label{s25}
A=1+4\pi G(t)(\omega_{1}\rho_{b}\frac{t}{H}-\frac{\theta}{3H}),
\end{equation}
\begin{equation}\label{s26}
B=-\frac{\theta}{2}\left(e^{-tH}+8 (1+b) \pi G(t)\right),
\end{equation}
and,
\begin{equation}\label{s27}
C=-\frac{\rho_{b}}{2}\left(1-8 (\omega_{0}-b) \pi G(t)\right).
\end{equation}
Therefore, we can obtain the following equation,
\begin{equation}\label{s6}
\dot{G}=-\frac{\dot{\rho}_{b}+\theta\dot{H}e^{-tH}-\theta H^{2} \dot{H} e^{-tH}} {8\pi (\omega(t)\rho_{b}+\theta H)}.
\end{equation}
We solved this equation numerically and find $G$ as an increasing
function of time (see Fig. 1).
\begin{figure}[h]
 \begin{center}$
 \begin{array}{cccc}
\includegraphics[width=50 mm]{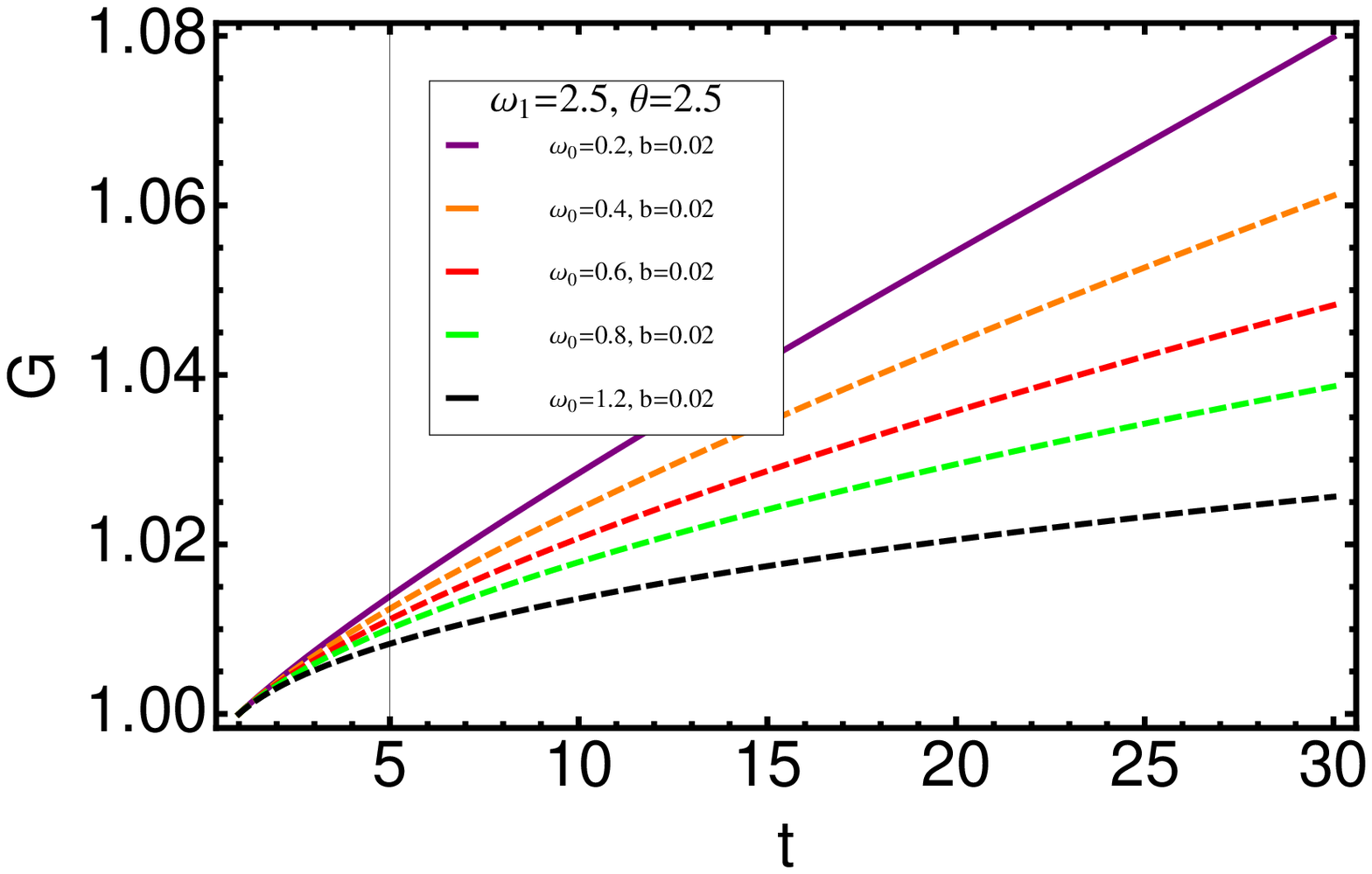} &
\includegraphics[width=50 mm]{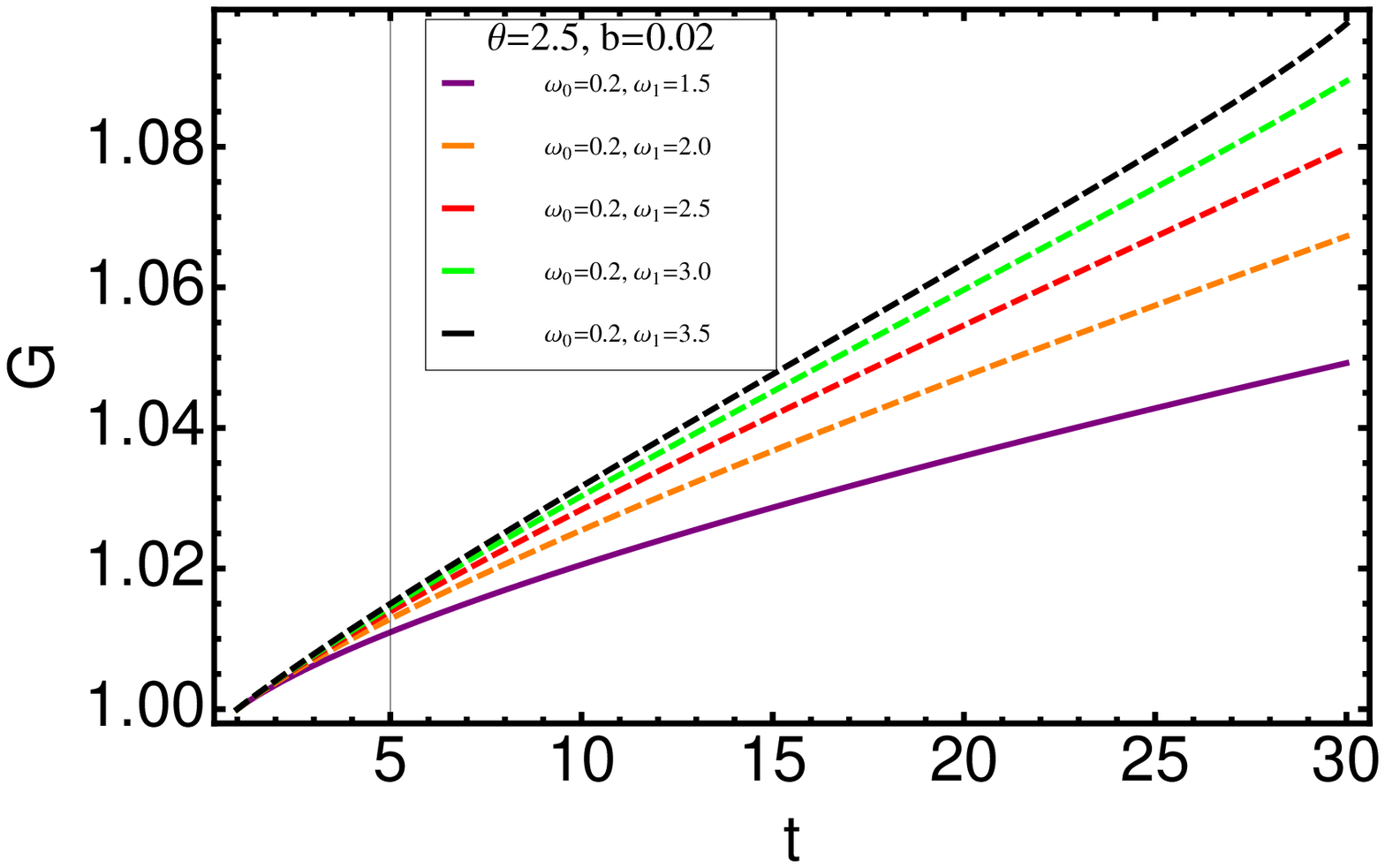}\\
\includegraphics[width=50 mm]{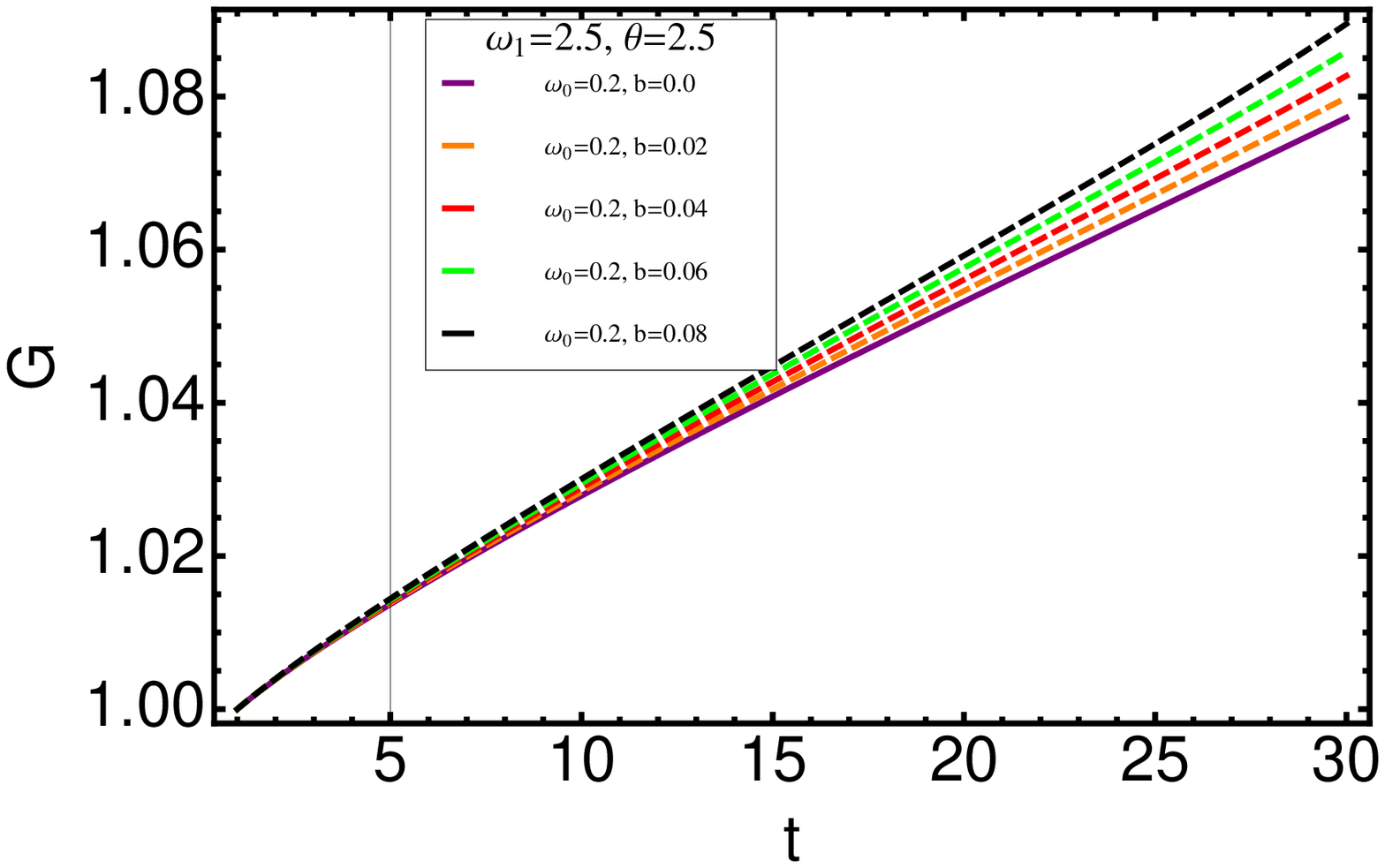} &
\includegraphics[width=50 mm]{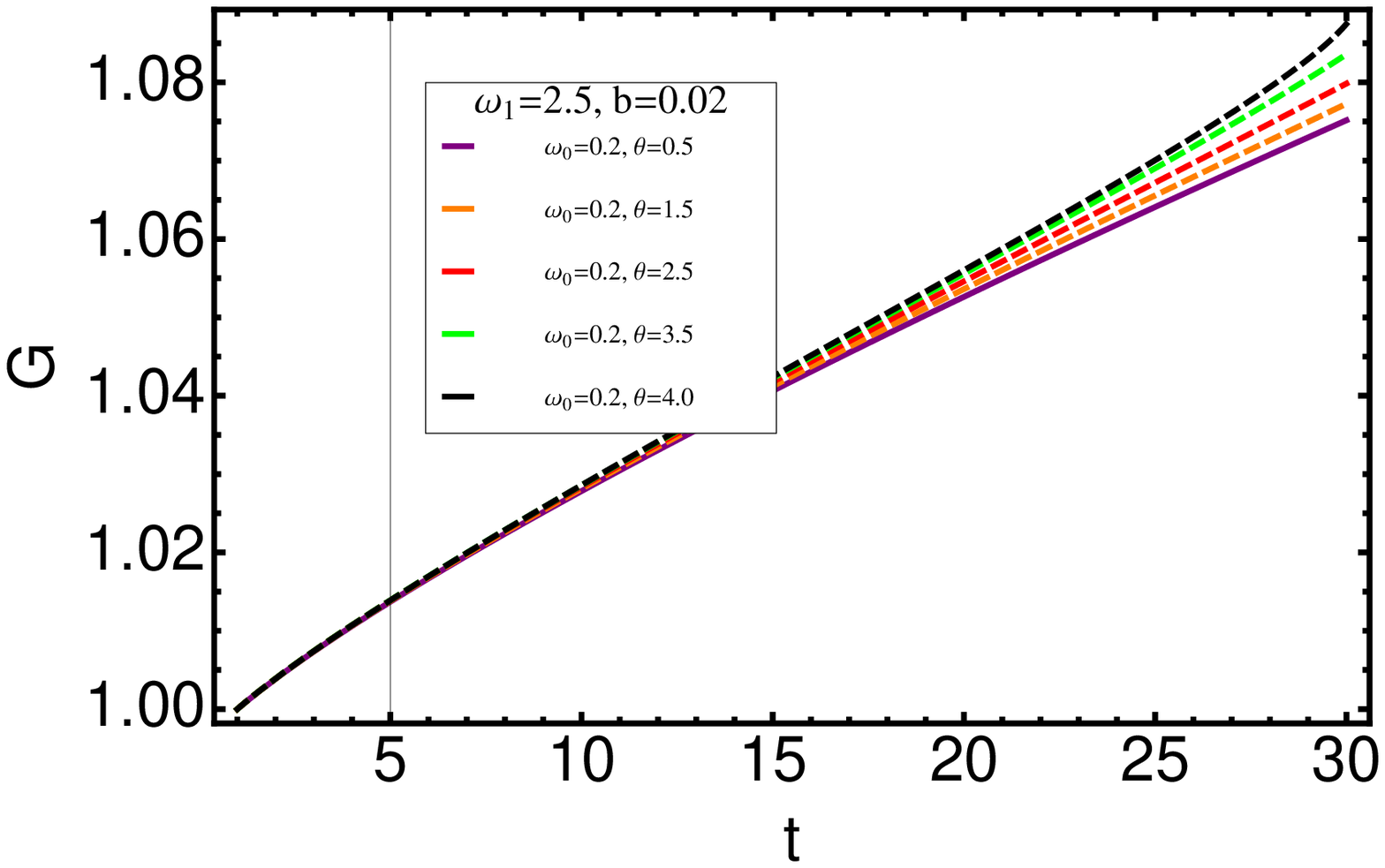}
 \end{array}$
 \end{center}
\caption{Model 1}
 \label{fig:1}
\end{figure}

\begin{figure}[h]
 \begin{center}$
 \begin{array}{cccc}
\includegraphics[width=50 mm]{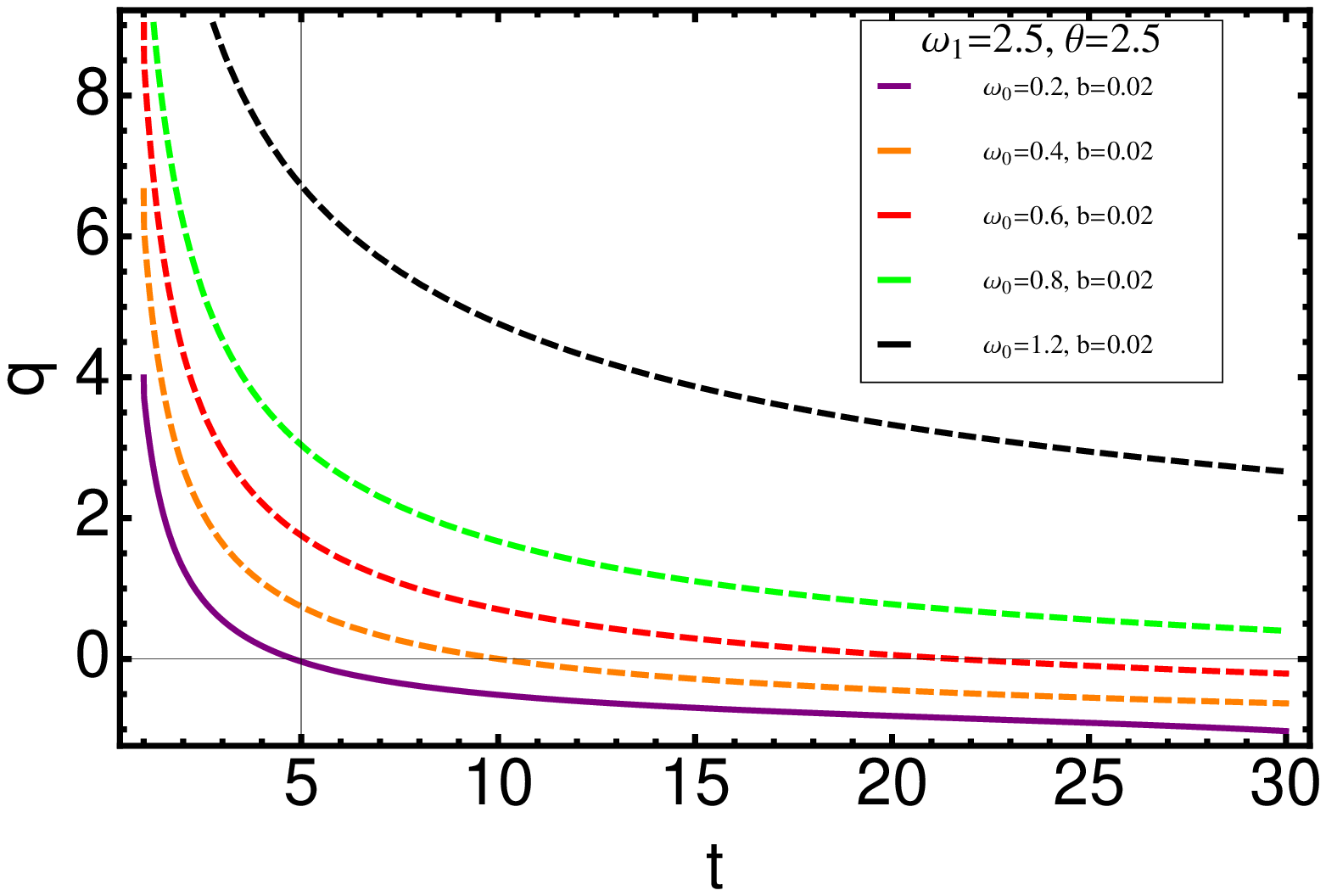} &
\includegraphics[width=50 mm]{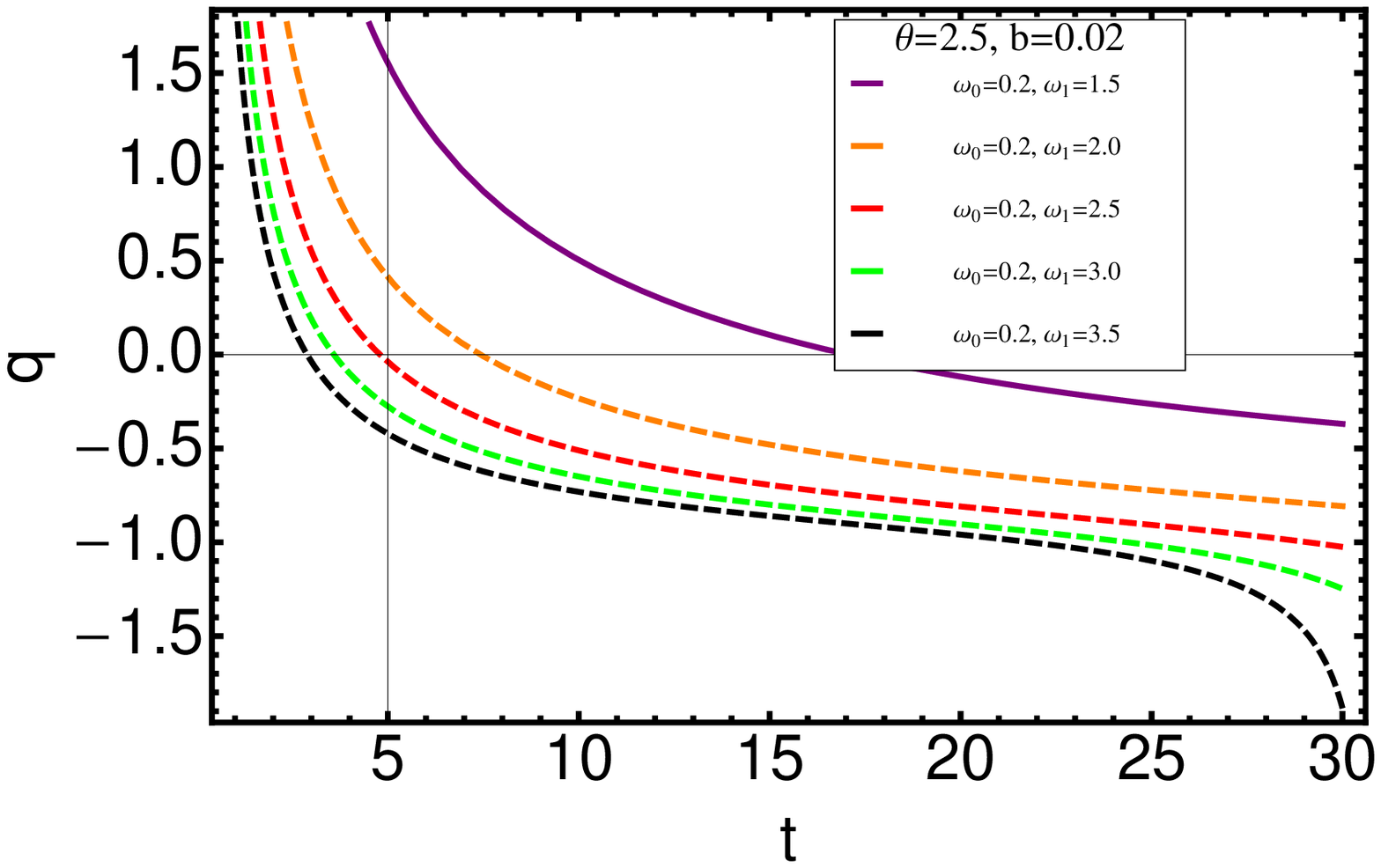}\\
\includegraphics[width=50 mm]{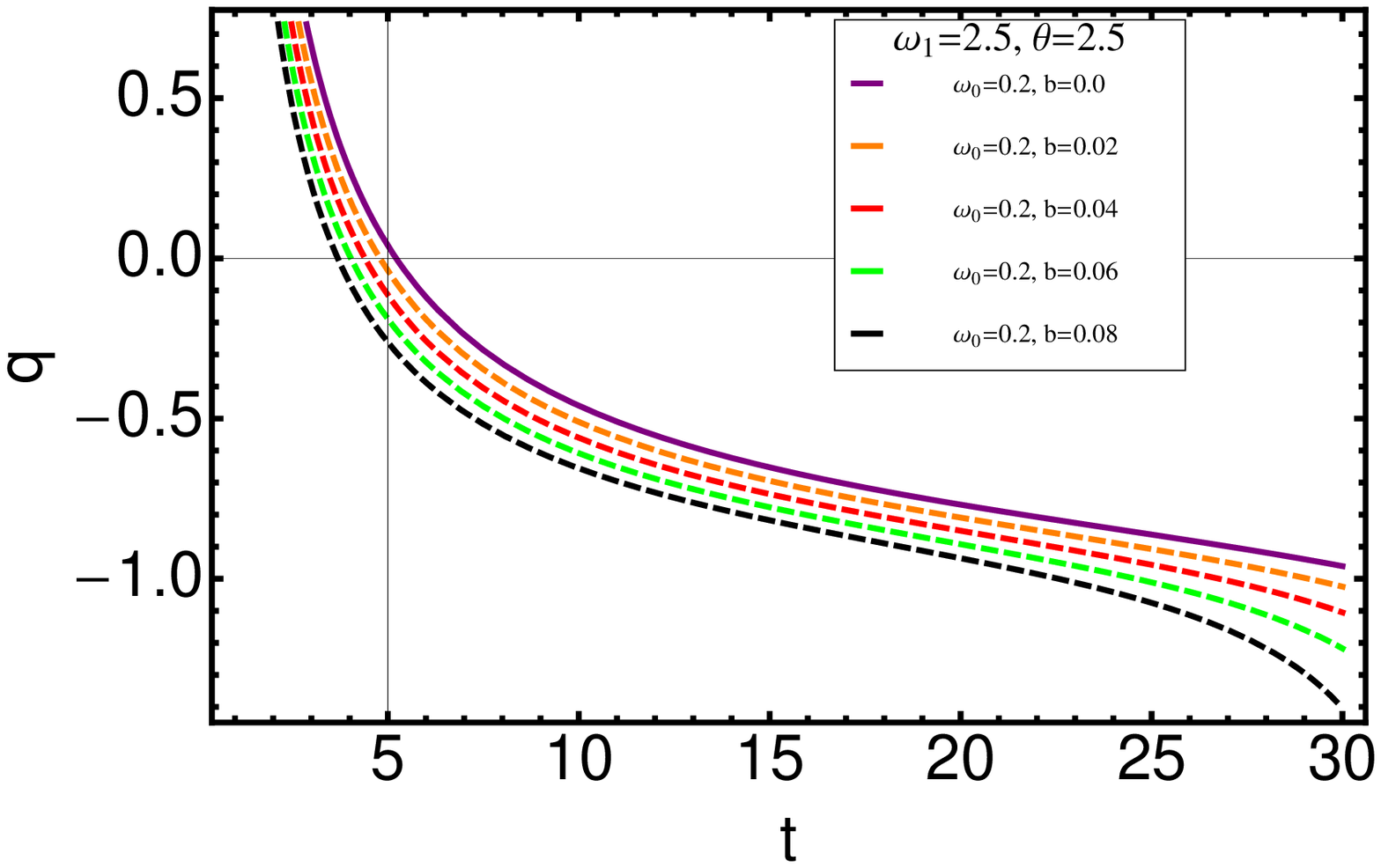} &
\includegraphics[width=50 mm]{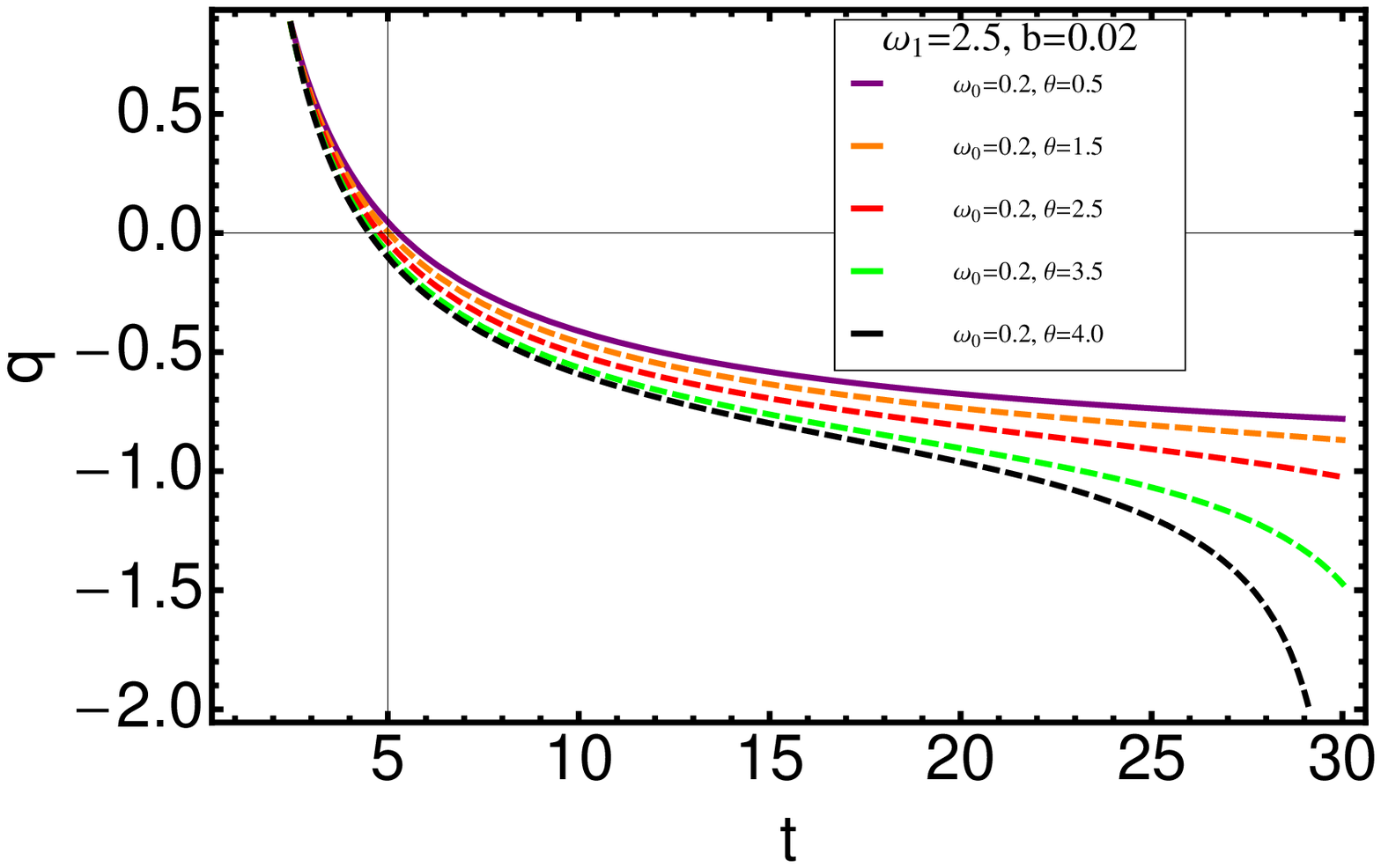}
 \end{array}$
 \end{center}
\caption{Model 1}
 \label{fig:2}
\end{figure}

\begin{figure}[h]
 \begin{center}$
 \begin{array}{cccc}
\includegraphics[width=48 mm]{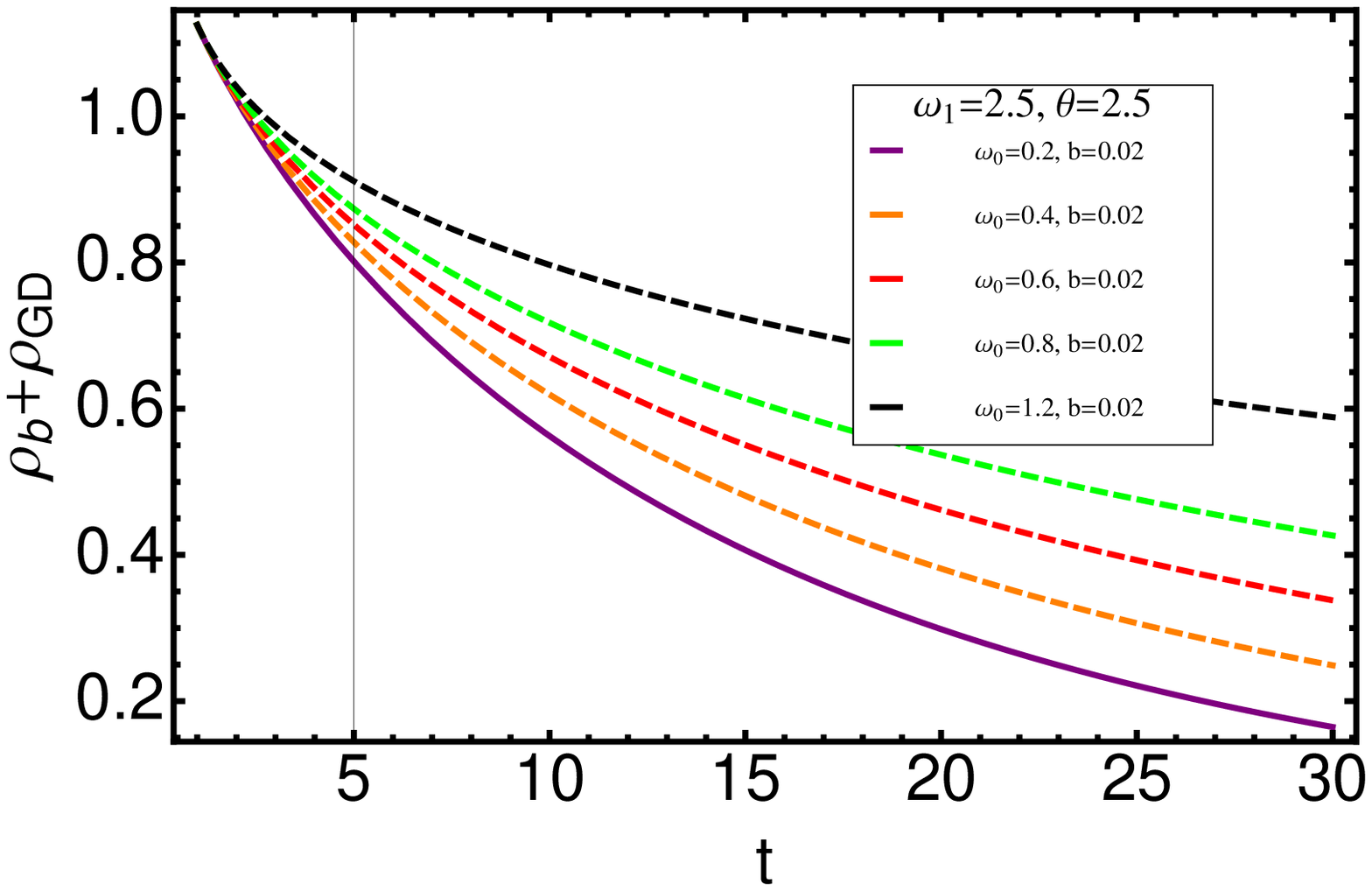} &
\includegraphics[width=48 mm]{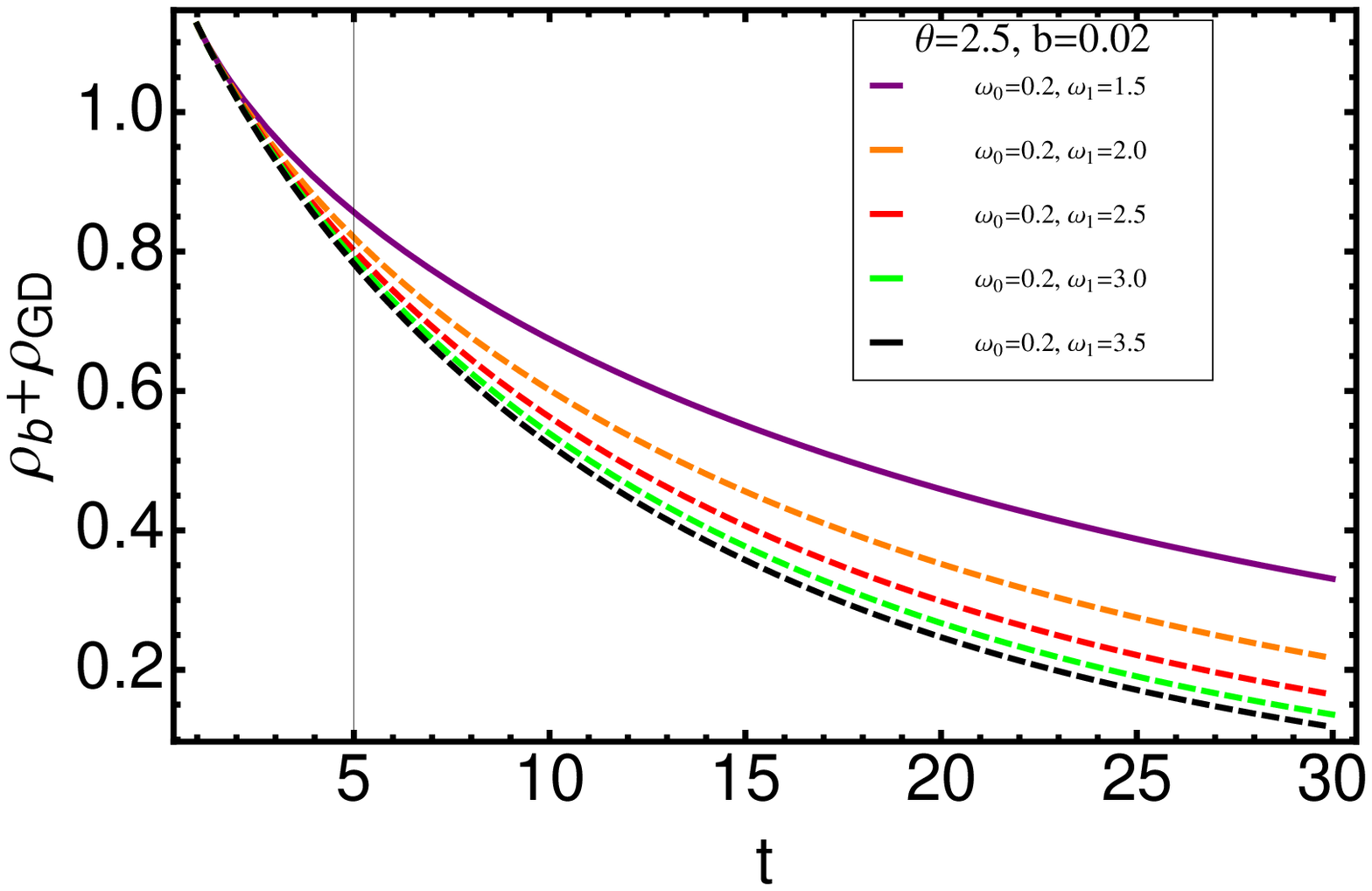}\\
\includegraphics[width=48 mm]{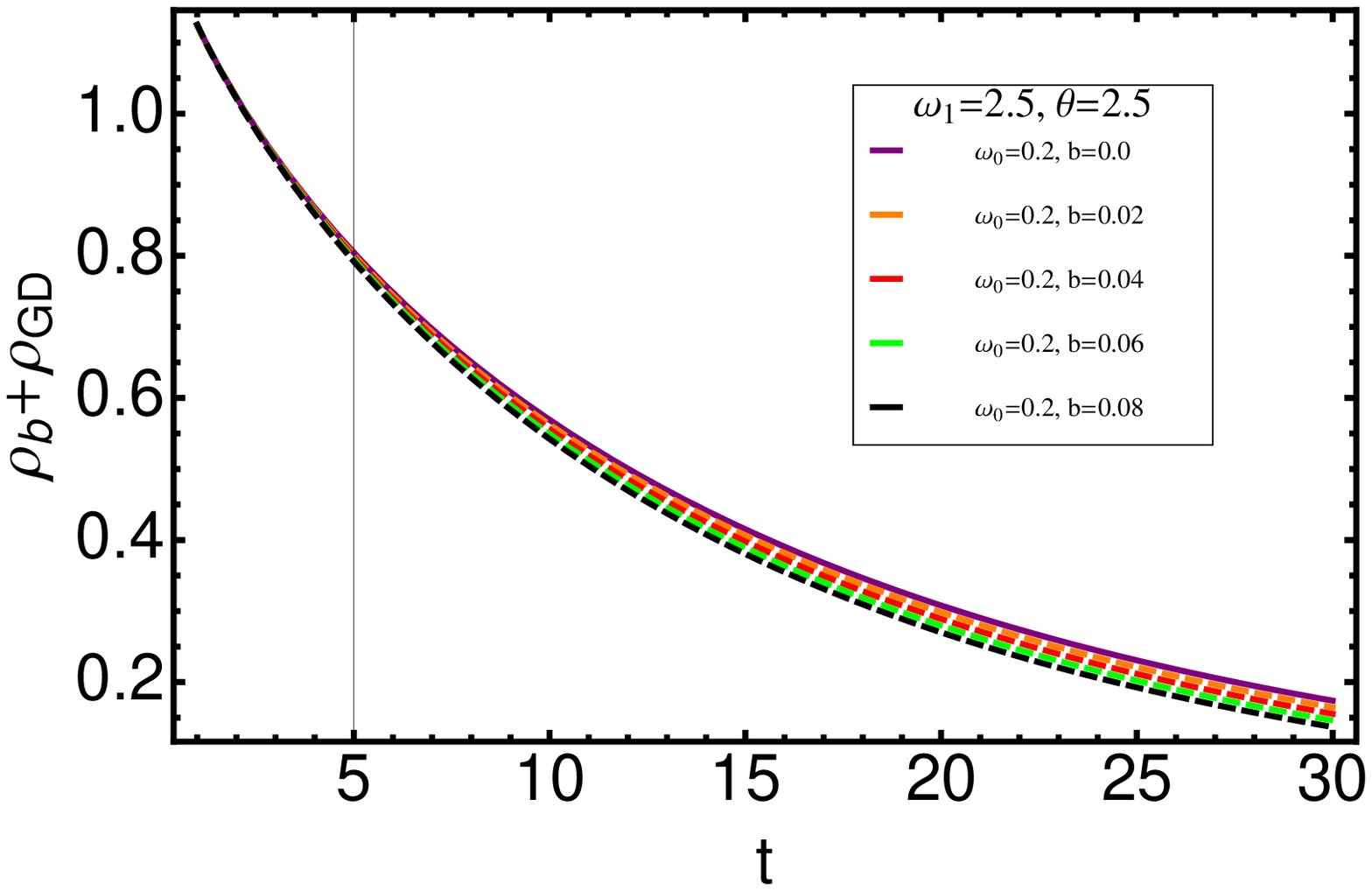} &
\includegraphics[width=48 mm]{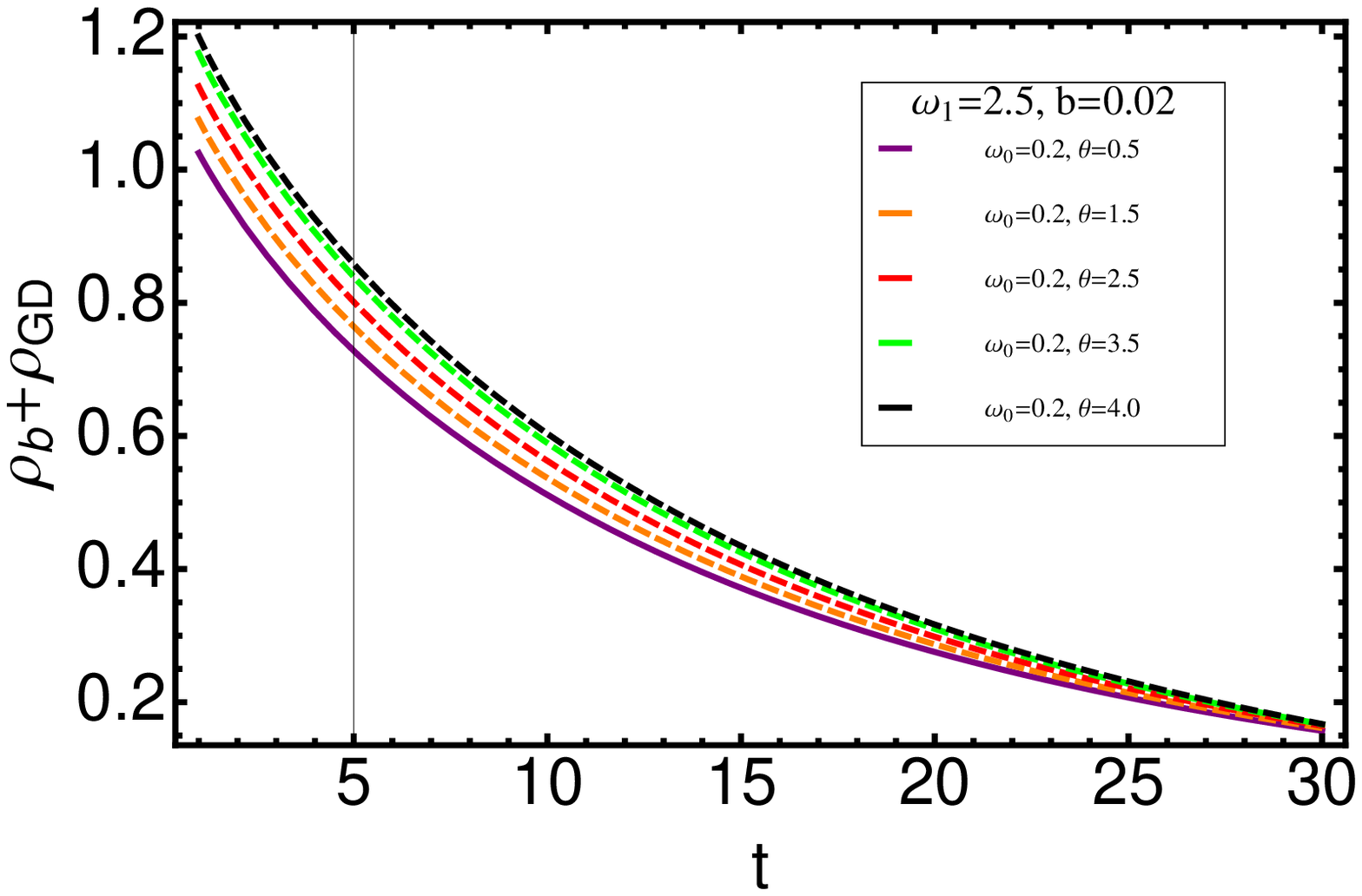}
 \end{array}$
 \end{center}
\caption{Model 1}
 \label{fig:3}
\end{figure}

\begin{figure}[h]
 \begin{center}$
 \begin{array}{cccc}
\includegraphics[width=48 mm]{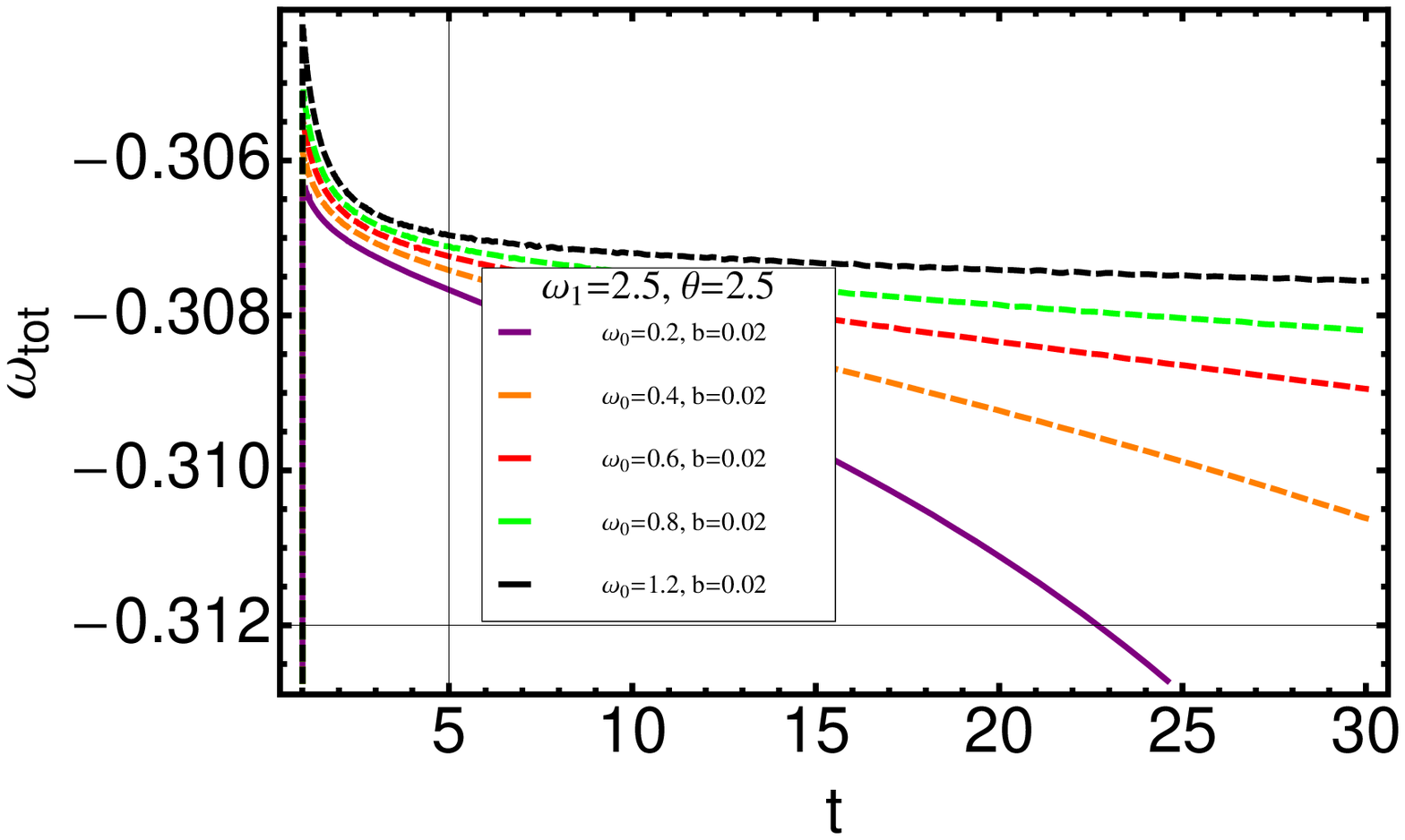} &
\includegraphics[width=48 mm]{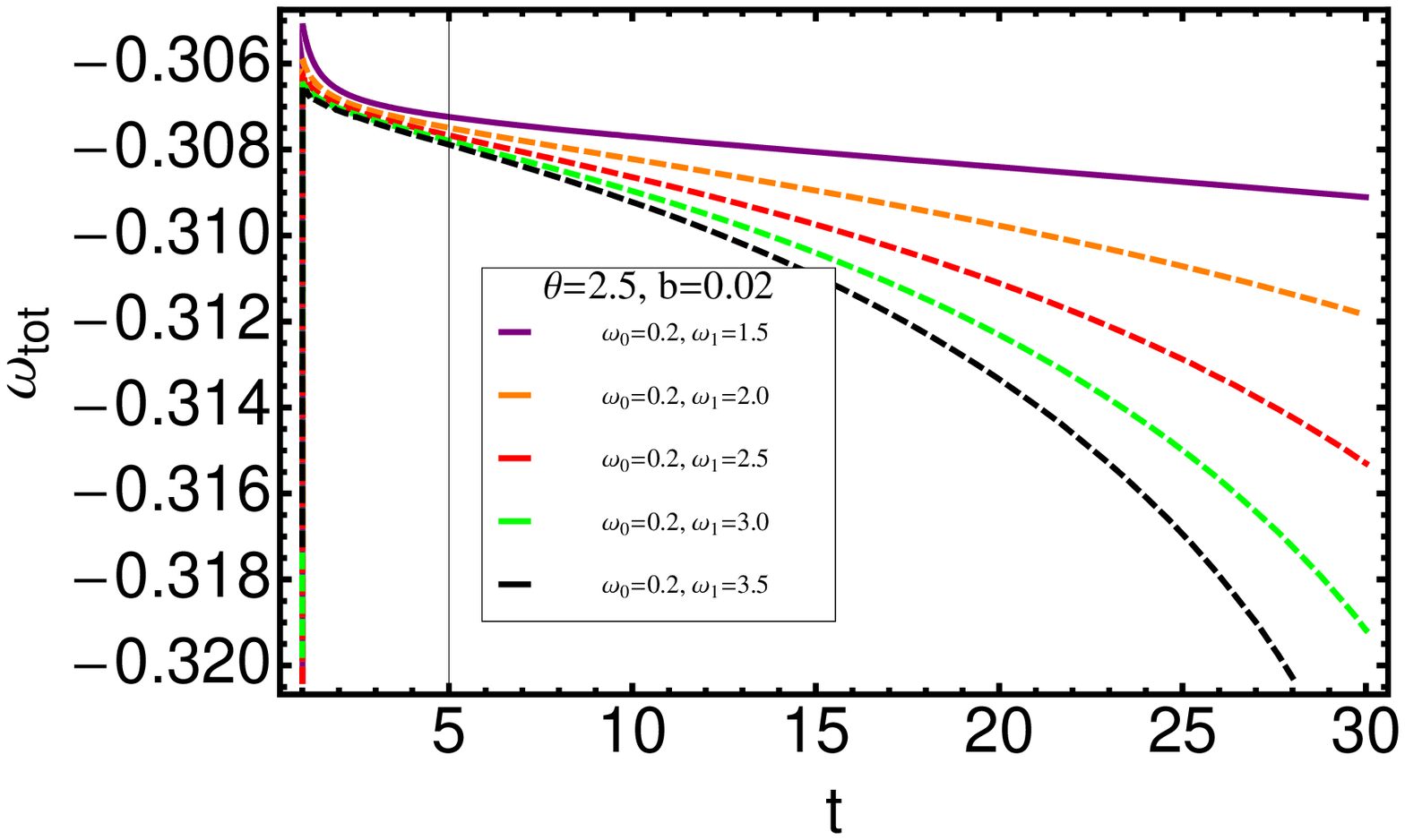}\\
\includegraphics[width=48 mm]{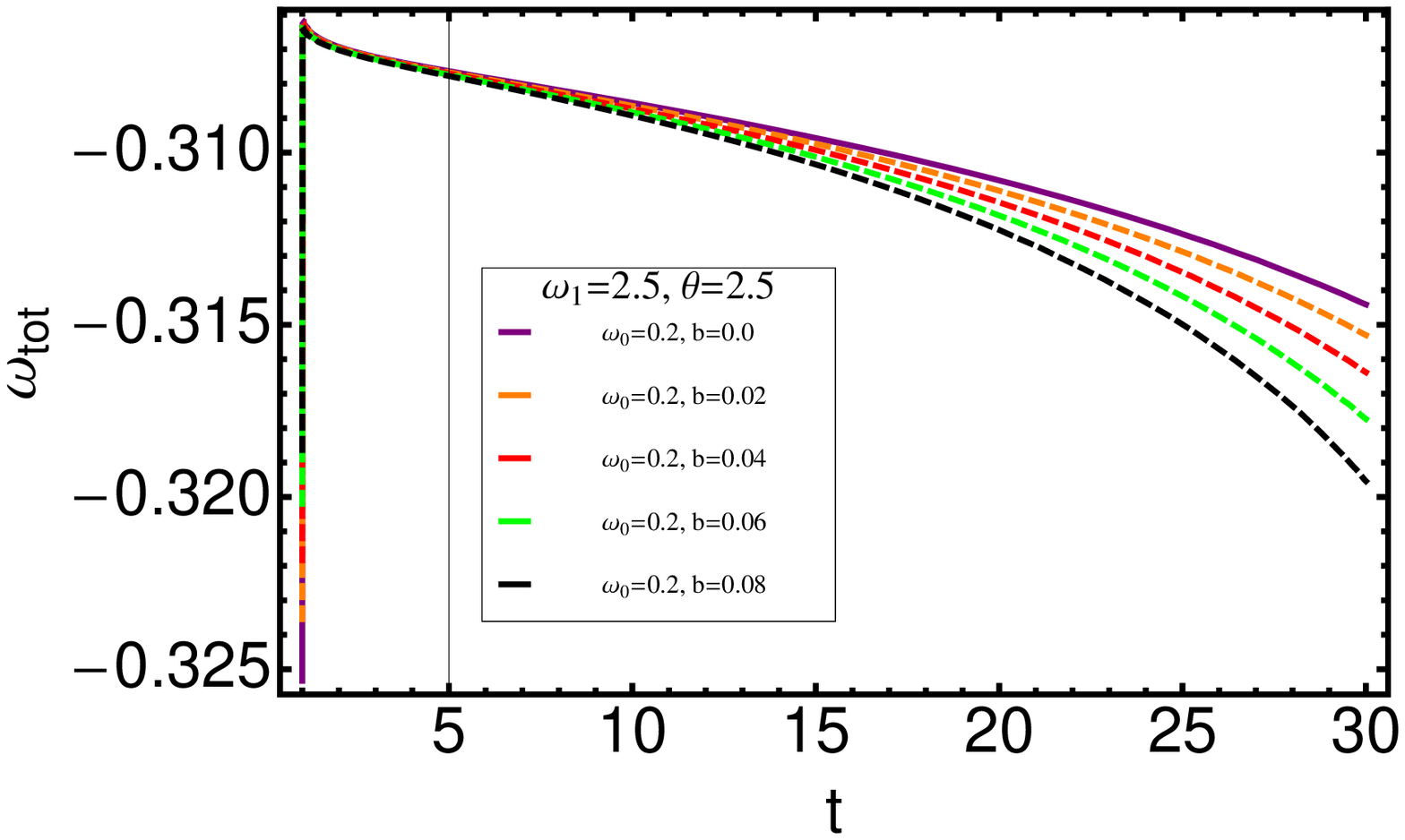} &
\includegraphics[width=48 mm]{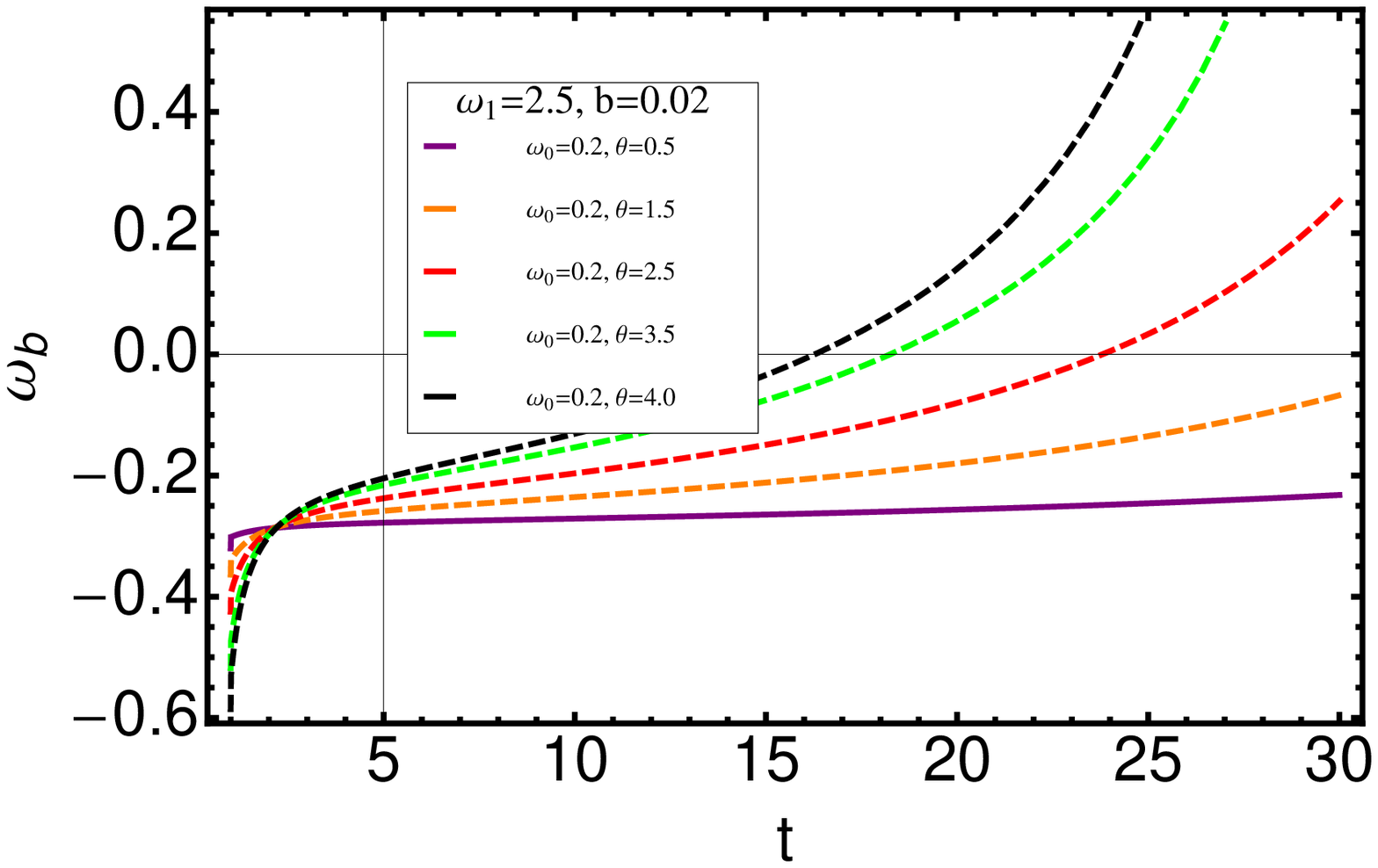}
 \end{array}$
 \end{center}
\caption{Model 1}
 \label{fig:4}
\end{figure}

Using the equations (8) and (14) we can investigate deceleration
parameter numerically. We draw this parameter in the Fig. 2 and find
that $q$ is increasing function of time. In some cases we see that
$q\rightarrow-1$. we know that all forms of matter have the condition $q\geq-1$ which is satisfied.\\
Also, energy density $\rho=\rho_{b}+\rho_{GD}$ illustrated in the
Fig. 3 and total EoS parameter drawn in the Fig. 4.

\subsection*{\large{Model 2}}
In the second model we consider,
\begin{equation}\label{29}
\Lambda(t)=H^{2}+(\rho_{b}+\rho_{GD})e^{-tH}.
\end{equation}
Using this relation in Friedmann equation together with the results
of the previous section we can obtain the following equation
describing dynamics of Hubble parameter,
\begin{equation}\label{s7}
A\dot{H}+H^{2}+BH+D=0,
\end{equation}
where $A$ and $B$ are given by the relations (25) and (26), but $D$
is defined as the following,
\begin{equation}\label{31}
D=-\frac{\rho_{b}}{2}\left(e^{-tH}-8 (\omega_{0}-b) \pi G(t)\right).
\end{equation}
Therefore, we can obtain the following equation,
\begin{equation}\label{s8}
\dot{G}=-\frac{2H\dot{H}+(\dot{\rho}_{b}+\theta\dot{H})e^{-tH} -(\rho_{b}+\theta H) H \dot{H}e^{-tH}} {8\pi (\omega(t)\rho_{b}+\theta H)}.
\end{equation}

\begin{figure}[h]
 \begin{center}$
 \begin{array}{cccc}
\includegraphics[width=50 mm]{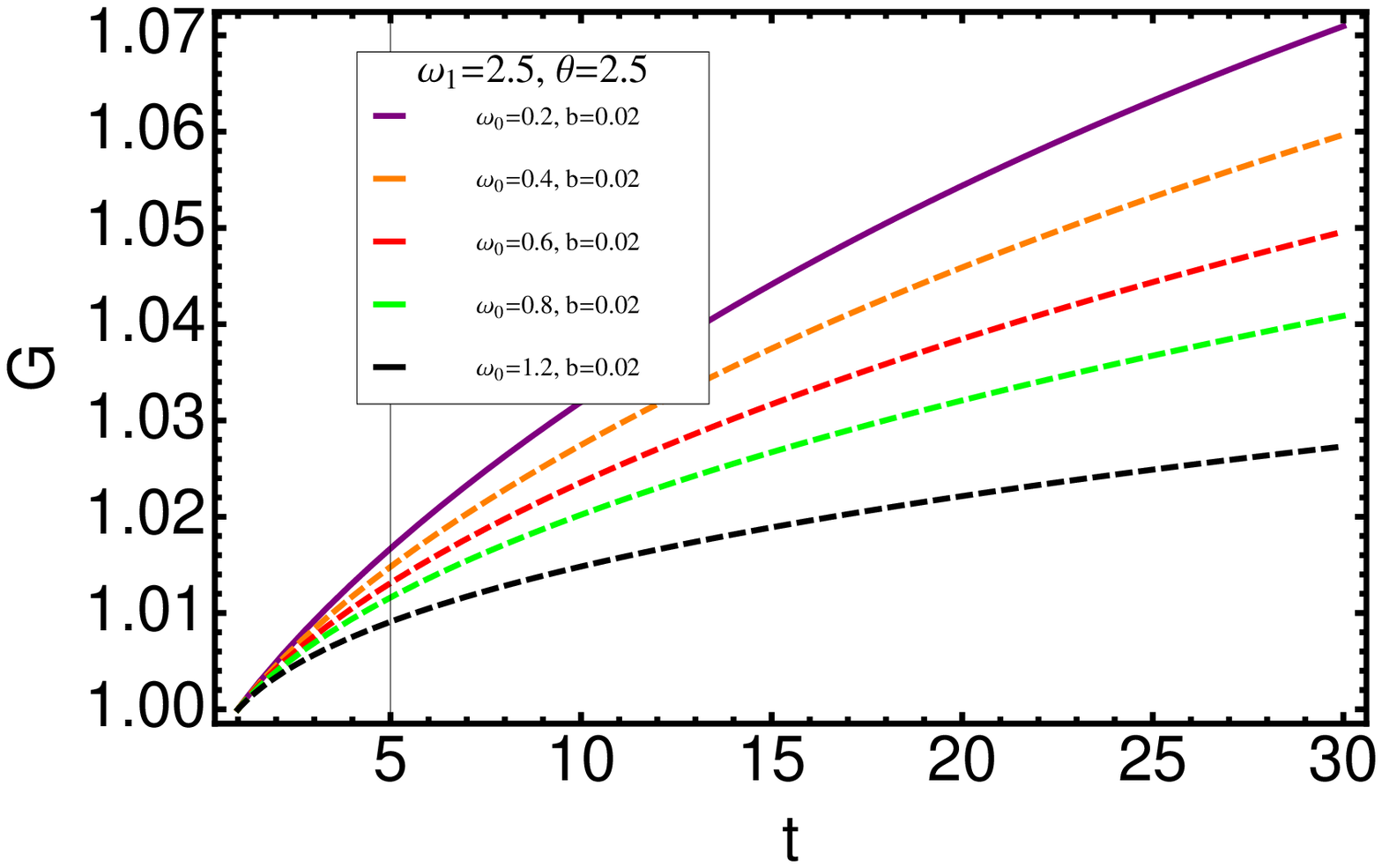} &
\includegraphics[width=50 mm]{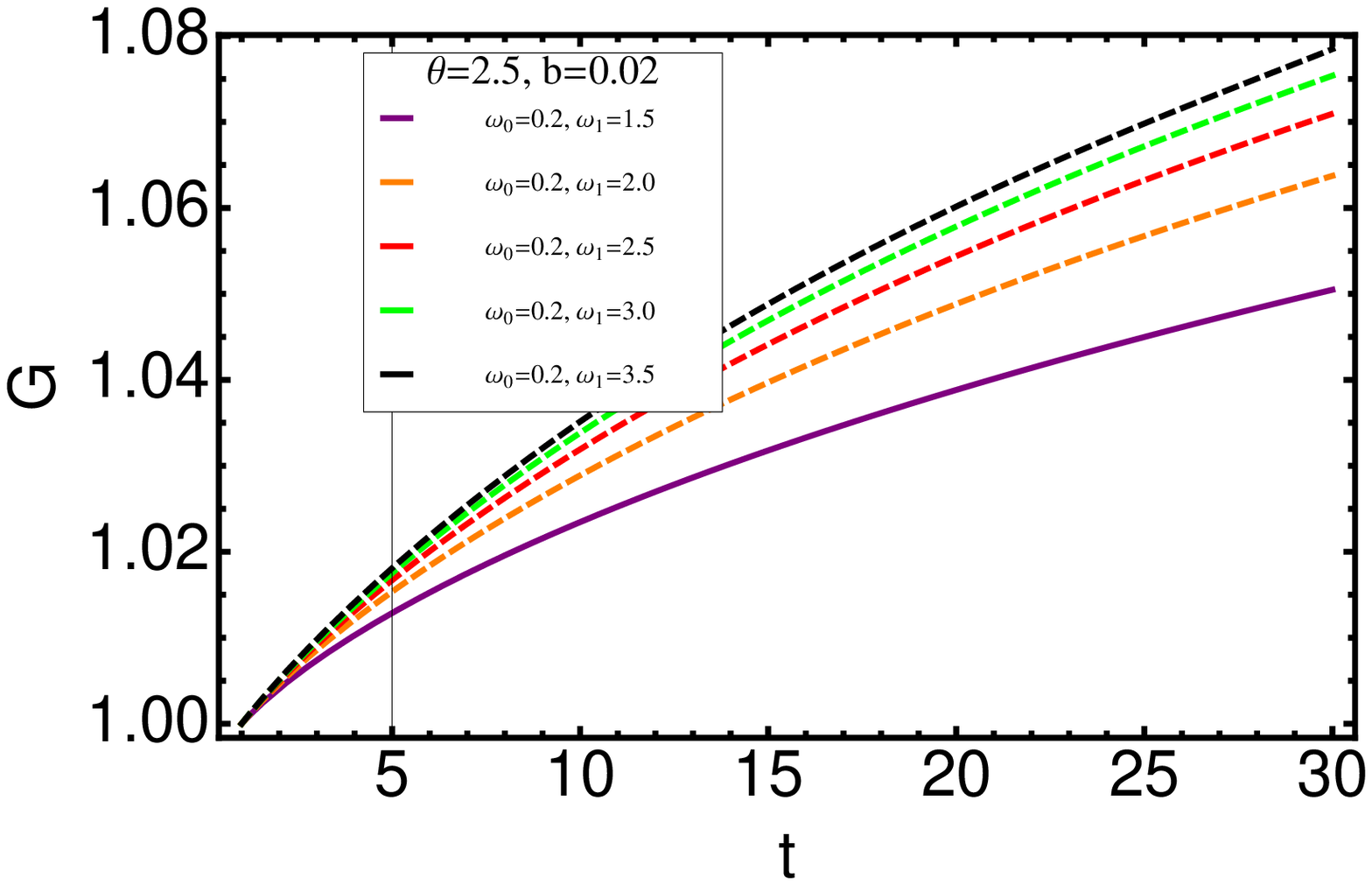}\\
\includegraphics[width=50 mm]{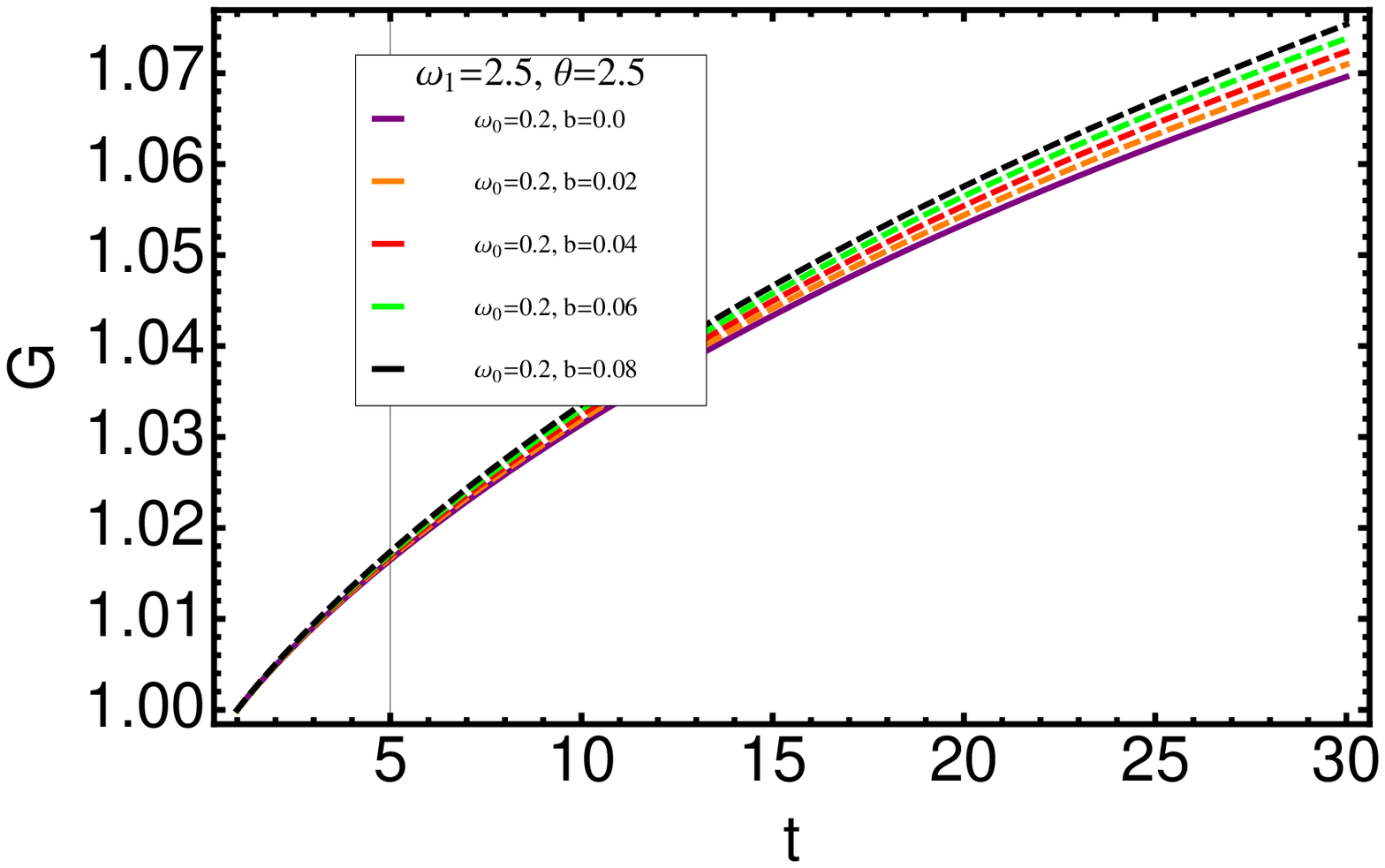} &
\includegraphics[width=50 mm]{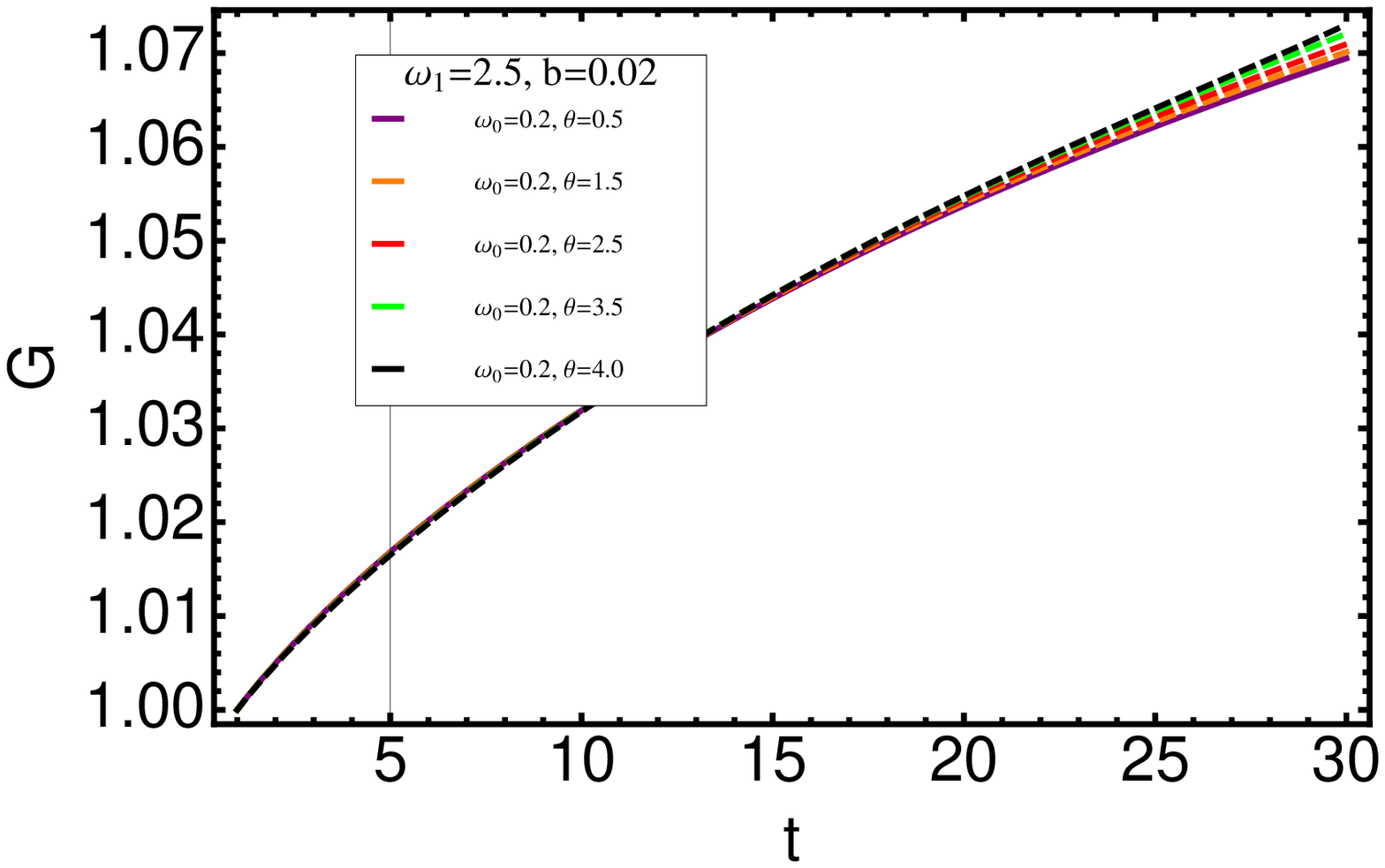}
 \end{array}$
 \end{center}
\caption{Model 2}
 \label{fig:5}
\end{figure}

\begin{figure}[h]
 \begin{center}$
 \begin{array}{cccc}
\includegraphics[width=50 mm]{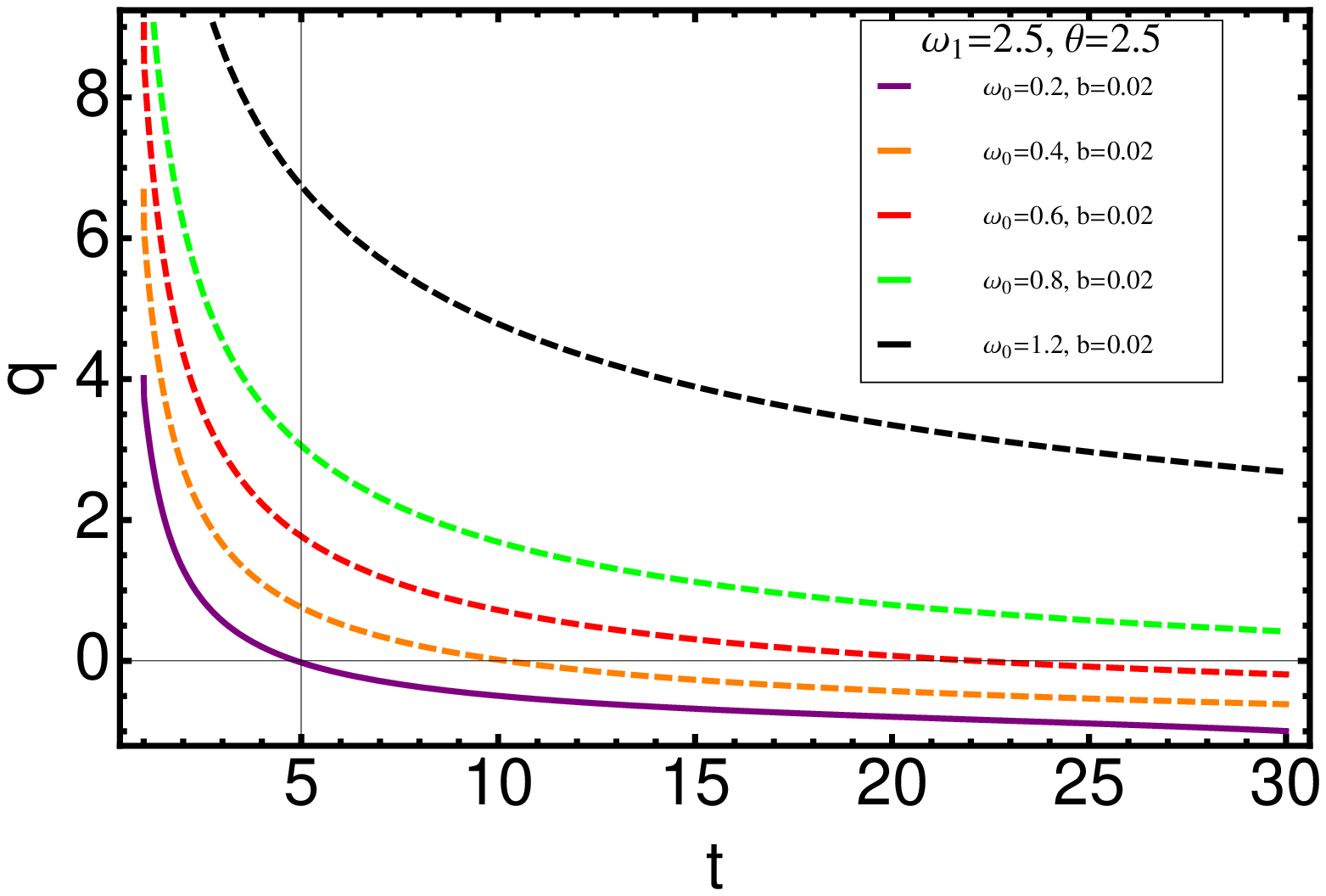} &
\includegraphics[width=50 mm]{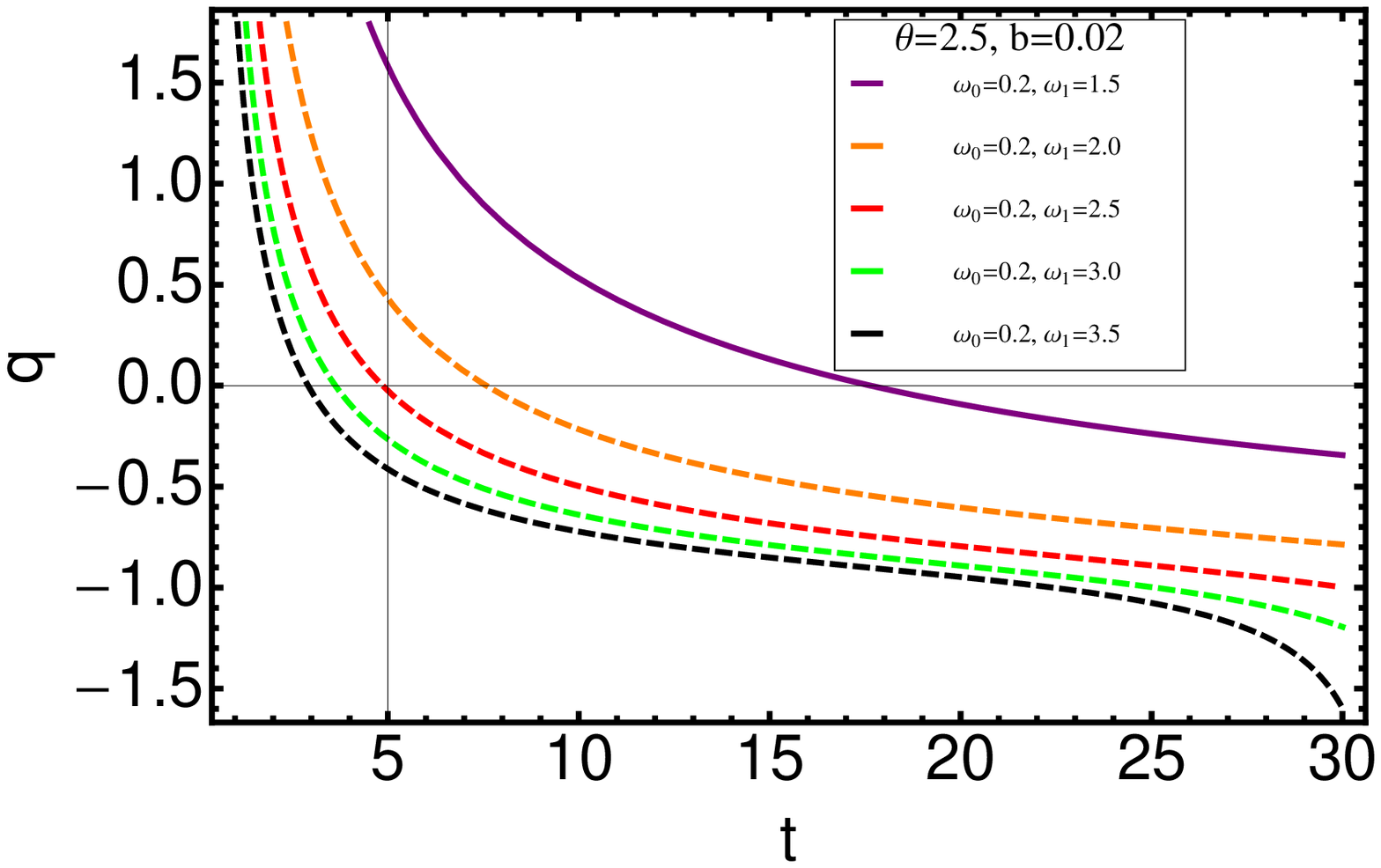}\\
\includegraphics[width=50 mm]{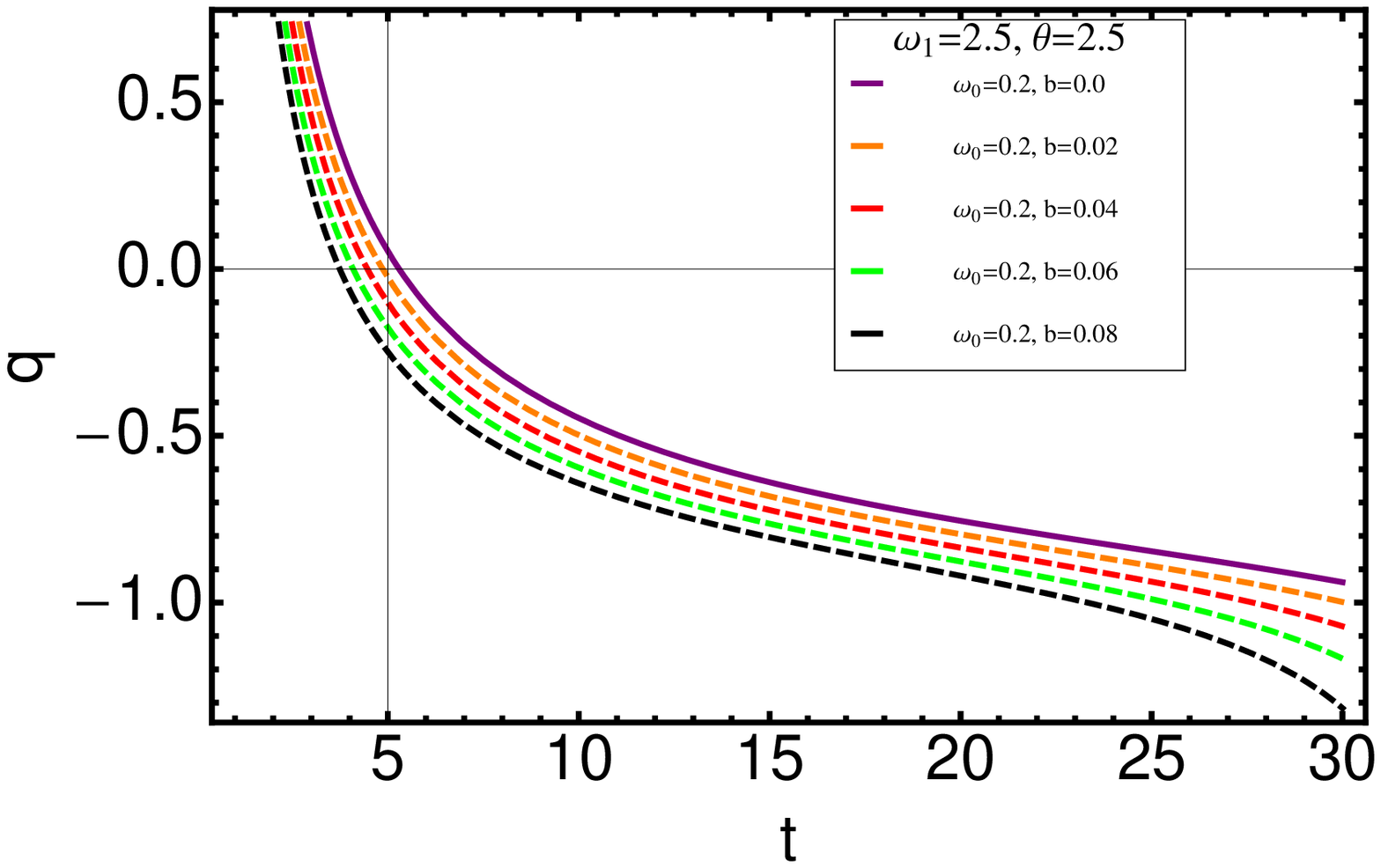} &
\includegraphics[width=50 mm]{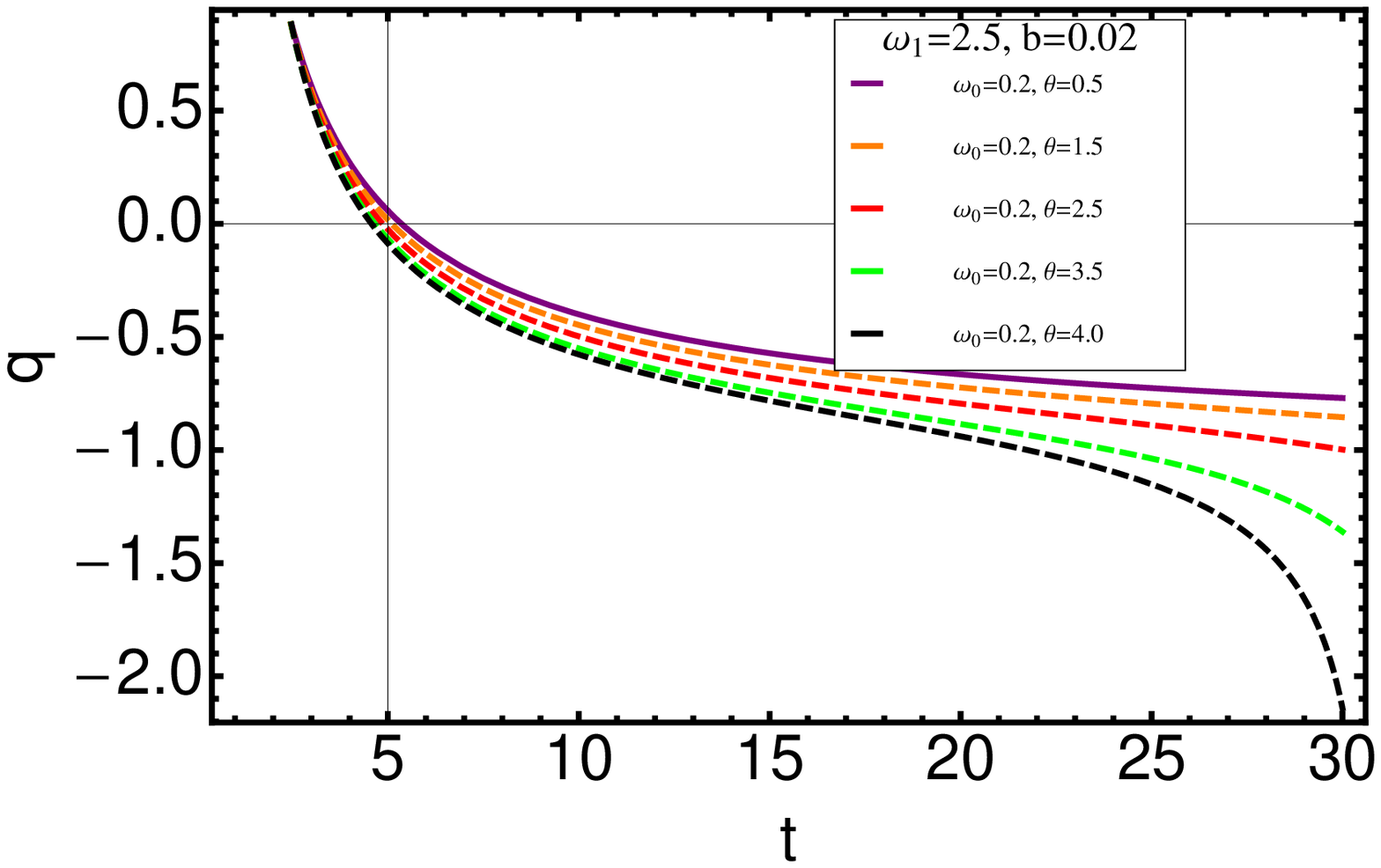}
 \end{array}$
 \end{center}
\caption{Model 2}
 \label{fig:6}
\end{figure}

\begin{figure}[h]
 \begin{center}$
 \begin{array}{cccc}
\includegraphics[width=48 mm]{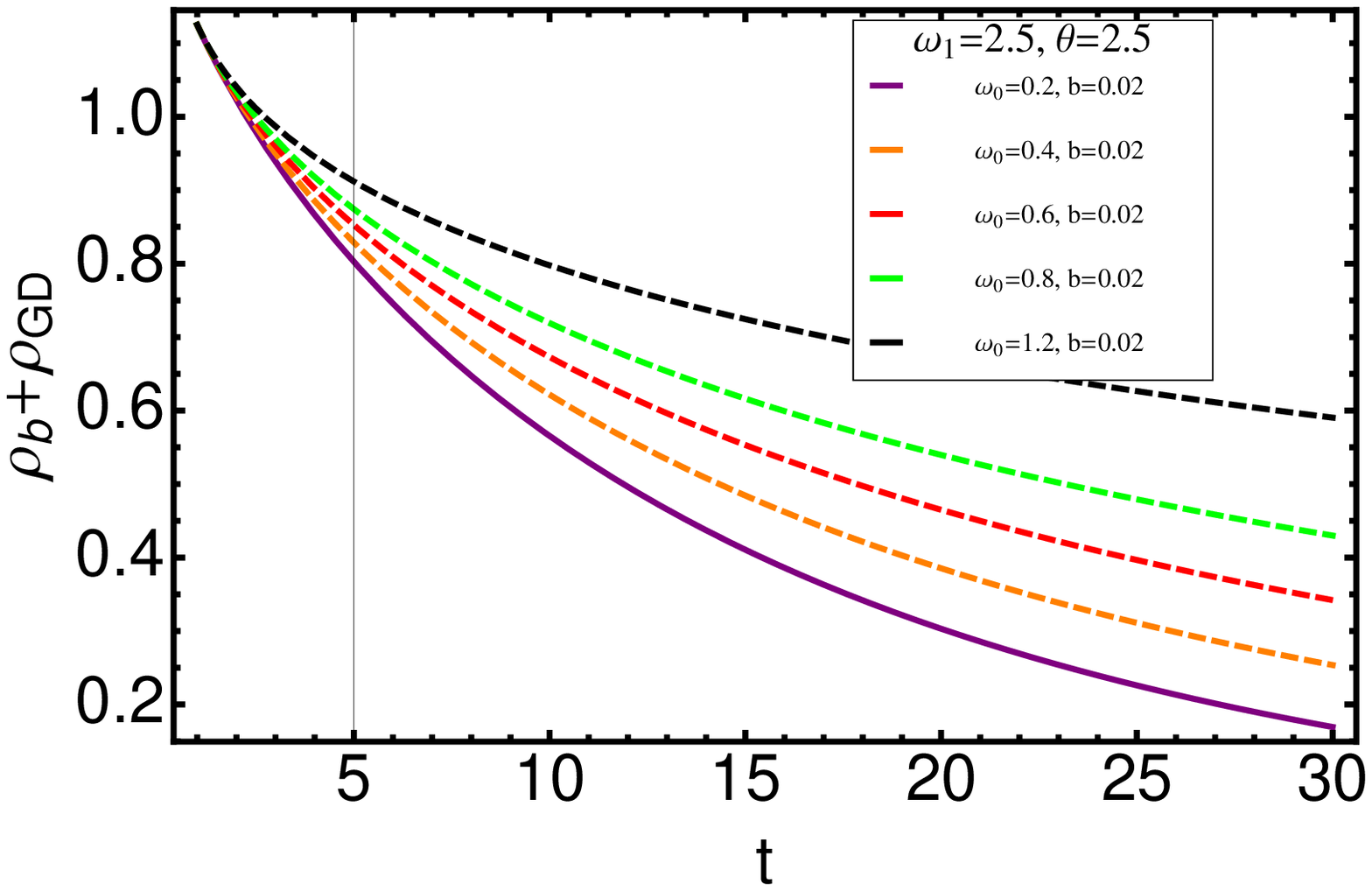} &
\includegraphics[width=48 mm]{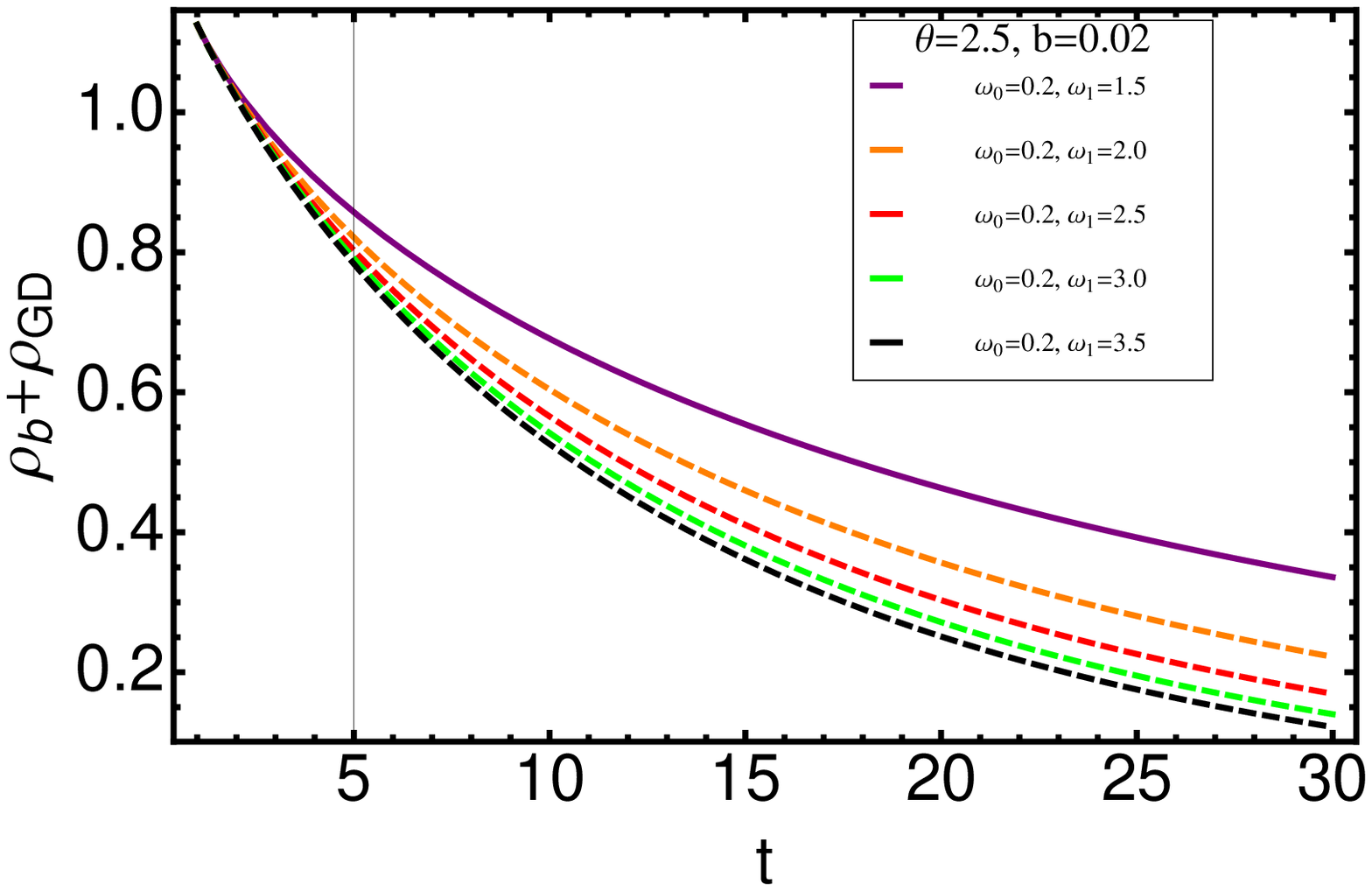}\\
\includegraphics[width=48 mm]{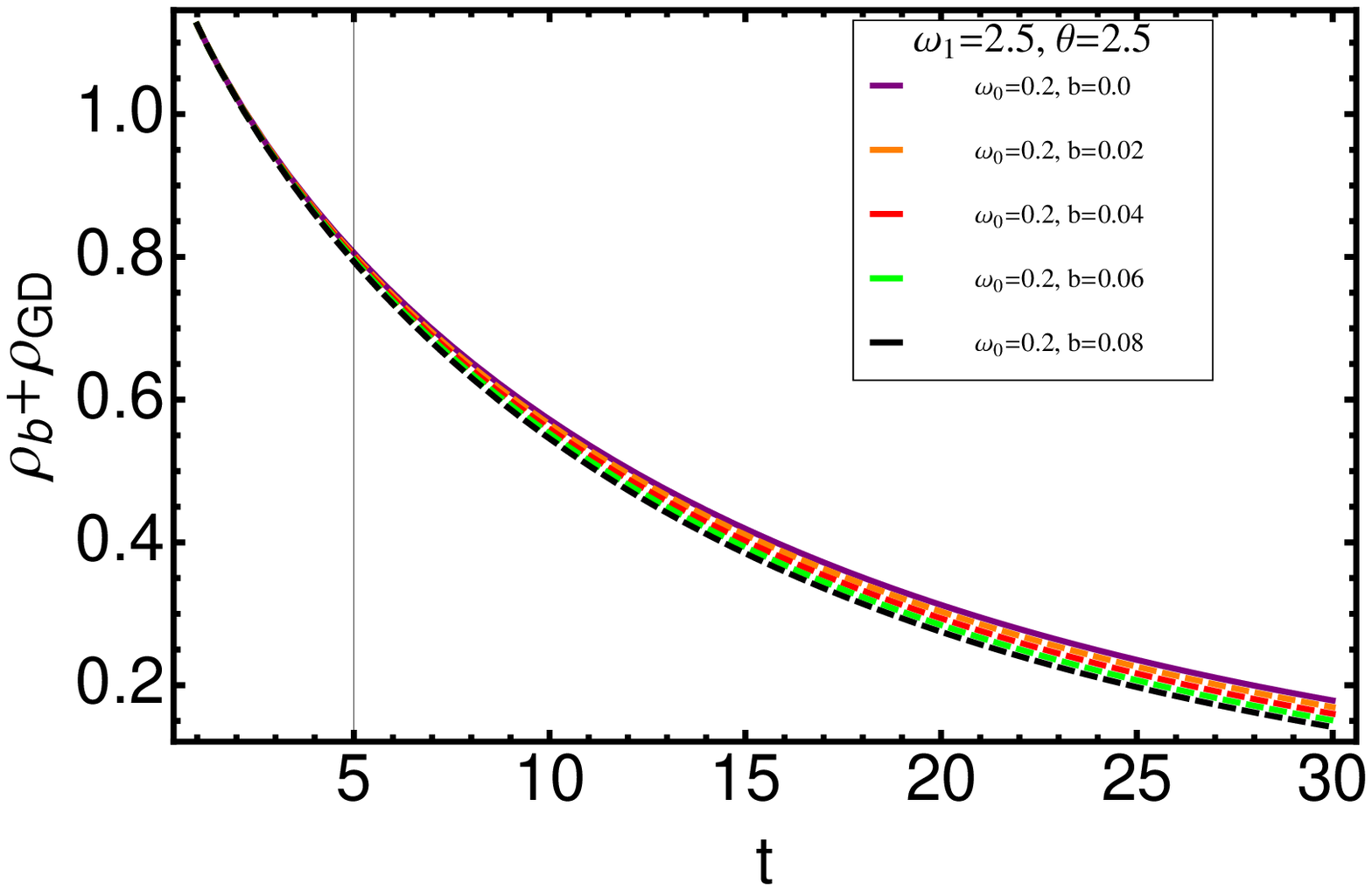} &
\includegraphics[width=48 mm]{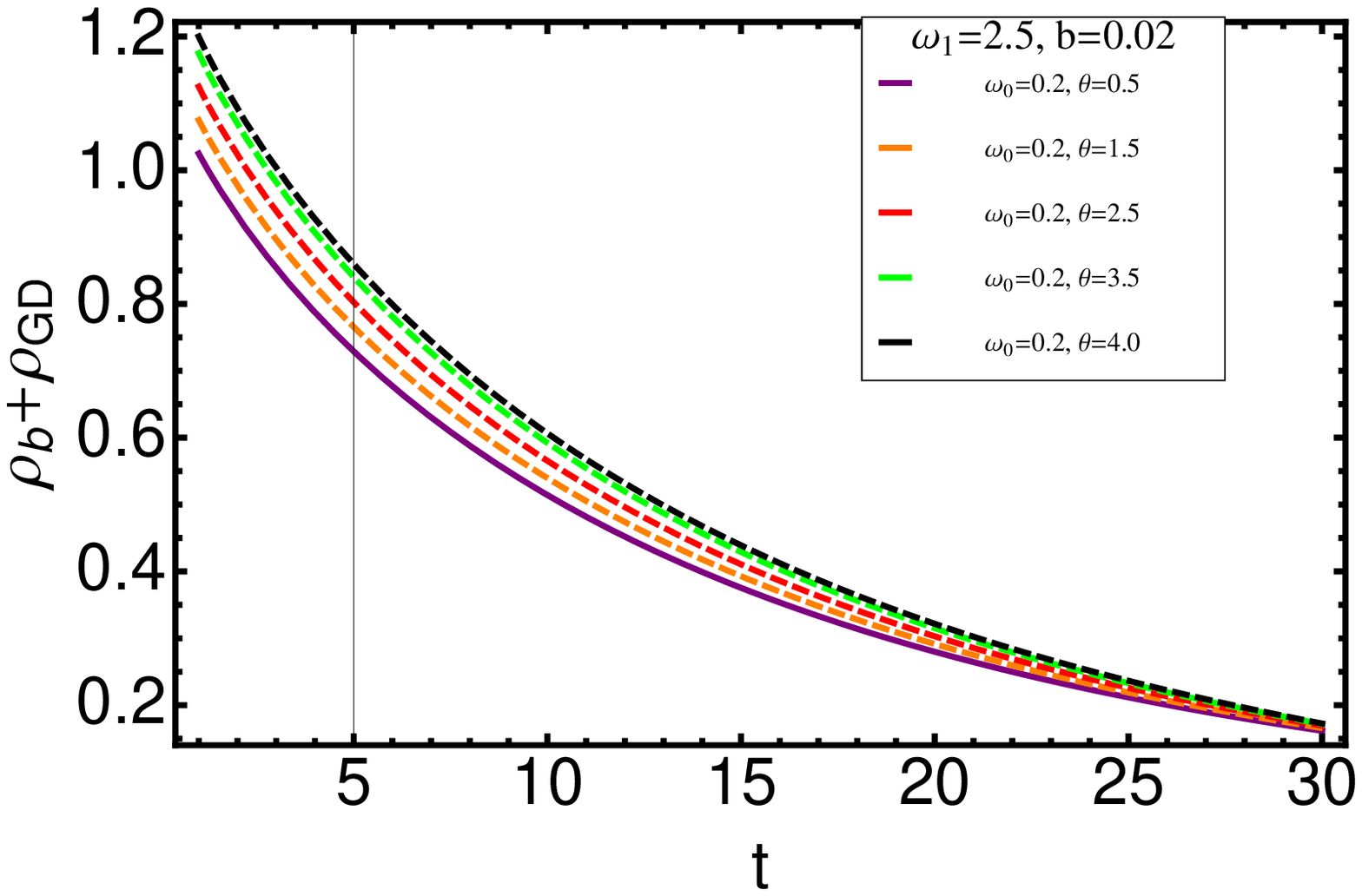}
 \end{array}$
 \end{center}
\caption{Model 2}
 \label{fig:7}
\end{figure}

\begin{figure}[h]
 \begin{center}$
 \begin{array}{cccc}
\includegraphics[width=48 mm]{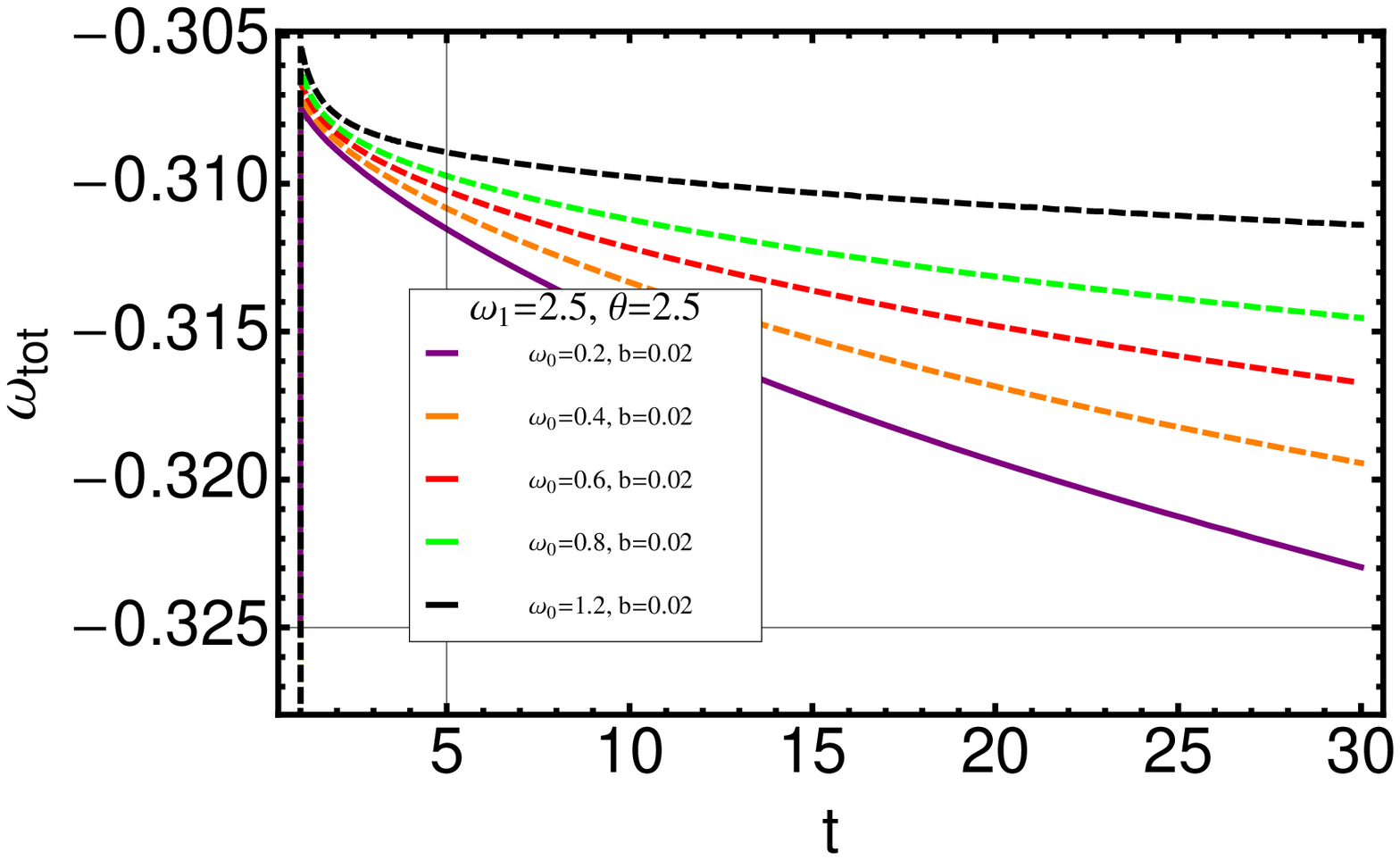} &
\includegraphics[width=48 mm]{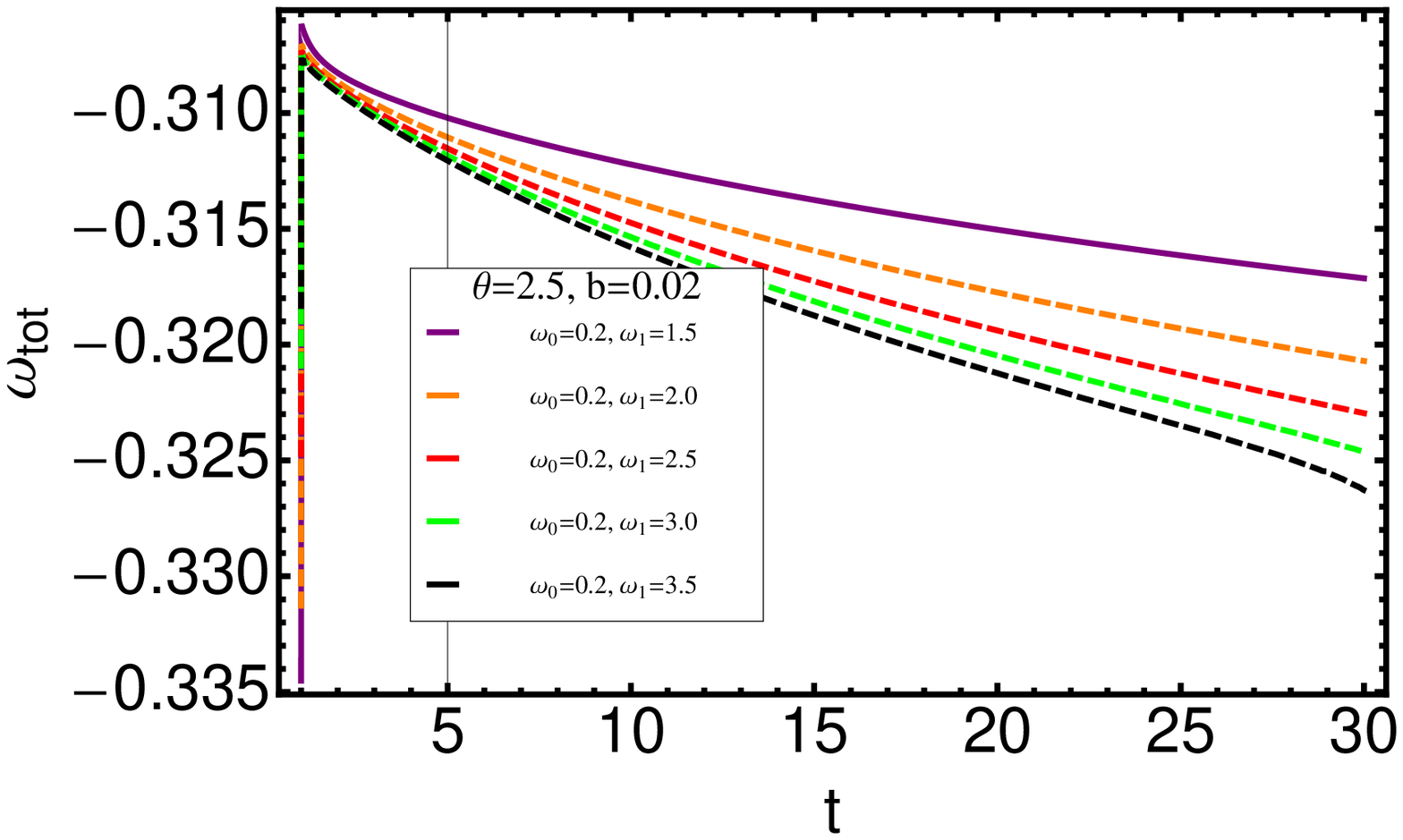}\\
\includegraphics[width=48 mm]{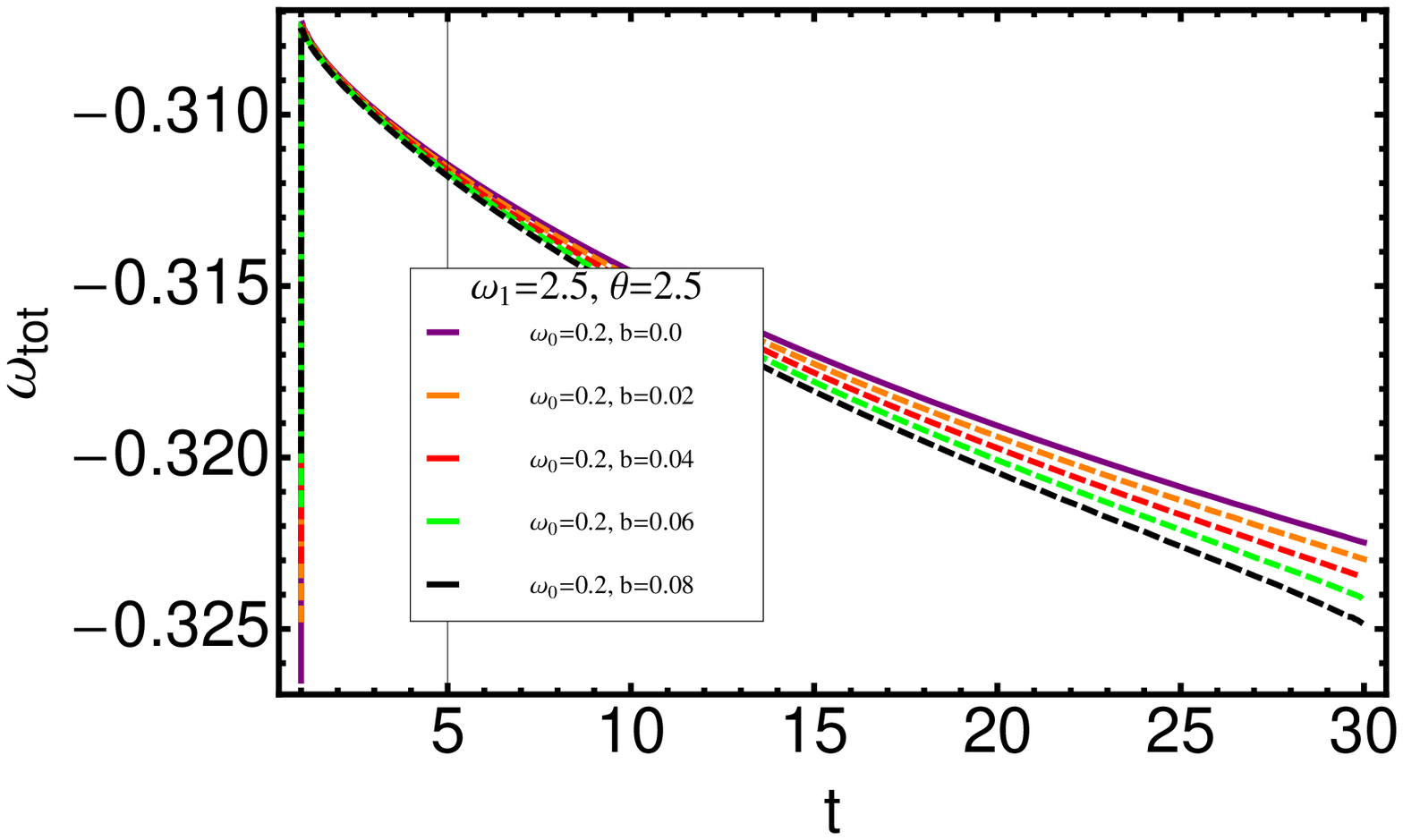} &
\includegraphics[width=48 mm]{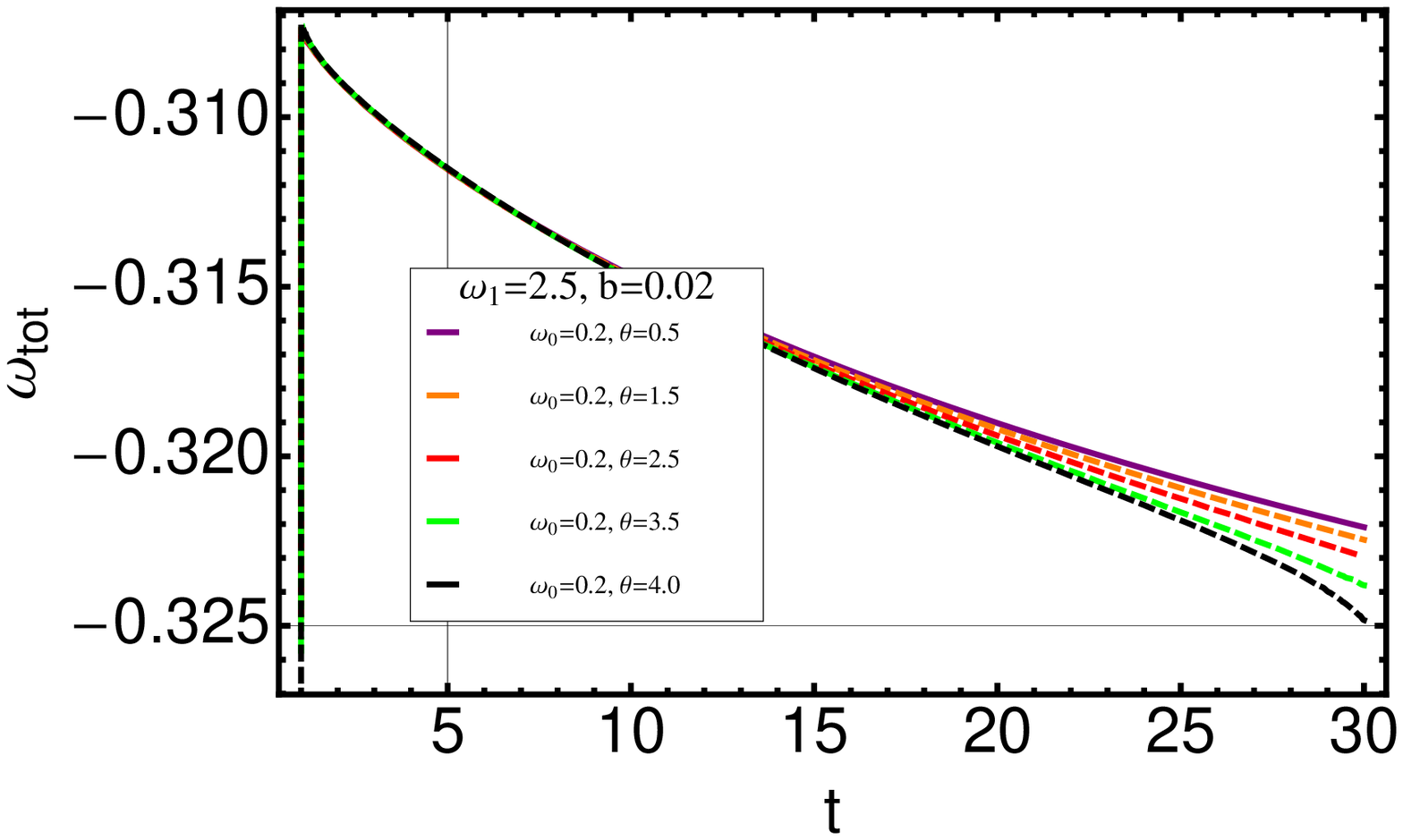}
 \end{array}$
 \end{center}
\caption{Model 2}
 \label{fig:8}
\end{figure}

We solved this equation numerically and find $G$ as an increasing
function of time (see Fig. 5). It shows that interaction parameters increase $G$.
It is clear that variation of $\theta$ at the initial time has no important effect on $G$.
Also, similar to the model 1 we see that increasing $\omega_{0}$ decreases $G$ but increasing $\omega_{1}$ increases the value of $G$.\\
Using the equations (8) and (14) we can investigate deceleration
parameter numerically. We draw this parameter in the Fig. 6, and
similar to the previous model, find that $q$ is increasing function
of time which shows that $q\geq-1$ satisfied. We can choose
parameters as $\omega_{0}=0.2$, $\omega_{1}=2.5$, $b=0.02$ and
$\theta=2.5$ to obtain $q\rightarrow-1$ at the late time.\\
Also, energy density $\rho=\rho_{b}+\rho_{GD}$ illustrated in the
Fig. 7 and total EoS parameter drawn in the Fig. 8. We can see that
$\rho$ and $\omega_{tot}$ are decreasing function of time.

\subsection*{\large{Model 3}}
In the third model we consider,
\begin{equation}\label{33}
\Lambda(t)=t^{-2}+(\rho_{b}-\rho_{GD})e^{-tH}.
\end{equation}
Using this relation in Friedmann equation together with the results
of the previous section we can study dynamics of Hubble parameter,
\begin{equation}\label{s9}
A\dot{H}+\frac{3}{2}H^{2}+EH+F=0.
\end{equation}
where $A$ is given by the equation (25), but $E$ and $F$ are given
by the following relations,
\begin{equation}\label{35}
E=\frac{\theta}{2}\left(e^{-tH}-8 (1+b) \pi G(t)\right),
\end{equation}
and,
\begin{equation}\label{36}
F=-\frac{t^{-2}}{2}-\frac{\rho_{b}}{2}\left(e^{-tH}-8 (\omega_{0}-b)
\pi G(t)\right).
\end{equation}
For $\dot{G}$ we can give the following expression,
\begin{equation}\label{s10}
\dot{G}=-\frac{2t^{-2}-(\dot{\rho}_{b}+\theta\dot{H})e^{-tH} -(\rho_{b}-\theta H)H\dot{H} e^{-tH}} {8\pi (\omega(t)\rho_{b}+\theta H)}.
\end{equation}

\begin{figure}[h]
 \begin{center}$
 \begin{array}{cccc}
\includegraphics[width=48 mm]{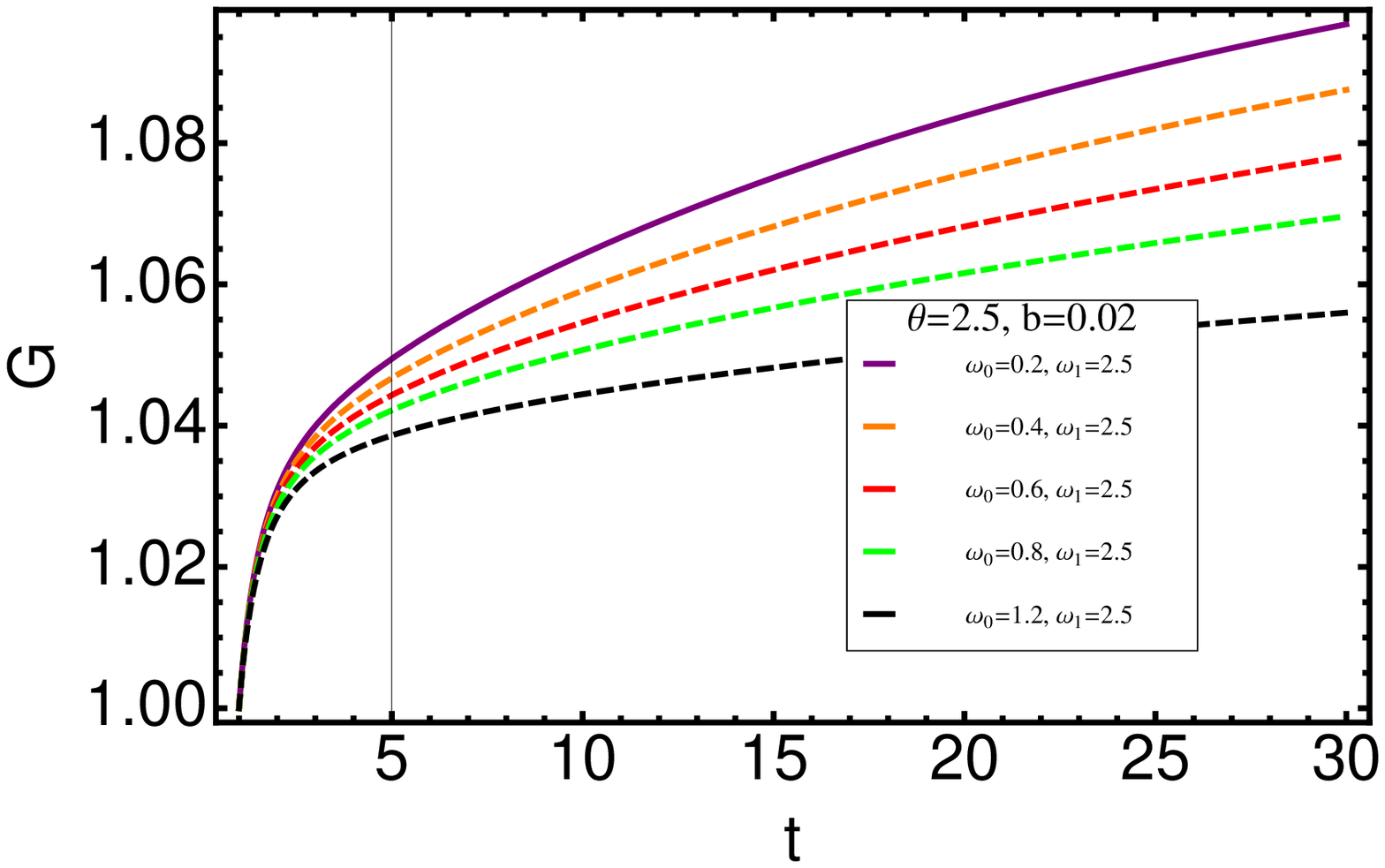} &
\includegraphics[width=48 mm]{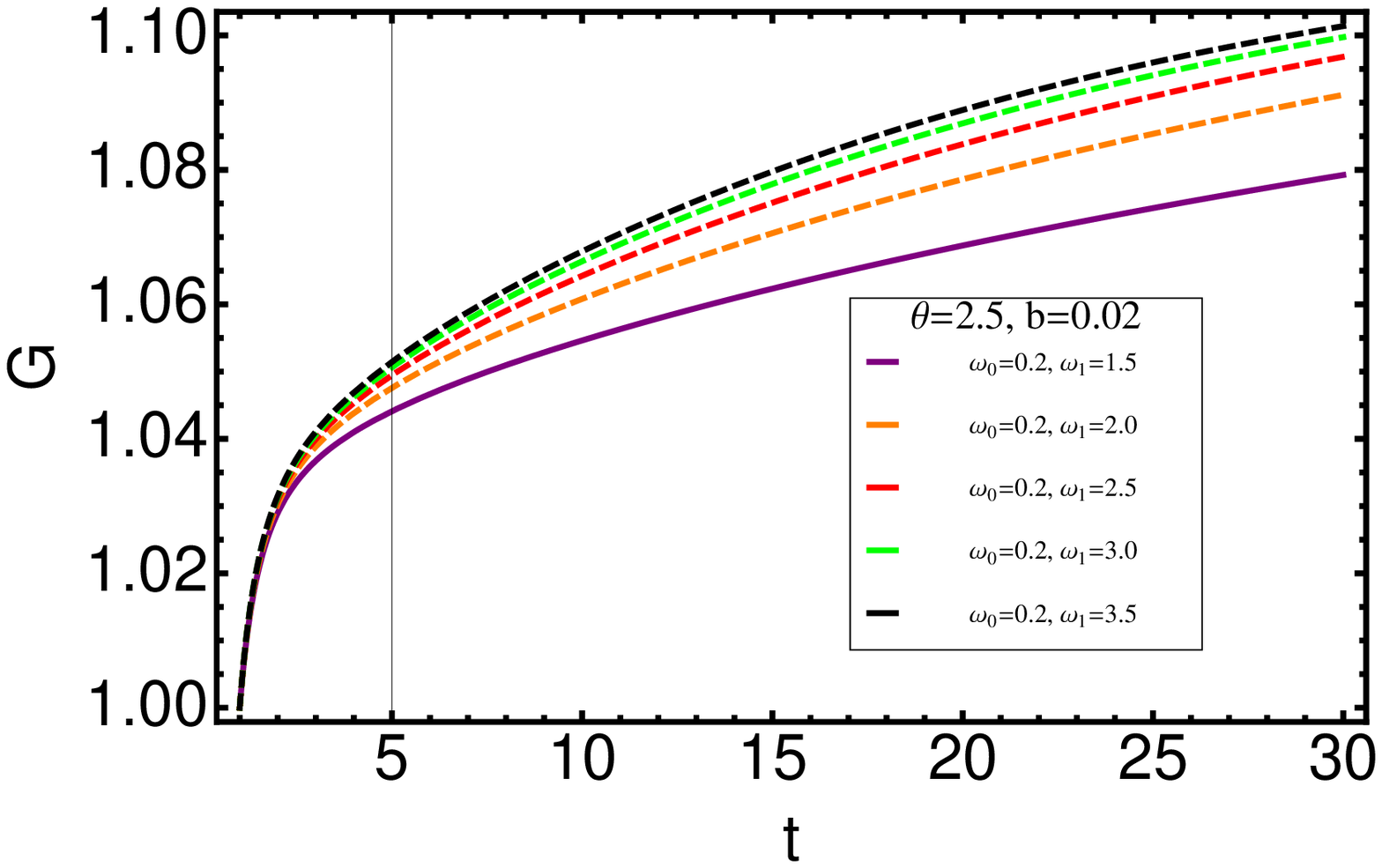}\\
\includegraphics[width=48 mm]{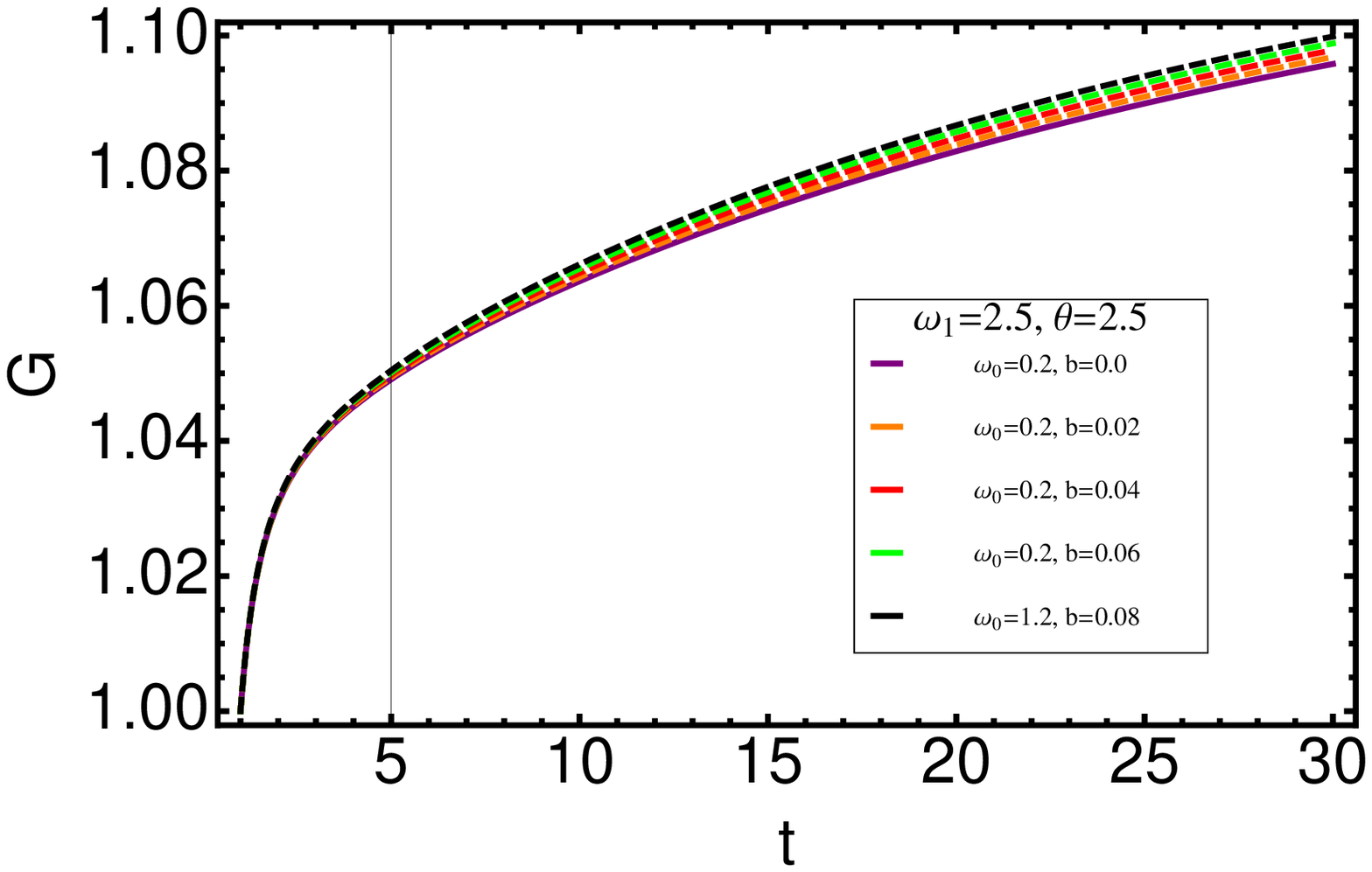} &
\includegraphics[width=48 mm]{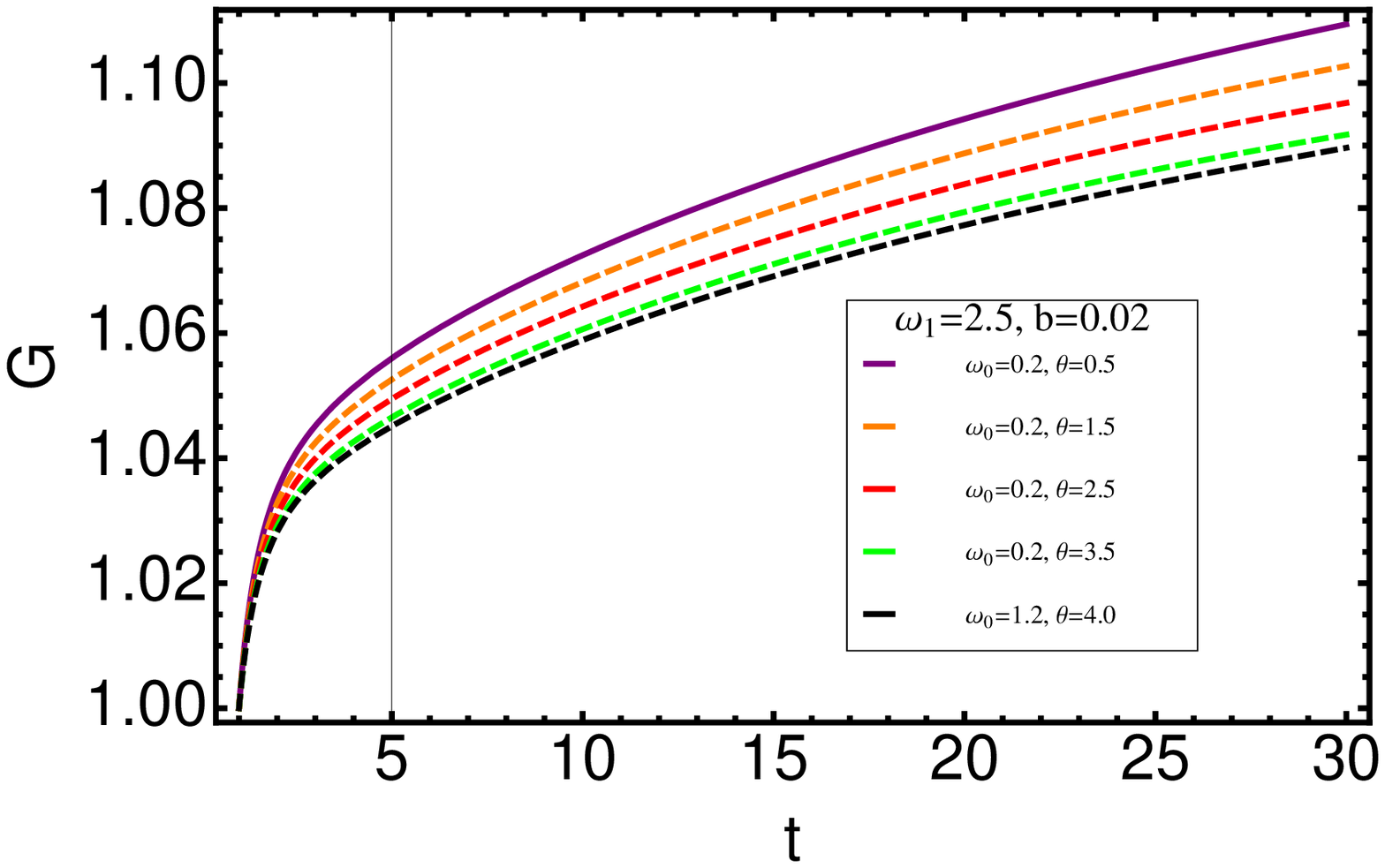}
 \end{array}$
 \end{center}
\caption{Model 3}
 \label{fig:9}
\end{figure}

\begin{figure}[h]
 \begin{center}$
 \begin{array}{cccc}
\includegraphics[width=48 mm]{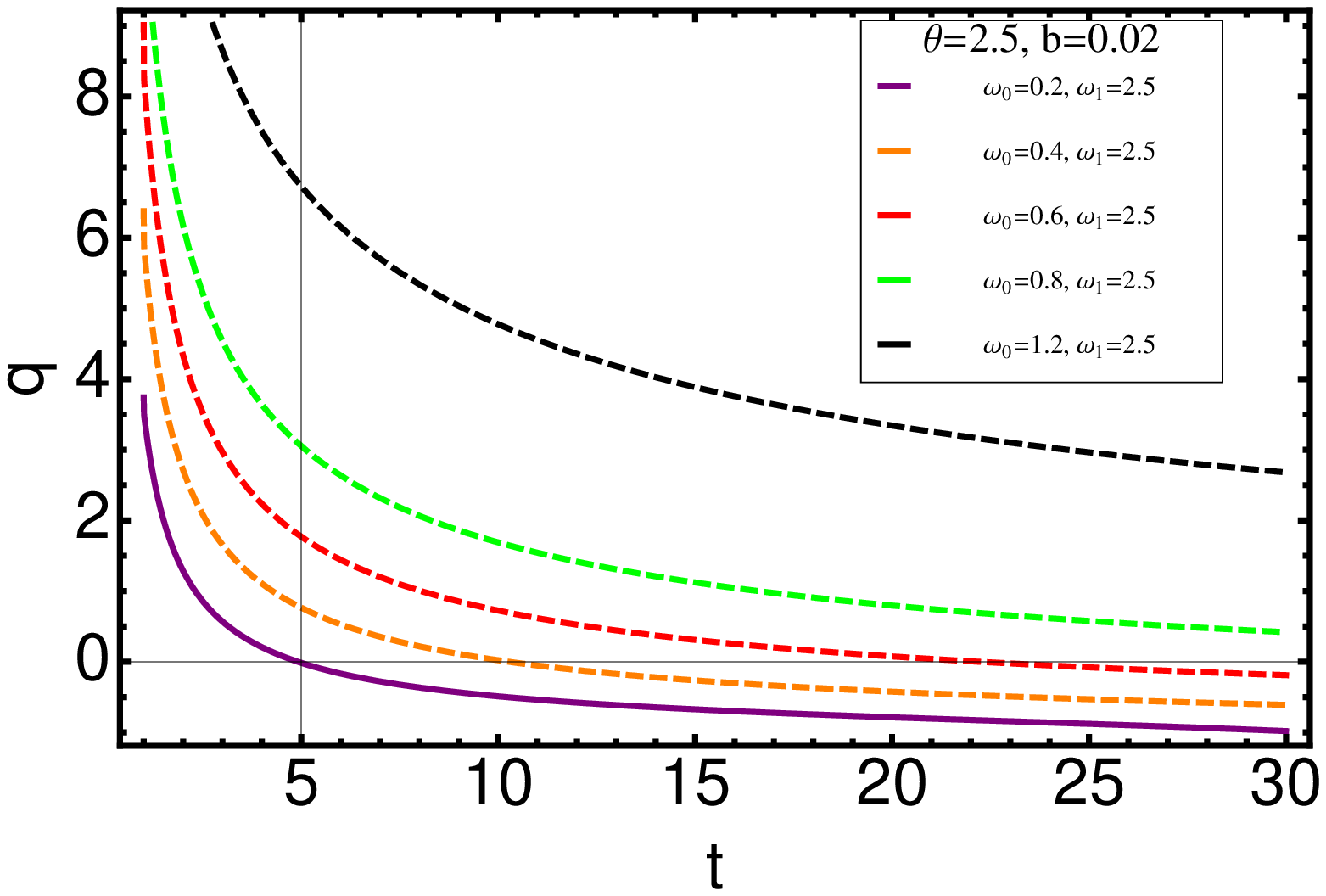} &
\includegraphics[width=48 mm]{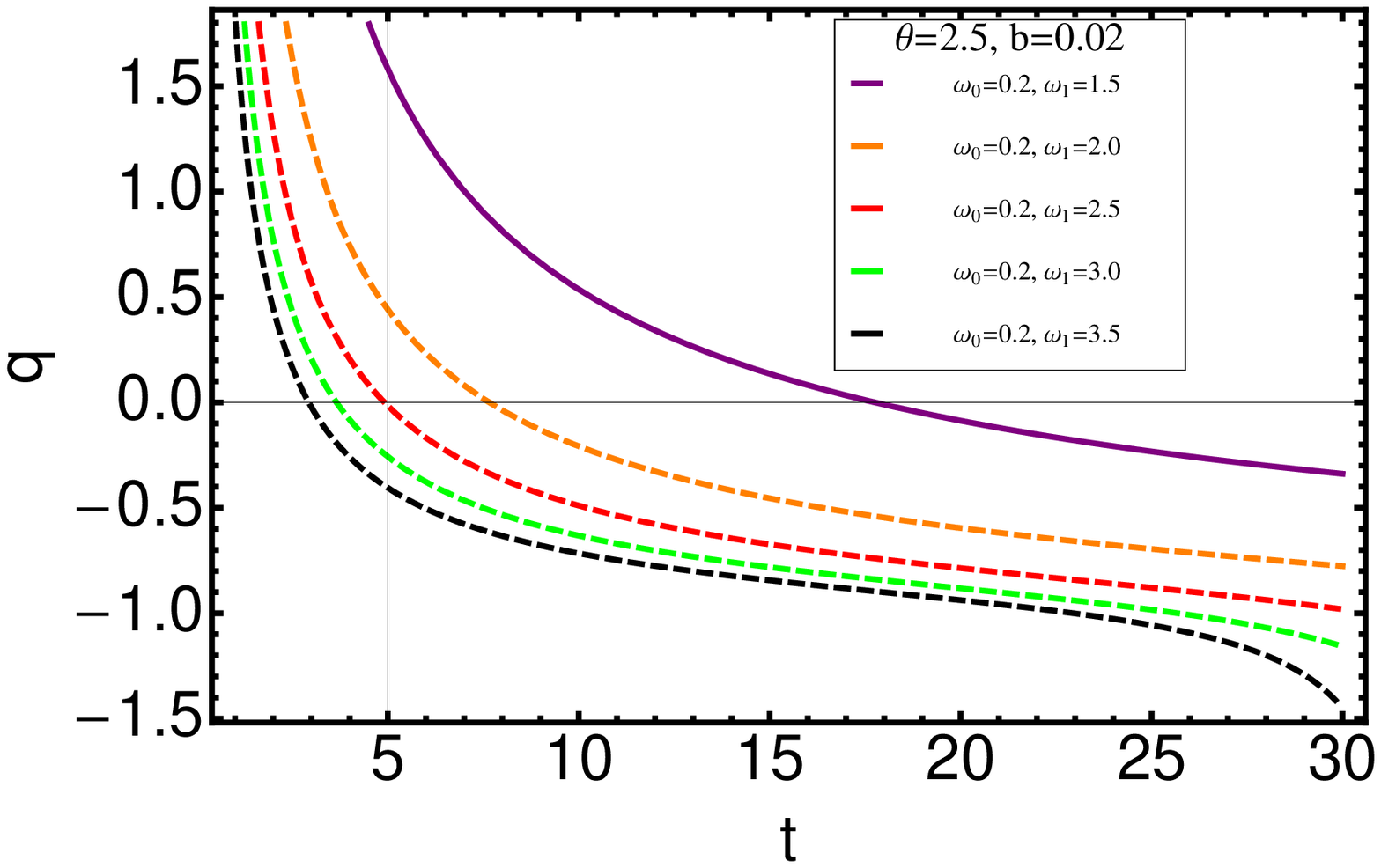}\\
\includegraphics[width=48 mm]{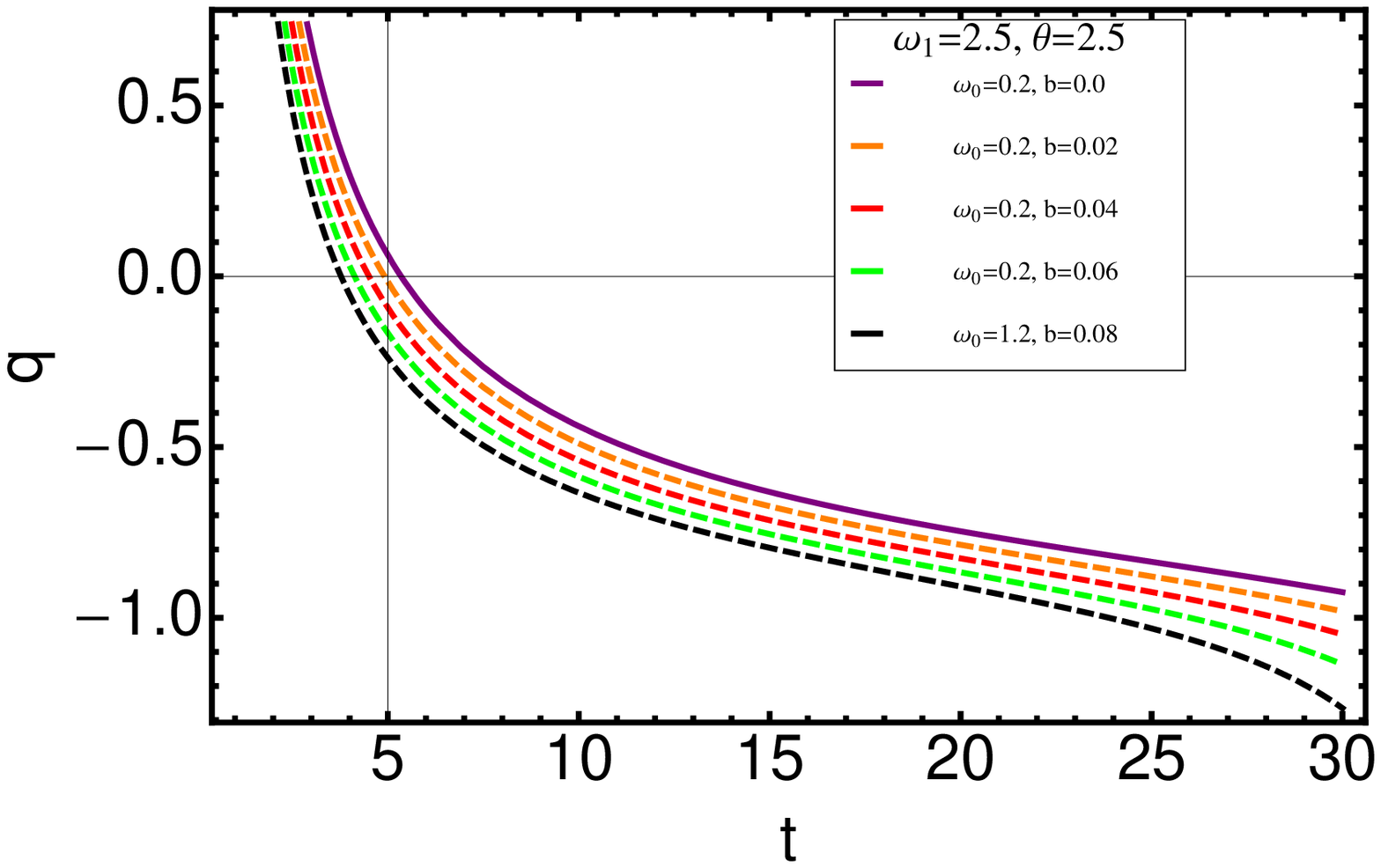} &
\includegraphics[width=48 mm]{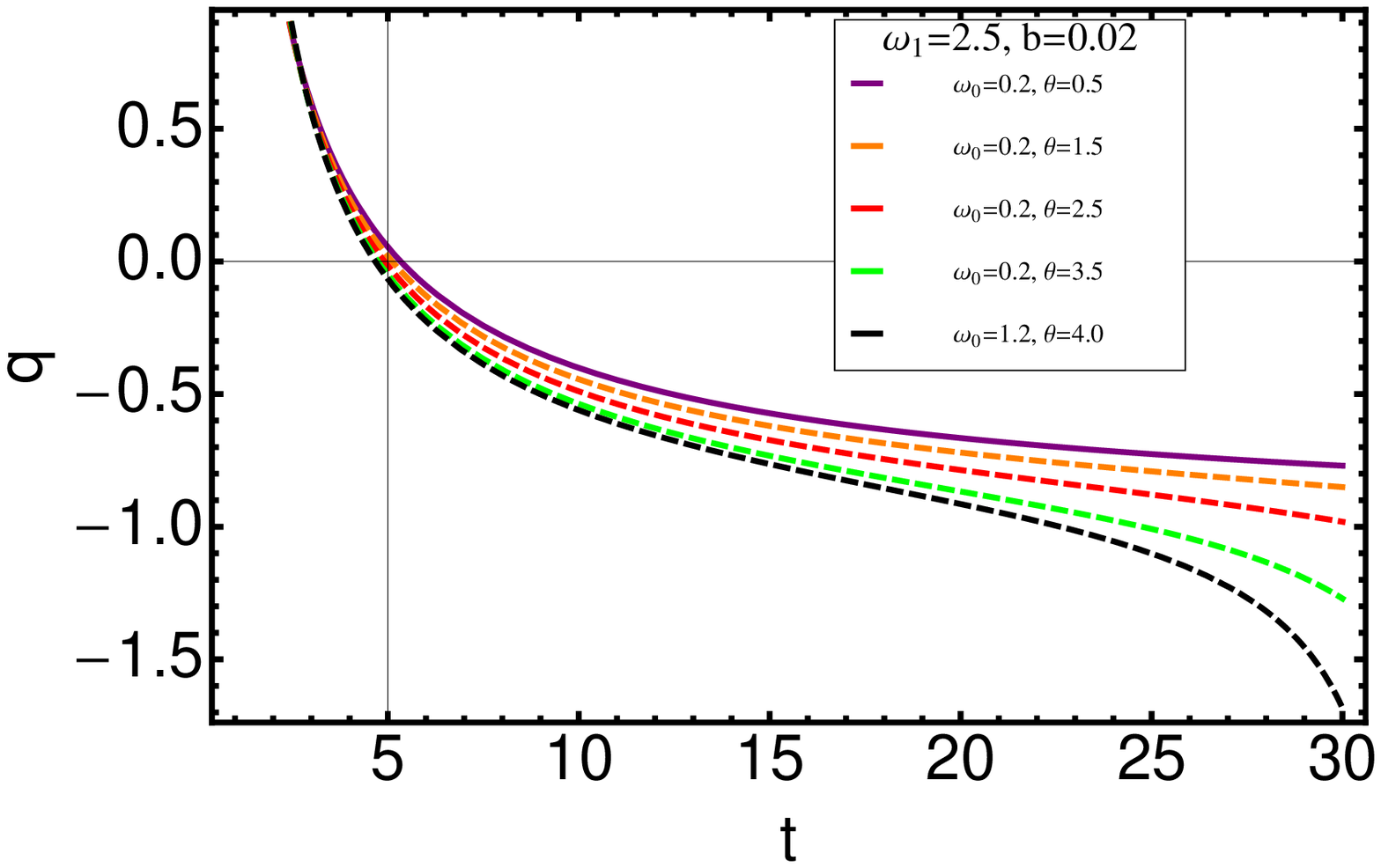}
 \end{array}$
 \end{center}
\caption{Model 3}
 \label{fig:10}
\end{figure}

\begin{figure}[h]
 \begin{center}$
 \begin{array}{cccc}
\includegraphics[width=48 mm]{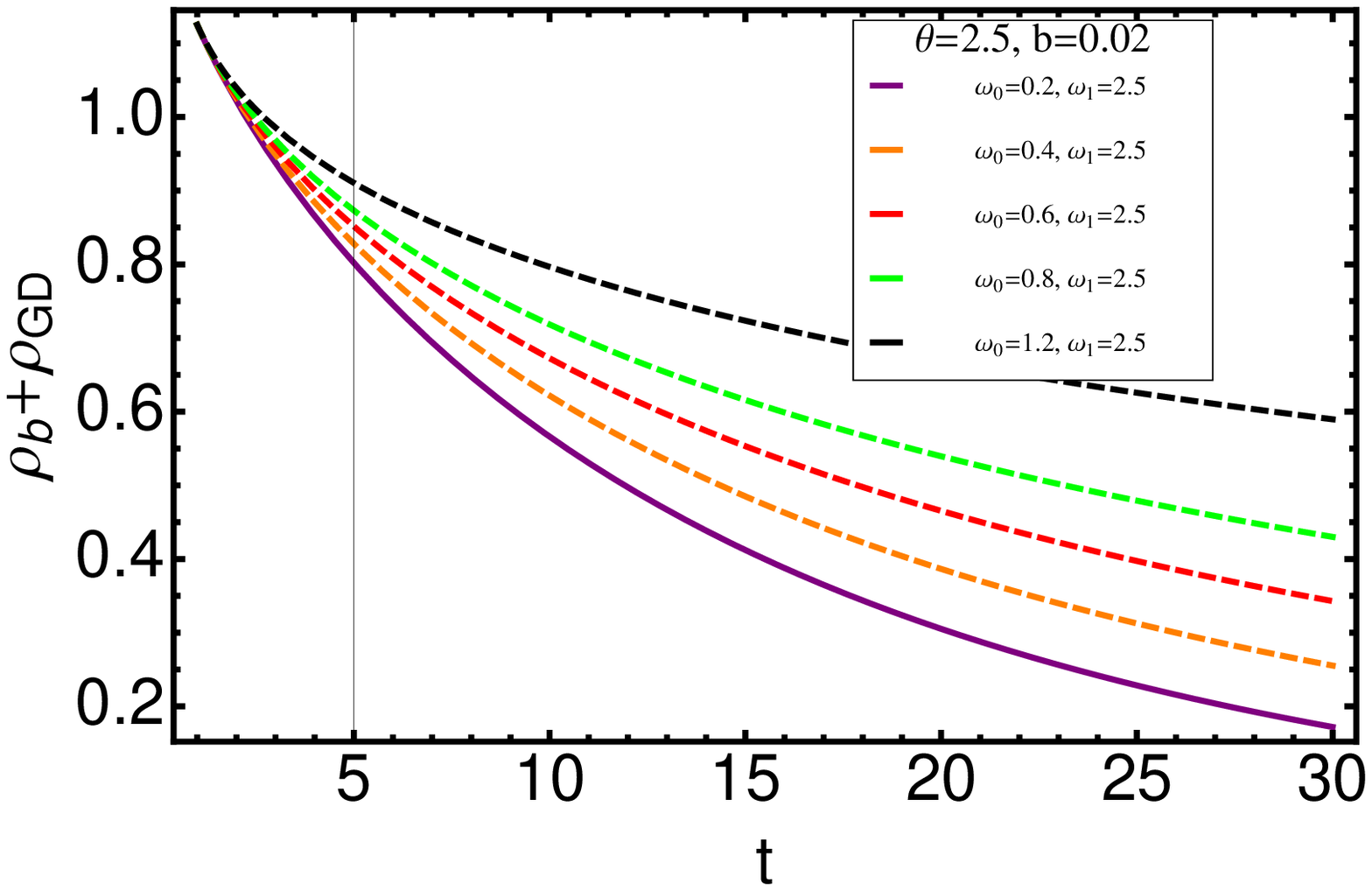} &
\includegraphics[width=48 mm]{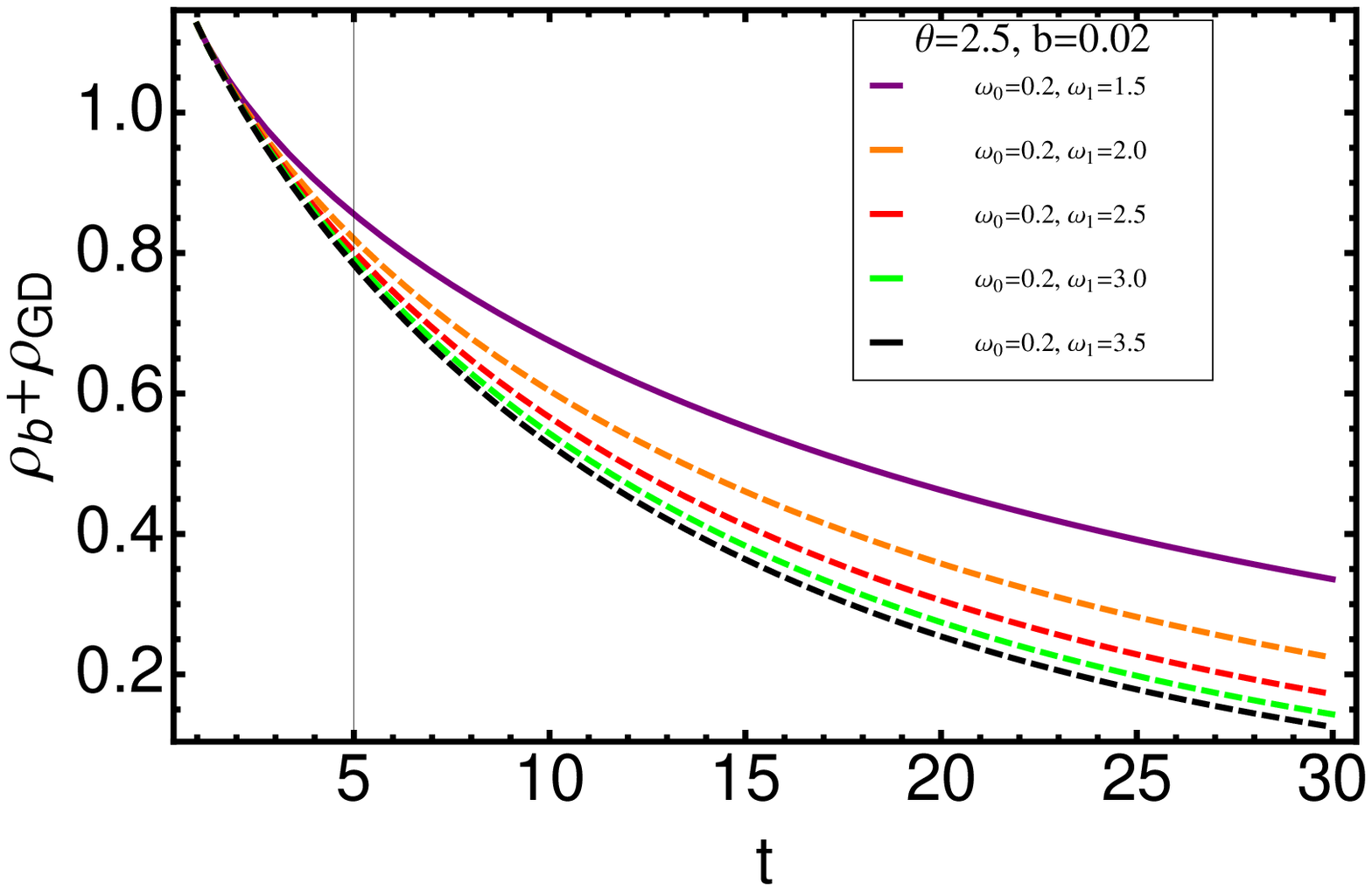}\\
\includegraphics[width=48 mm]{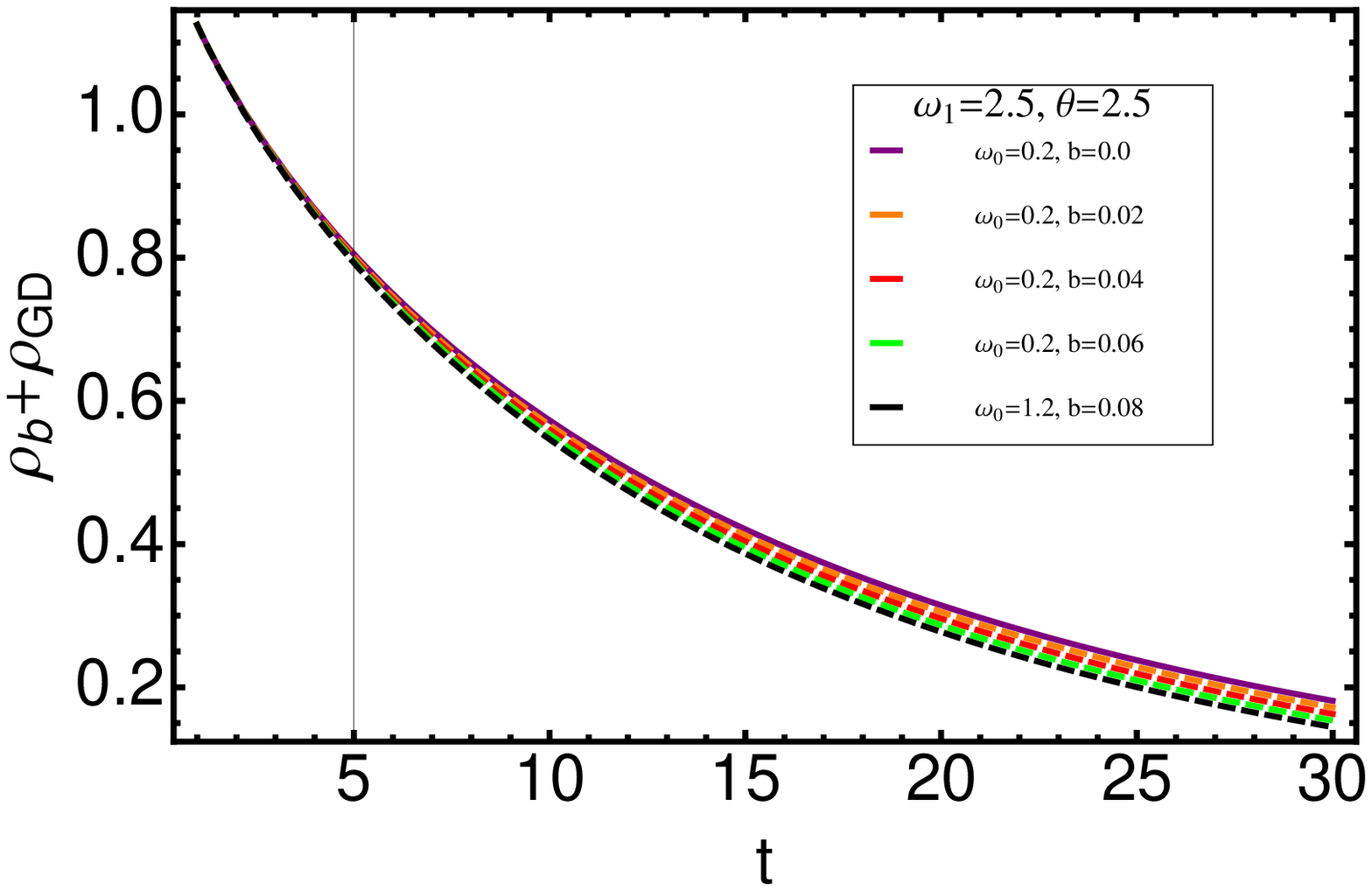} &
\includegraphics[width=48 mm]{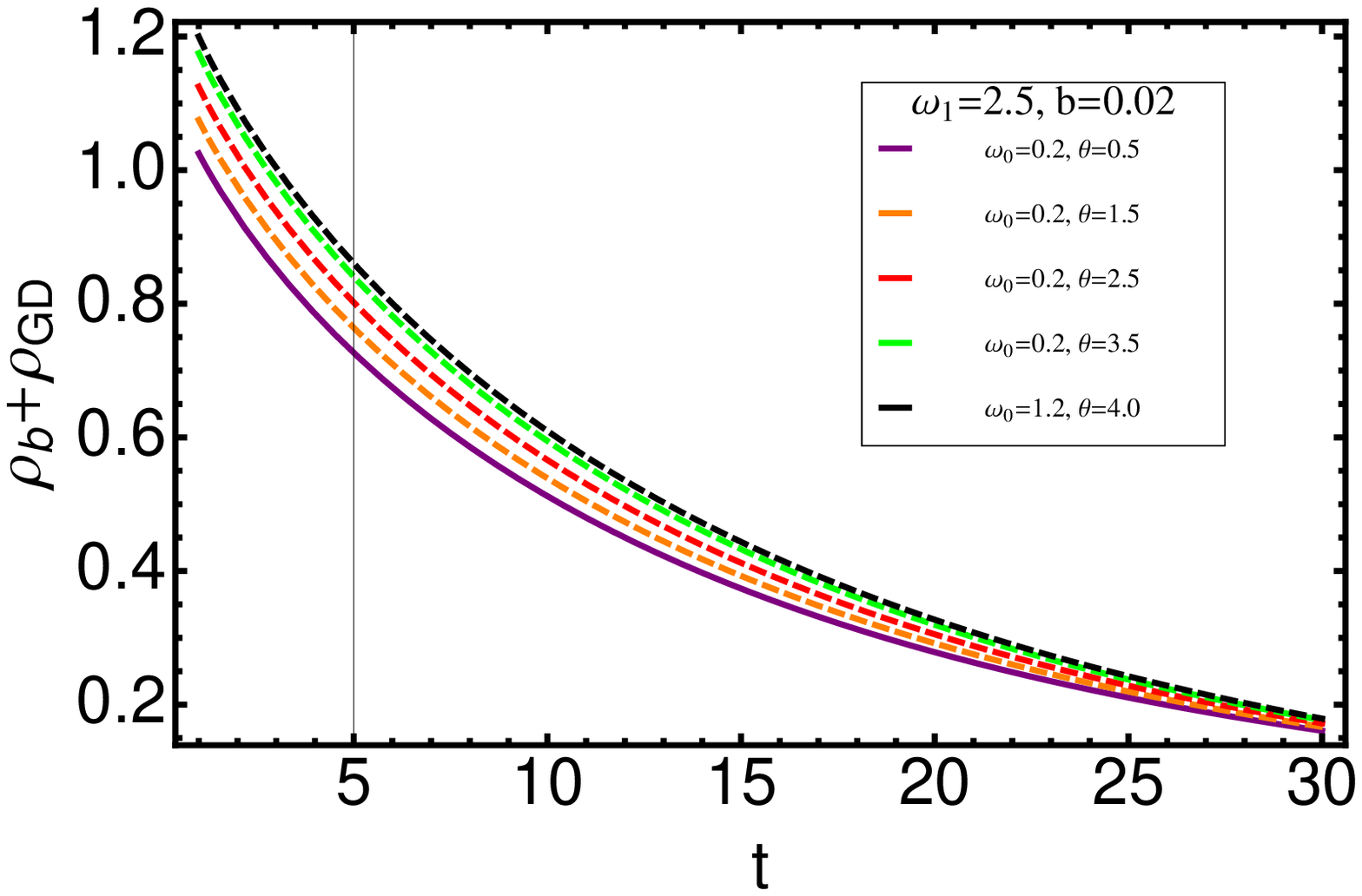}
 \end{array}$
 \end{center}
\caption{Model 3}
 \label{fig:11}
\end{figure}

\begin{figure}[h]
 \begin{center}$
 \begin{array}{cccc}
\includegraphics[width=48 mm]{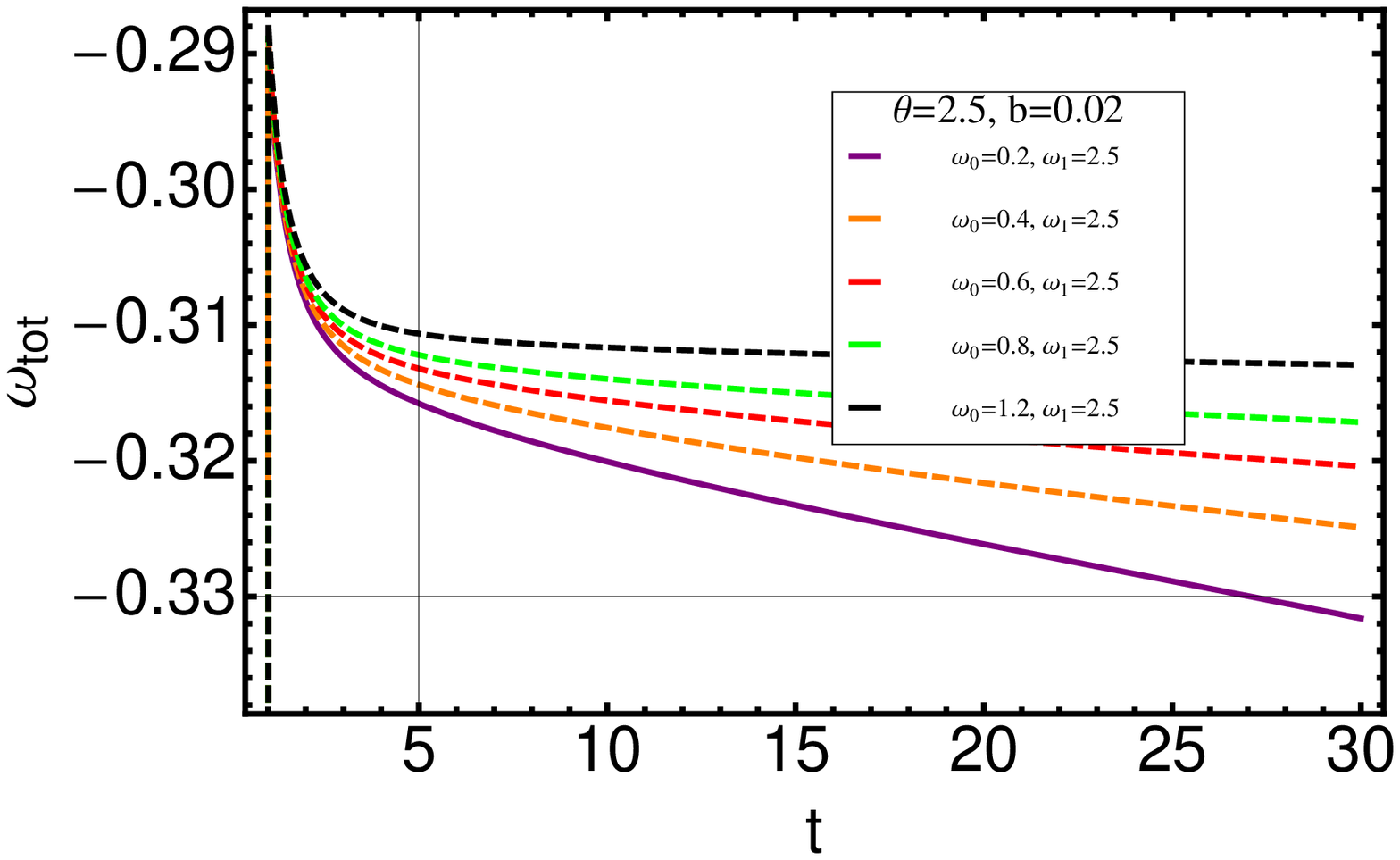} &
\includegraphics[width=48 mm]{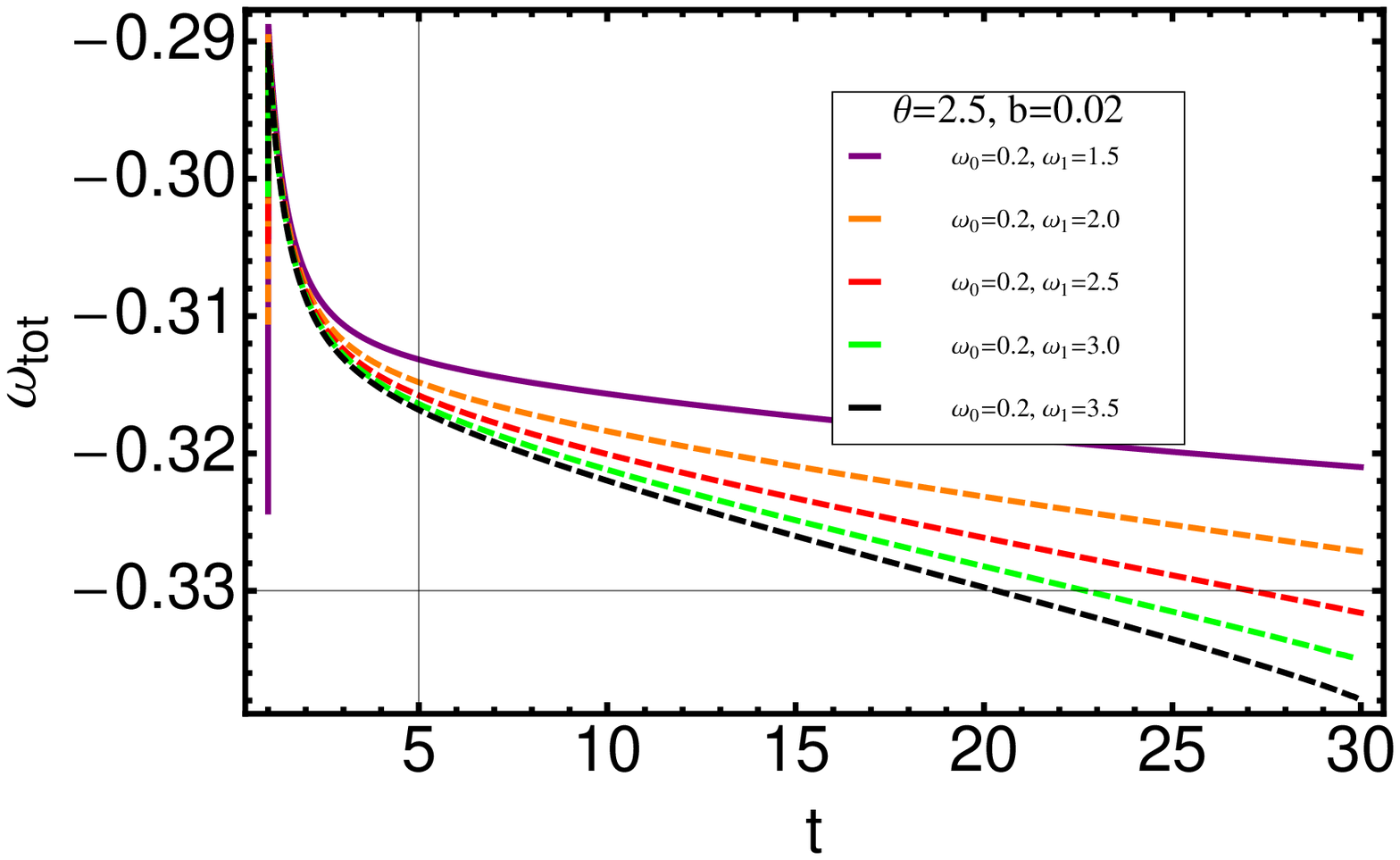}\\
\includegraphics[width=48 mm]{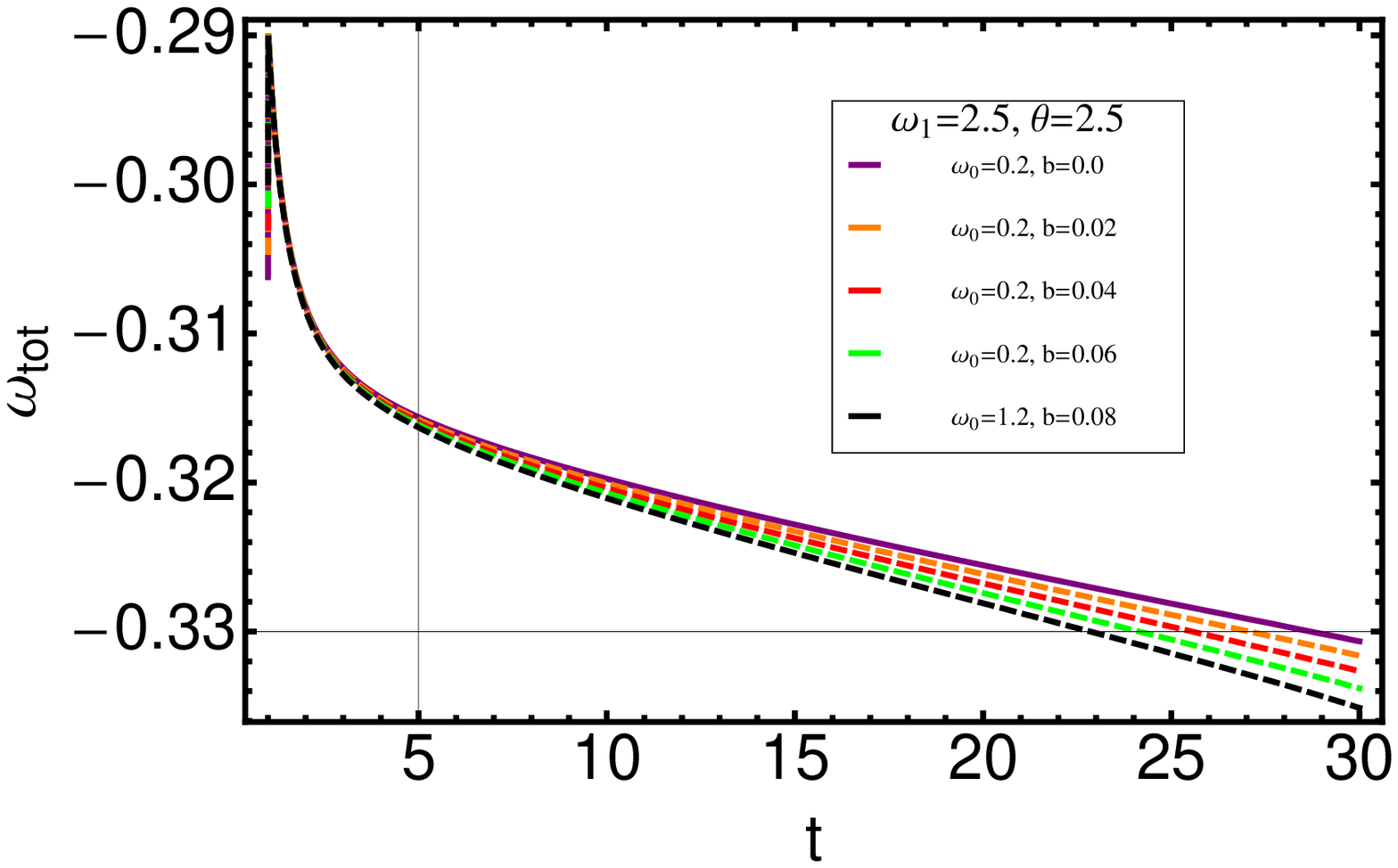} &
\includegraphics[width=48 mm]{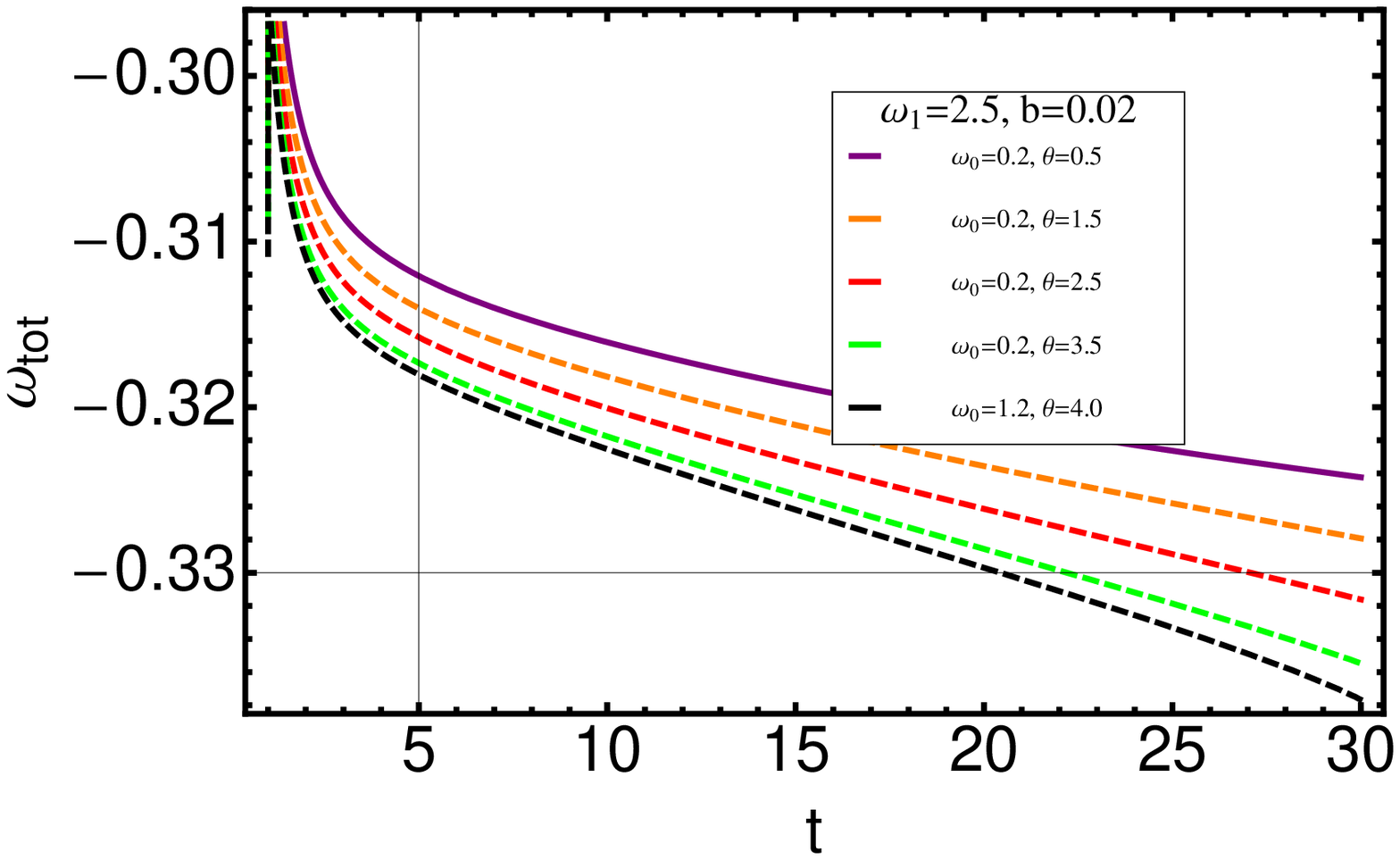}
 \end{array}$
 \end{center}
\caption{Model 3}
 \label{fig:12}
\end{figure}

Equation (37) may be solve numerically to find behavior of $G$ as an
increasing function of time (see Fig. 9).\\
Using the equations (8) and (14), deceleration parameter also drawn
in the Fig. 10, which has expected behavior.\\
Also, energy density $\rho=\rho_{b}+\rho_{GD}$ illustrated in the
Fig. 11 and total EoS parameter drawn in the Fig. 12. We can see
that $\rho$ and $\omega_{tot}$ are decreasing function of time.

\section*{\large{Conclusion}}
In this paper we studied several models of variable cosmological
constant and interacting two-component fluid Universe. Also we
assumed variable $G$. We considered ghost dark energy and barotropic
fluid as components of Universe with the interaction term of the
form of the equation (17). We solved Friedmann equations and
conservation law numerically and
obtained behavior of deceleration and EoS parameters.\\
In the first model, where cosmological constant is of the form of
the equation (23), the variable $G$ found increasing function of
time, as illustrated in the Fig. 1. The first plot of the Fig. 1
shows that increasing $\omega_{0}$ decreases the value of $G$. This
is opposite for the $\omega_{1}$ (see second plot of the Fig. 1).
Dependence of $G$ to interaction parameters $b$ and $\theta$
illustrated in the third and fourth plots of the Fig. 1. They have
shown that both $b$ and $\theta$ increase the value of $G$. We have
seen that variation of $G$ is approximately linear for time. For the
second model we found similar behavior to the first model (see Fig.
5). This situation has few differences for the third model. the
first and second plots of the Fig. 9 show similar behavior to the
previous models with variation of $\omega_{0}$ and $\omega_{1}$, but
there are no linearity in this case, at least initially. The
variable $G$ grows rapidly to $t=5$ and then has approximately
linear behavior. Also similar to first and second cases, increasing
$b$ increases $G$, but increasing $\theta$ decreases $G$ which is
opposite with the
previous cases.\\
Variation of the deceleration parameter with time for all models
illustrated in the Fig. 2, 6 and 10. They have shown that the
deceleration parameter is decreasing function of time. The first and
second plots of these figures have shown increasing with
$\omega_{0}$ and decreasing with $\omega_{1}$. Also we found that
increasing interaction parameters $b$ and $\theta$ decreases the
value of $q$. We know that $q\geq-1$ is important condition which
satisfied with some values of our parameters. The best selected
parameters to have $q\rightarrow-1$ at the late time are
$\omega_{0}=0.2$, $\omega_{1}=2.5$, $b=0.02$ and $\theta=2.5$ which
are identical for
all models.\\
Total energy density presented by the Figs. 3, 7 and 11 which shown
that increasing $\omega_{0}$ and $\theta$ increase value of $\rho$,
but increasing $\omega_{1}$ and $b$ decrease value of $\rho$. As
expected, energy densities of all models are decreasing function of
time.\\
Total EoS drawn by the Figs. 4, 8 and 12 which shown that increasing
$\omega_{0}$  increase value of $\omega_{tot}$, but increasing
$\omega_{1}$, $\theta$ and $b$ decrease value of $\omega_{tot}$. We
found that the total EoS corresponding to three models are
decreasing function of time.\\
We also studied scale factor, total pressure and barotropic EoS
parameter numerically. Plots of the Fig. 13 shown behavior of scale
factor in the first model which is increasing function of time. We
found that increasing $\omega_{1}$ and $b$ increase $a$, but
increasing $\omega_{0}$ decreases $a$. The last plot of the Fig. 13
shows that variation of $\theta$ is not important for the scale
factor.\\
Plots of the Fig. 14 show behavior of total pressure in the first
model which is increasing function of time. We found that increasing
$\omega_{1}$ and $b$ increase $P=P_{b}+P_{GD}$, but increasing
$\omega_{0}$ and $\theta$ decrease $P$.\\
$\omega_{b}$ of the first model plotted in the Fig. 15 which is
increasing function of time. The first plot of the Fig. 15 tells
that increasing $\omega_{0}$ decreases the value of $\omega_{b}$,
while increasing $\omega_{1}$, $\theta$ and $b$ increase
$\omega_{b}$.\\
Variation of the scale factor of the second model with respect to
$\omega_{0}$, $\omega_{1}$ and $b$ is similar to the first model
(see Fig. 16). The last plot of the Fig. 16 tells that increasing
$\theta$ increases scale factor for the late time, but has not
important effect at the early time.\\
in the Fig. 17 and Fig. 18 we see that total pressure and barotropic
EoS of the second model are similar to the first model.\\
Fig. 19 and 20 represent time-dependent scale factor and total
pressure of the third model, respectively, which are increasing
function of time and have similar manner with the previous models.\\
Finally, Fig. 21 includes plots of $\omega_{b}$ of the third model
and has similar description with previous models. We conclude that
all models of cosmological constant which introduced in this paper
yields to acceptable behavior of cosmological parameters.\\\\
\section*{Acknowledgments}
Martiros Khurshudyan has been supported by EU fonds in the frame of the program FP7-Marie Curie Initial Training Network INDEX NO.289968.

\newpage
\begin{figure}[h]
 \begin{center}$
 \begin{array}{cccc}
\includegraphics[width=50 mm]{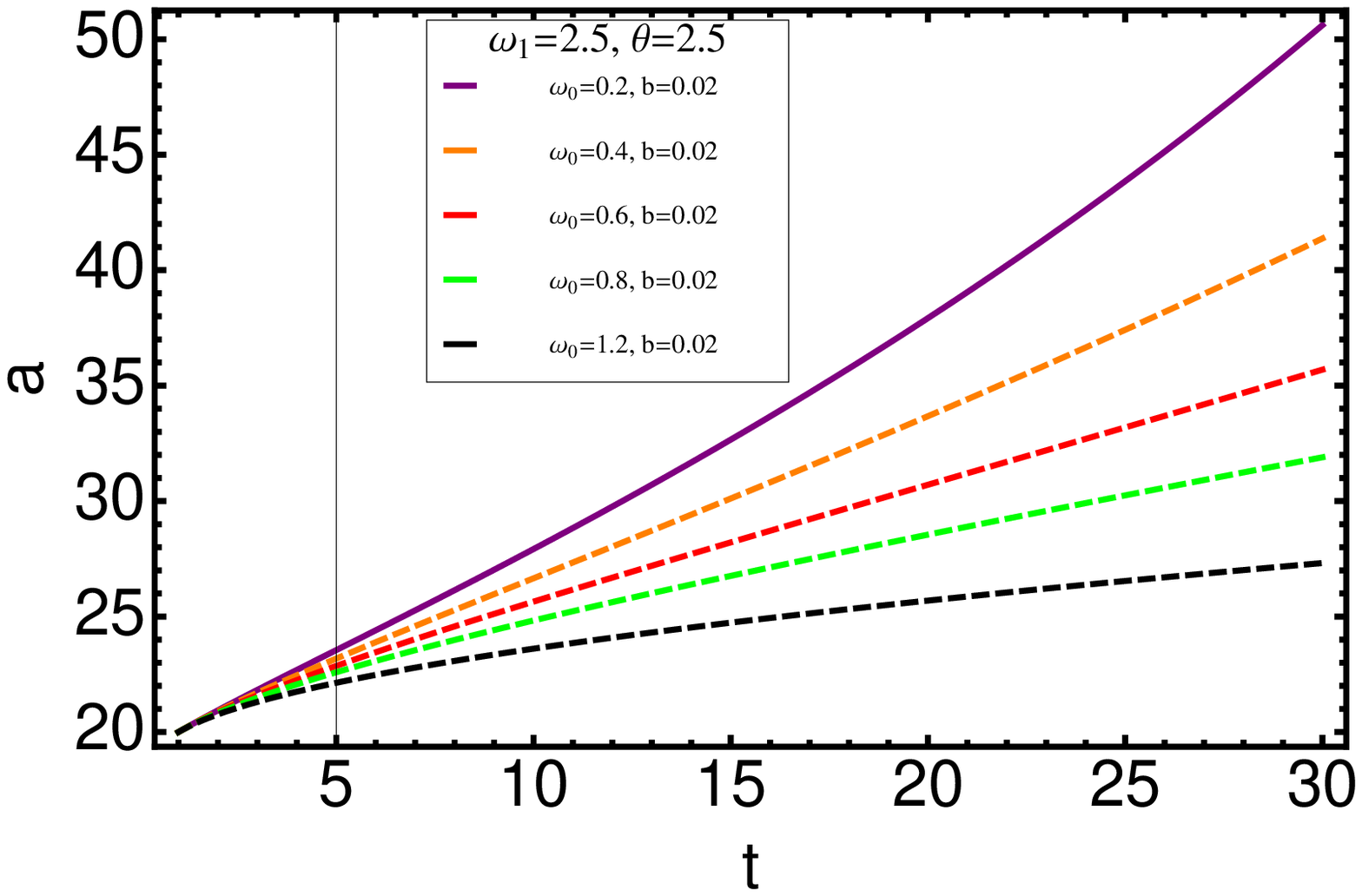} &
\includegraphics[width=50 mm]{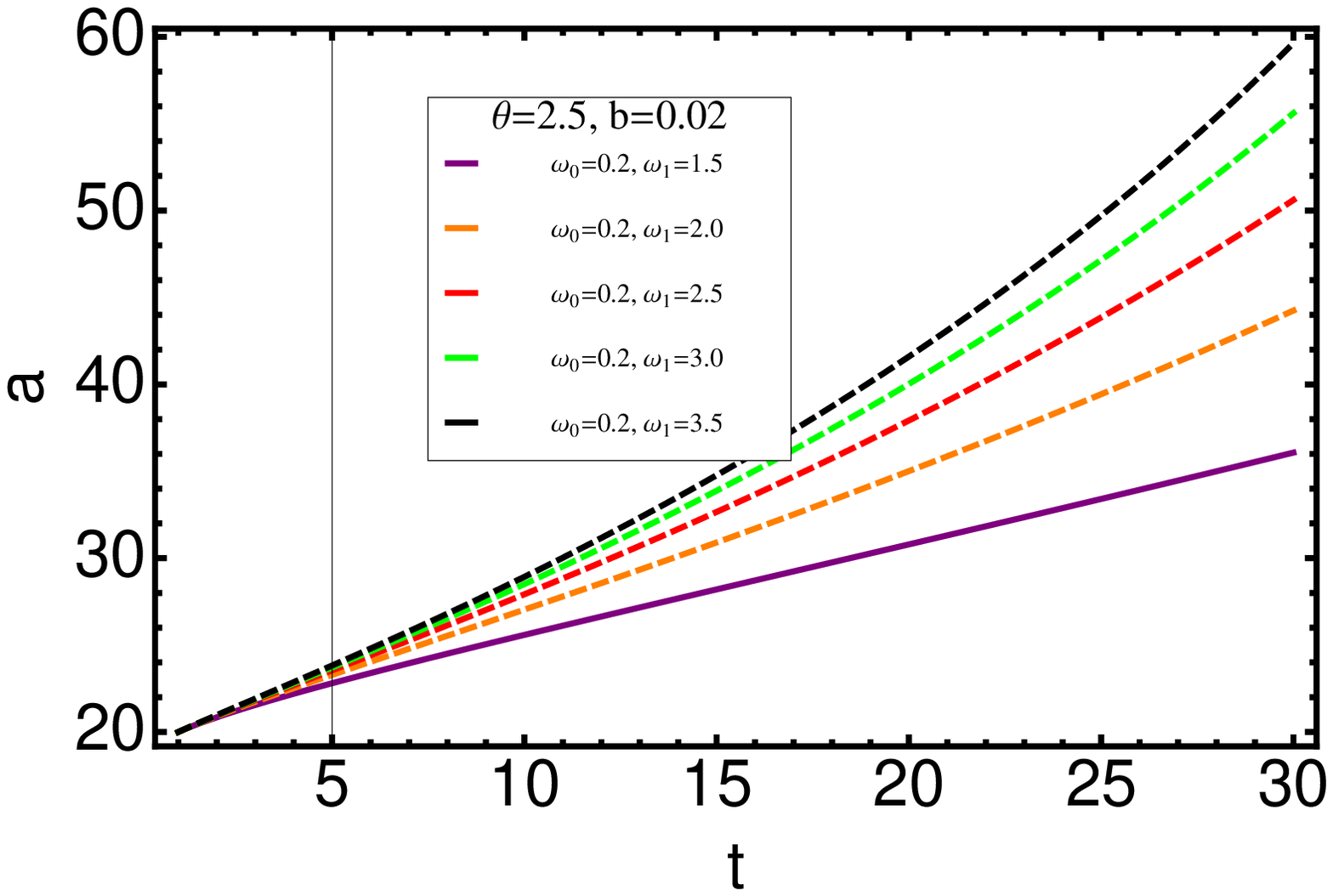}\\
\includegraphics[width=50 mm]{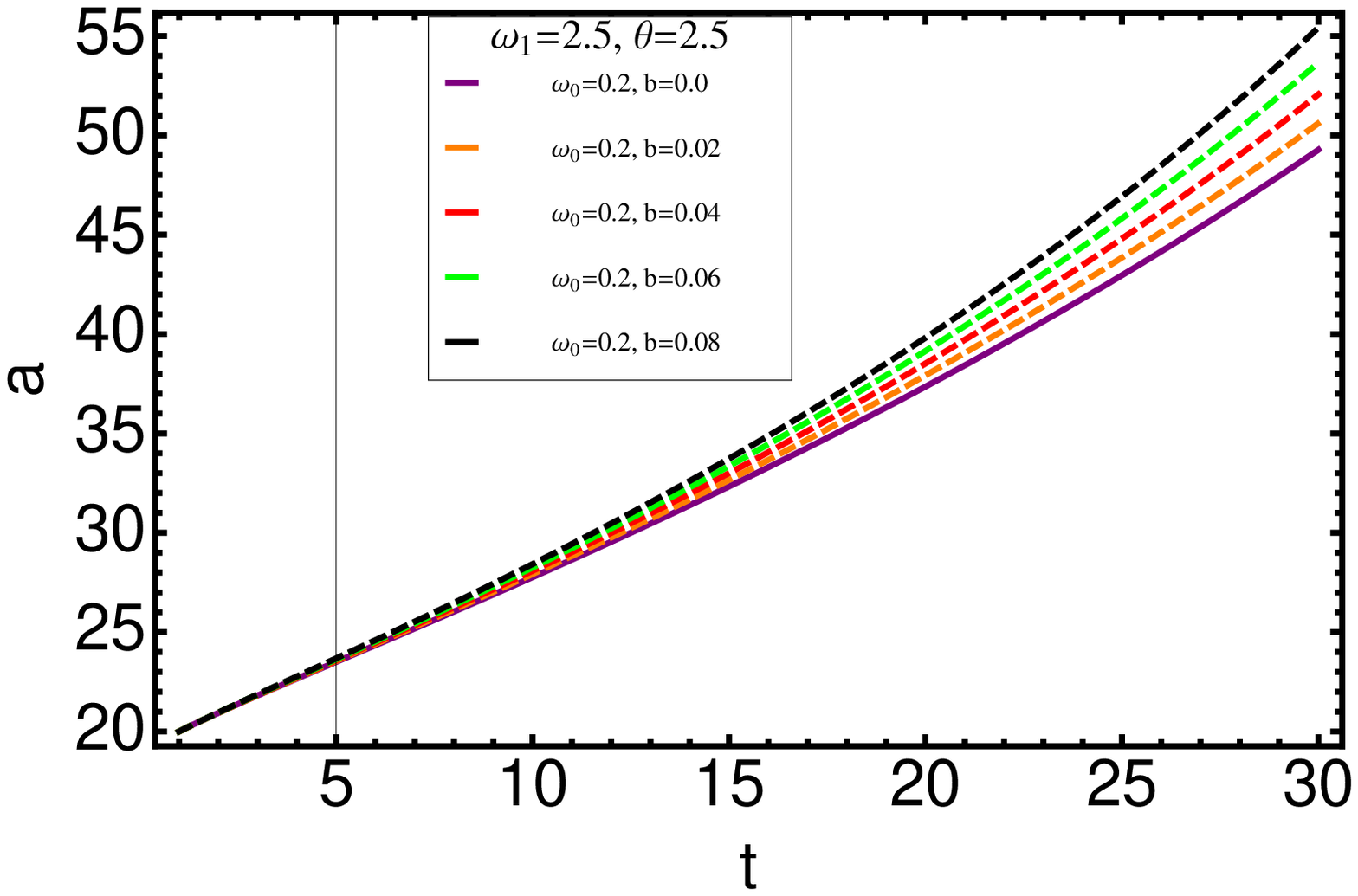} &
\includegraphics[width=50 mm]{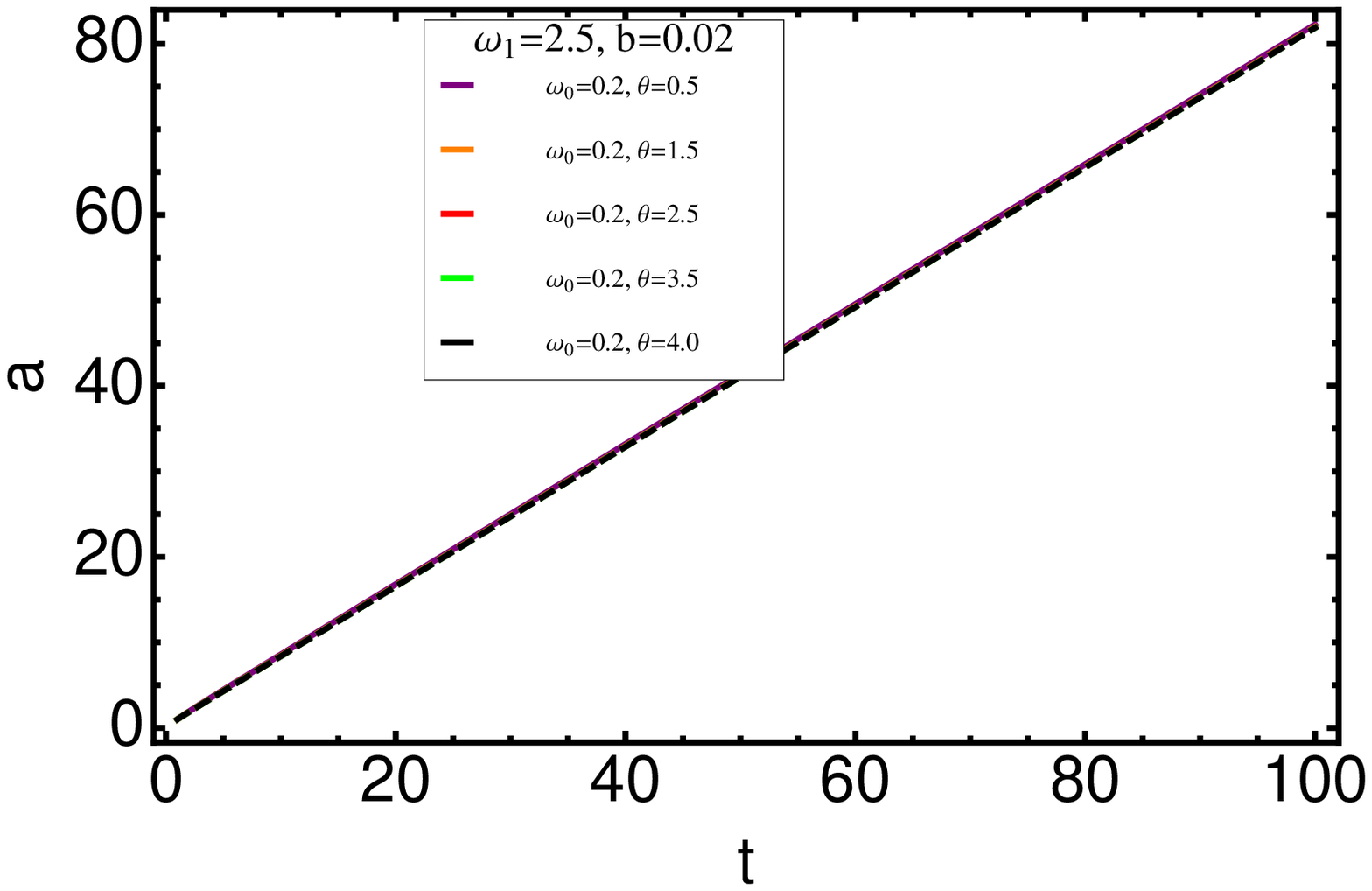}
 \end{array}$
 \end{center}
\caption{Model 1}
 \label{fig:13}
\end{figure}

\begin{figure}[h]
 \begin{center}$
 \begin{array}{cccc}
\includegraphics[width=50 mm]{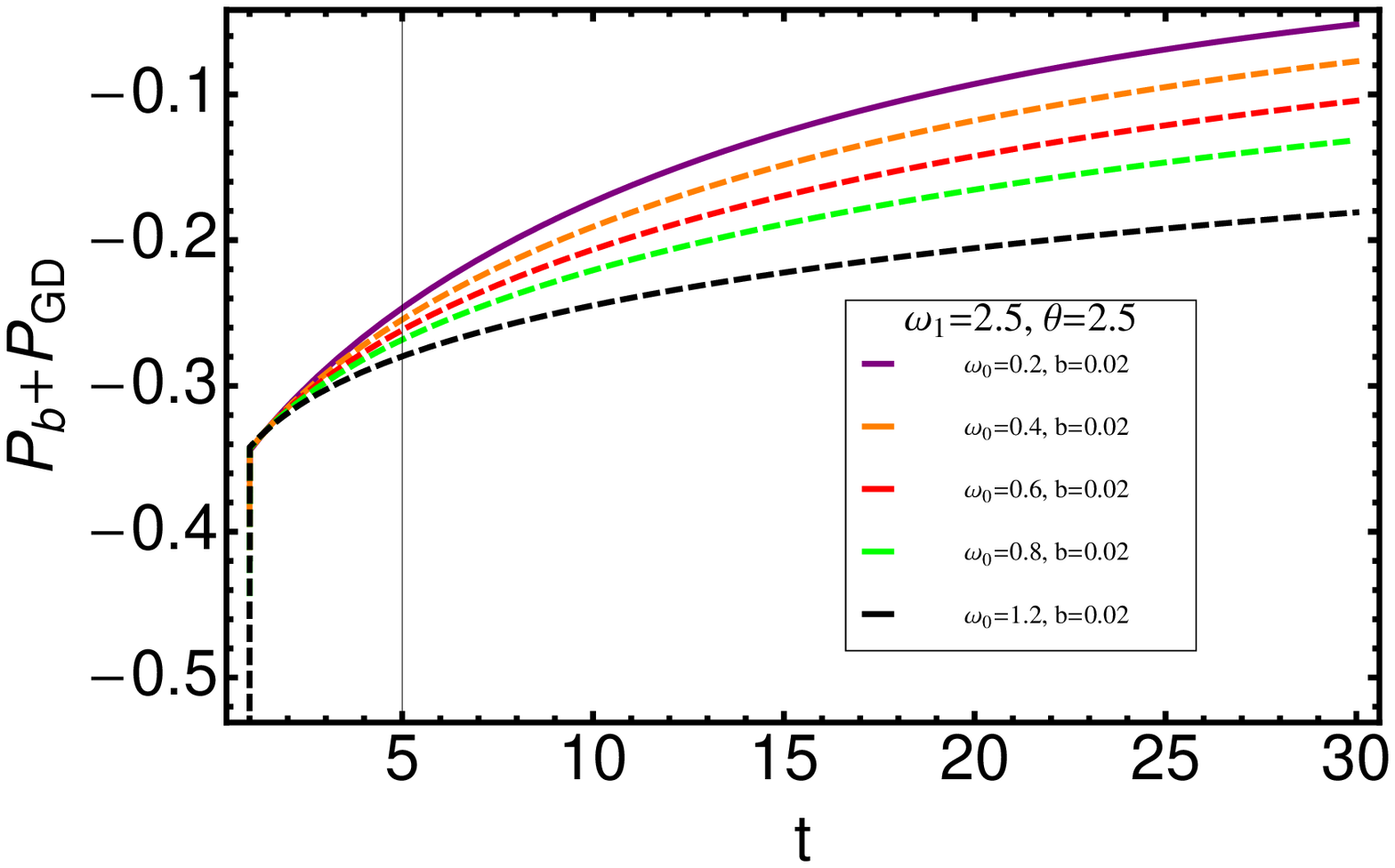} &
\includegraphics[width=50 mm]{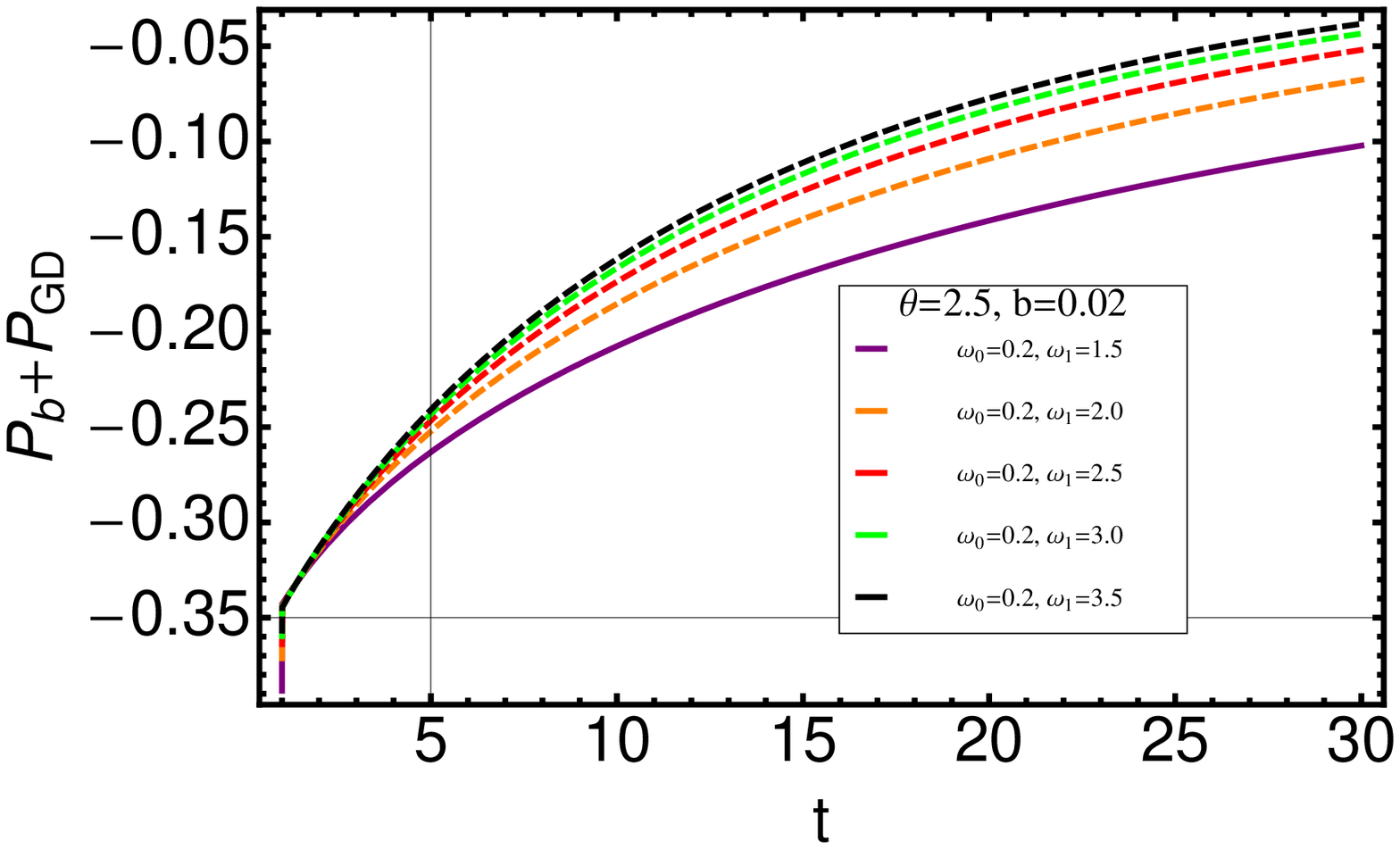}\\
\includegraphics[width=50 mm]{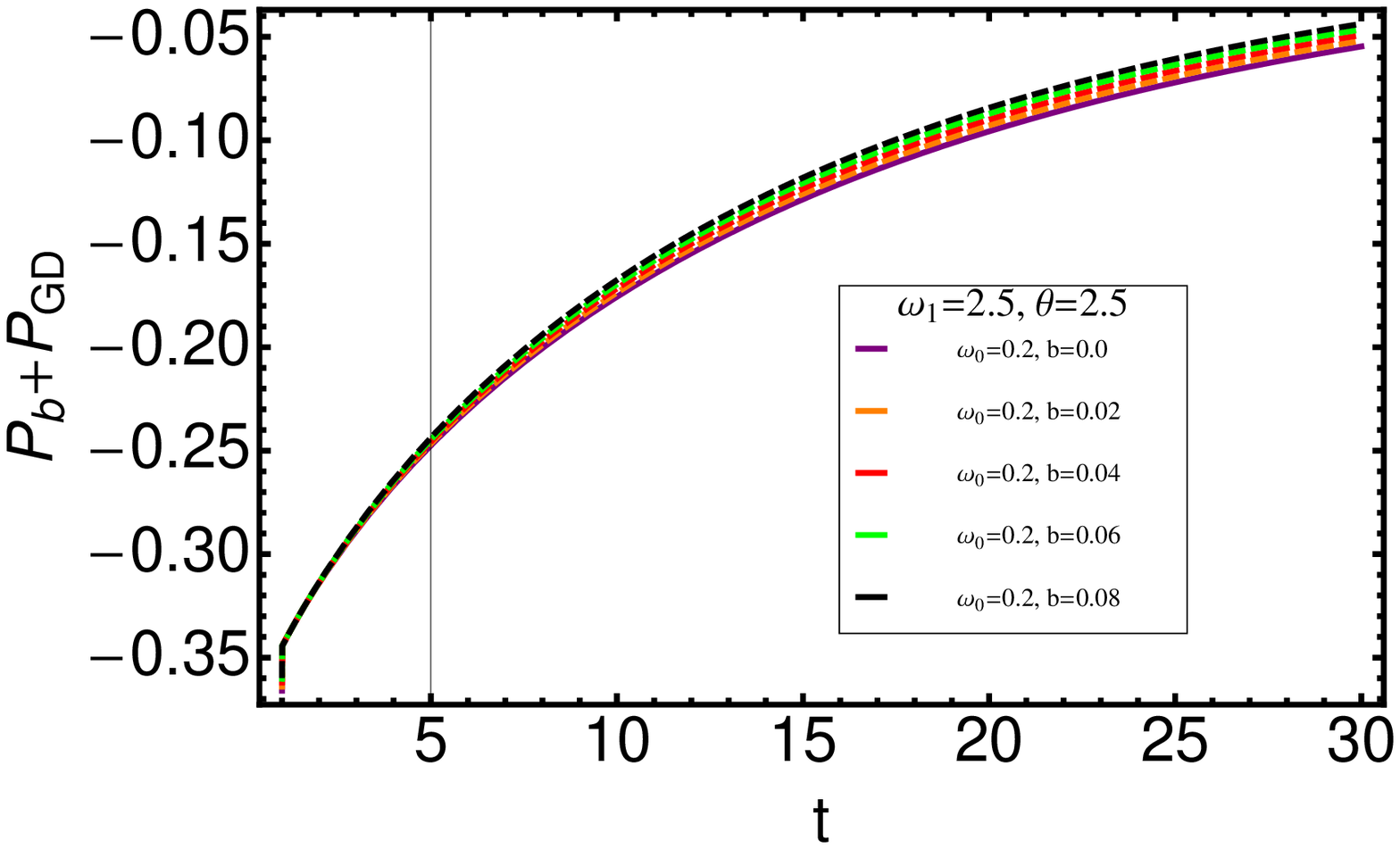} &
\includegraphics[width=50 mm]{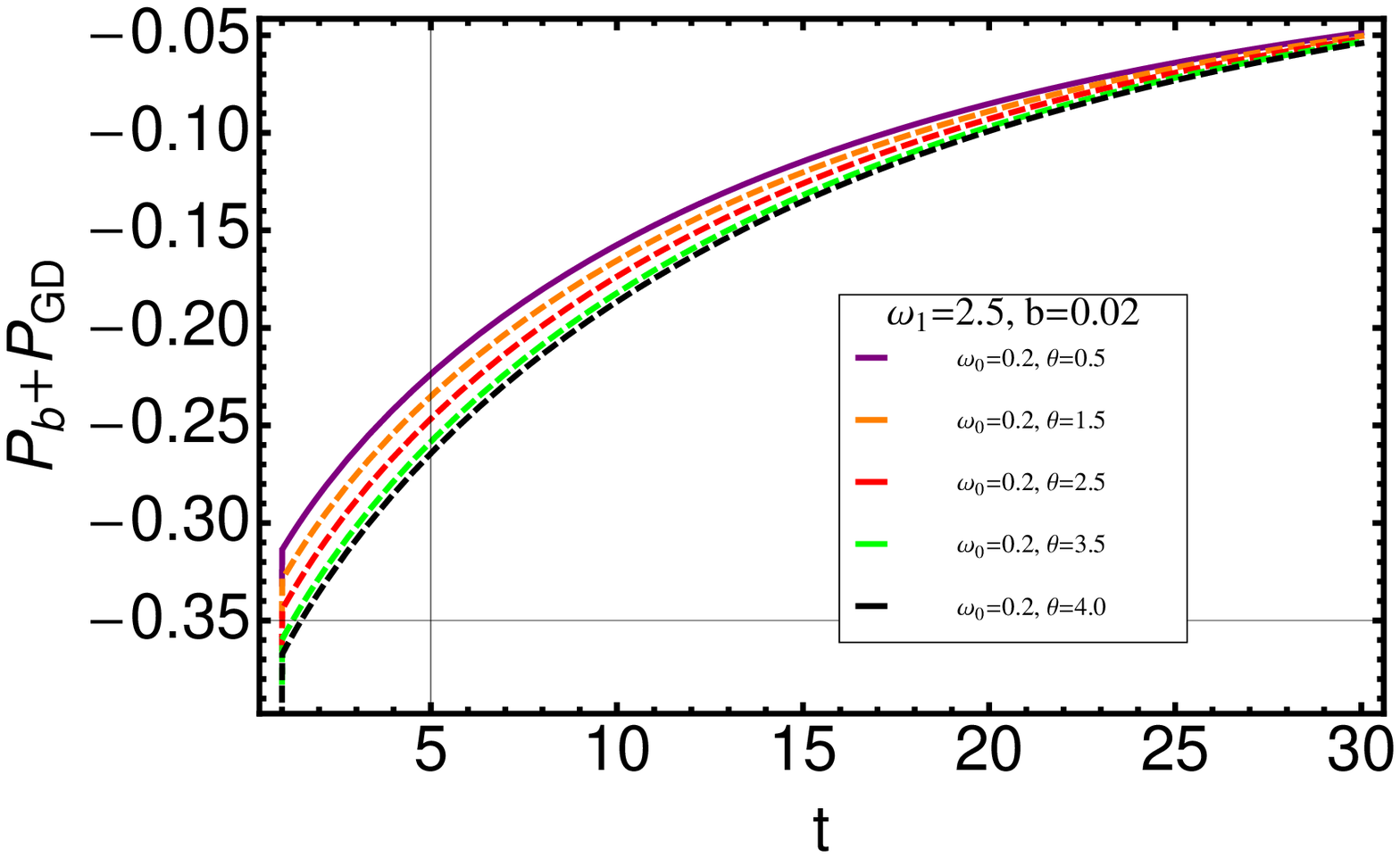}
 \end{array}$
 \end{center}
\caption{Model 1}
 \label{fig:14}
\end{figure}

\begin{figure}[h]
 \begin{center}$
 \begin{array}{cccc}
\includegraphics[width=50 mm]{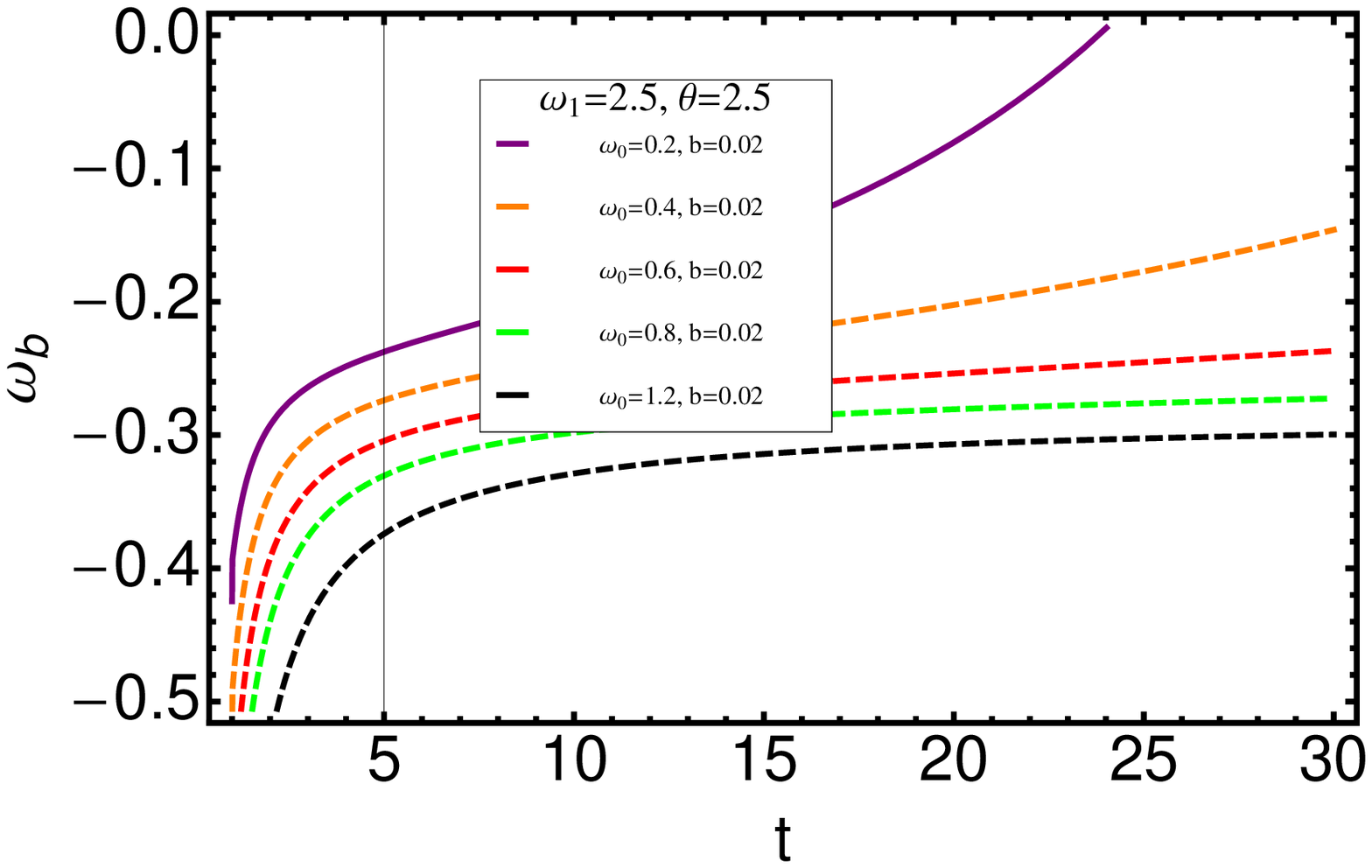} &
\includegraphics[width=50 mm]{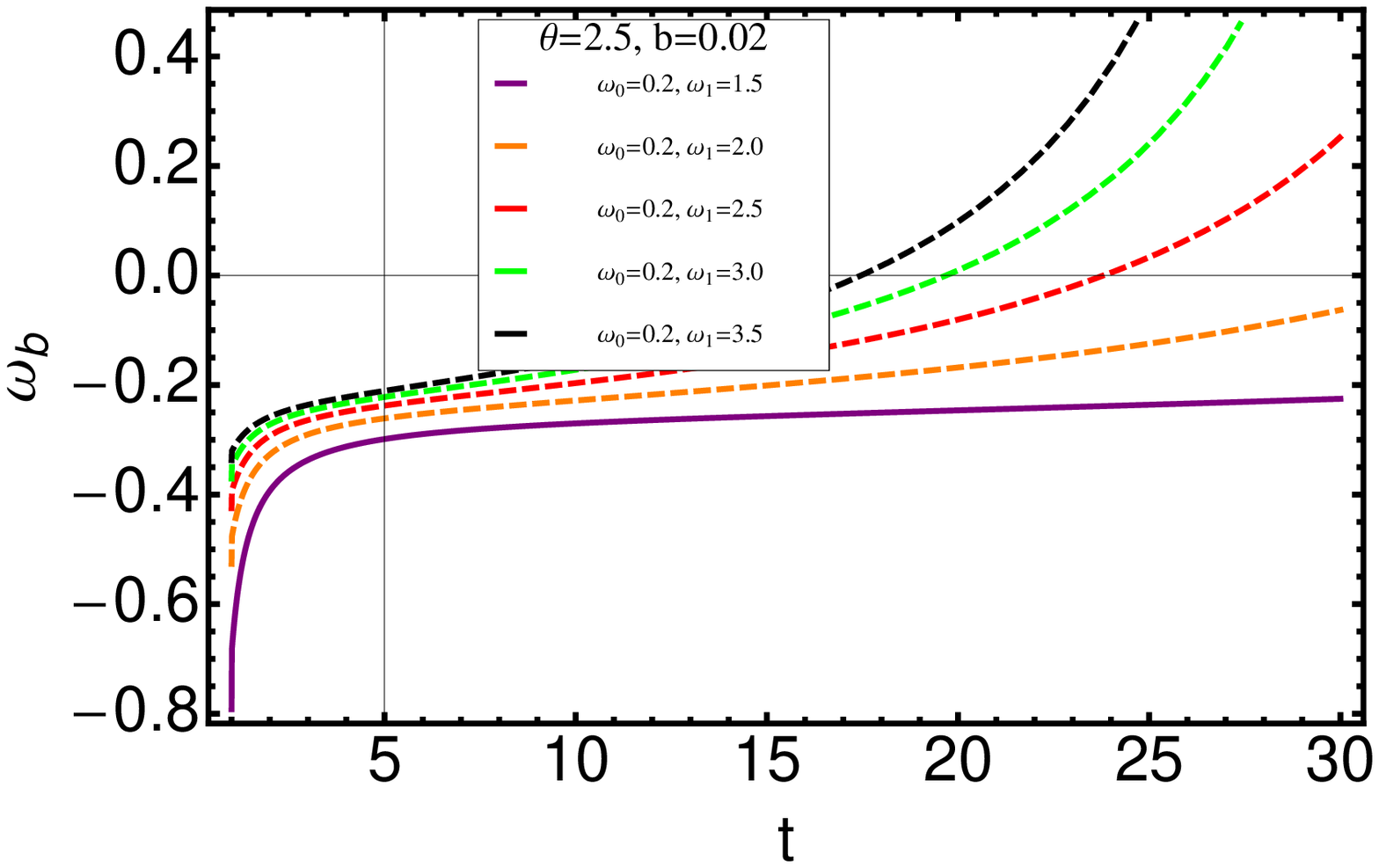}\\
\includegraphics[width=50 mm]{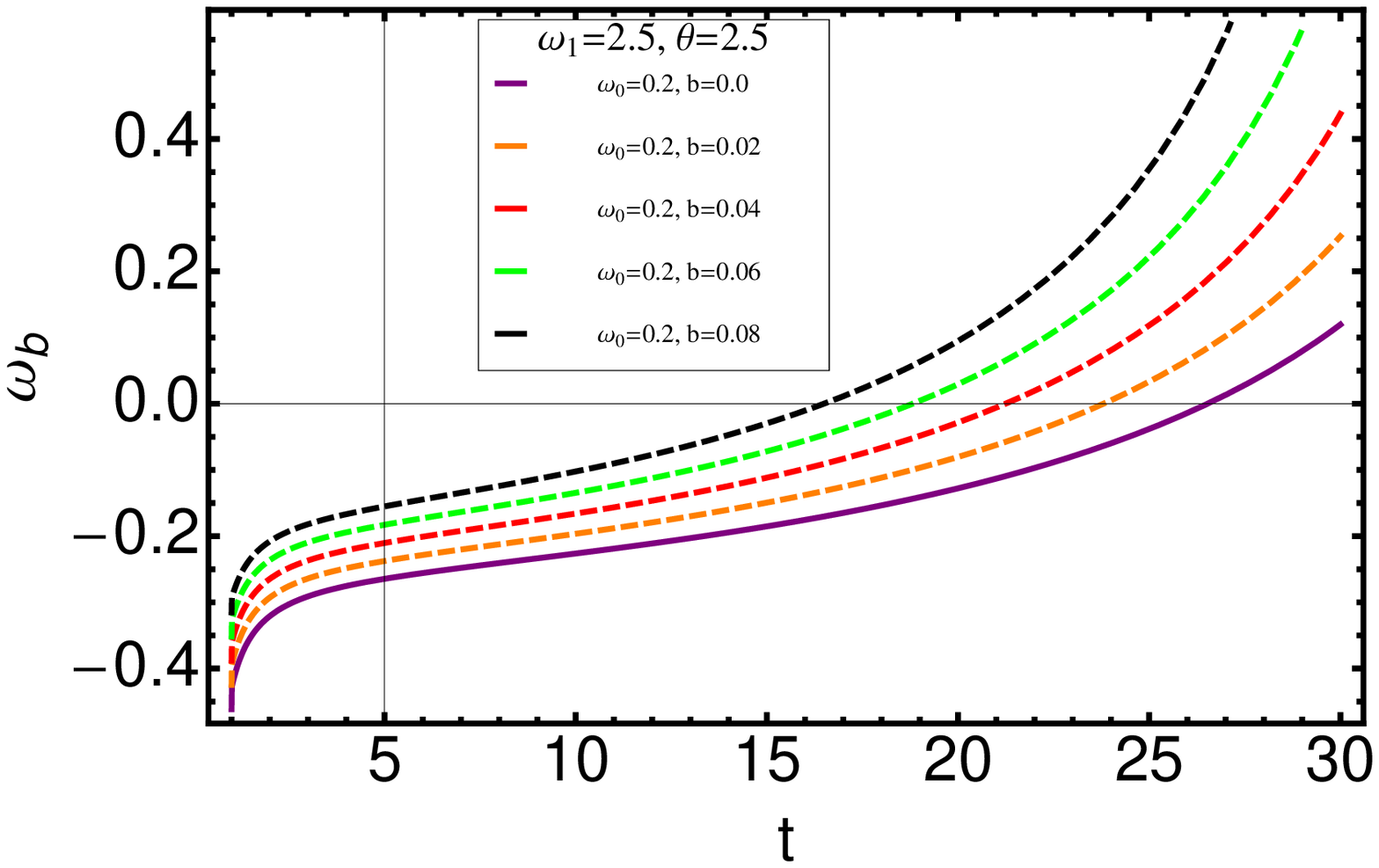} &
\includegraphics[width=50 mm]{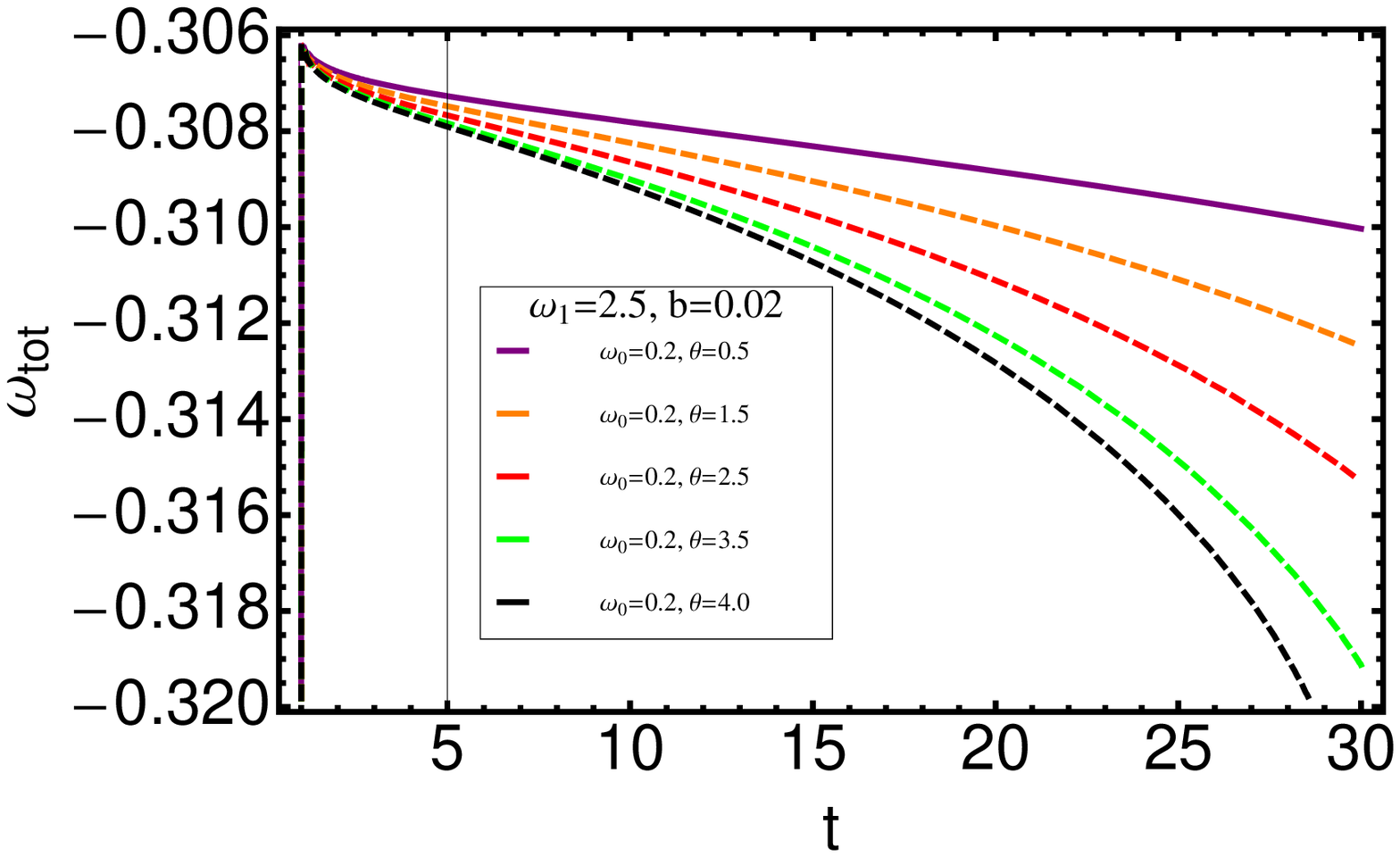}
 \end{array}$
 \end{center}
\caption{Model 1}
 \label{fig:15}
\end{figure}

\begin{figure}[h]
 \begin{center}$
 \begin{array}{cccc}
\includegraphics[width=50 mm]{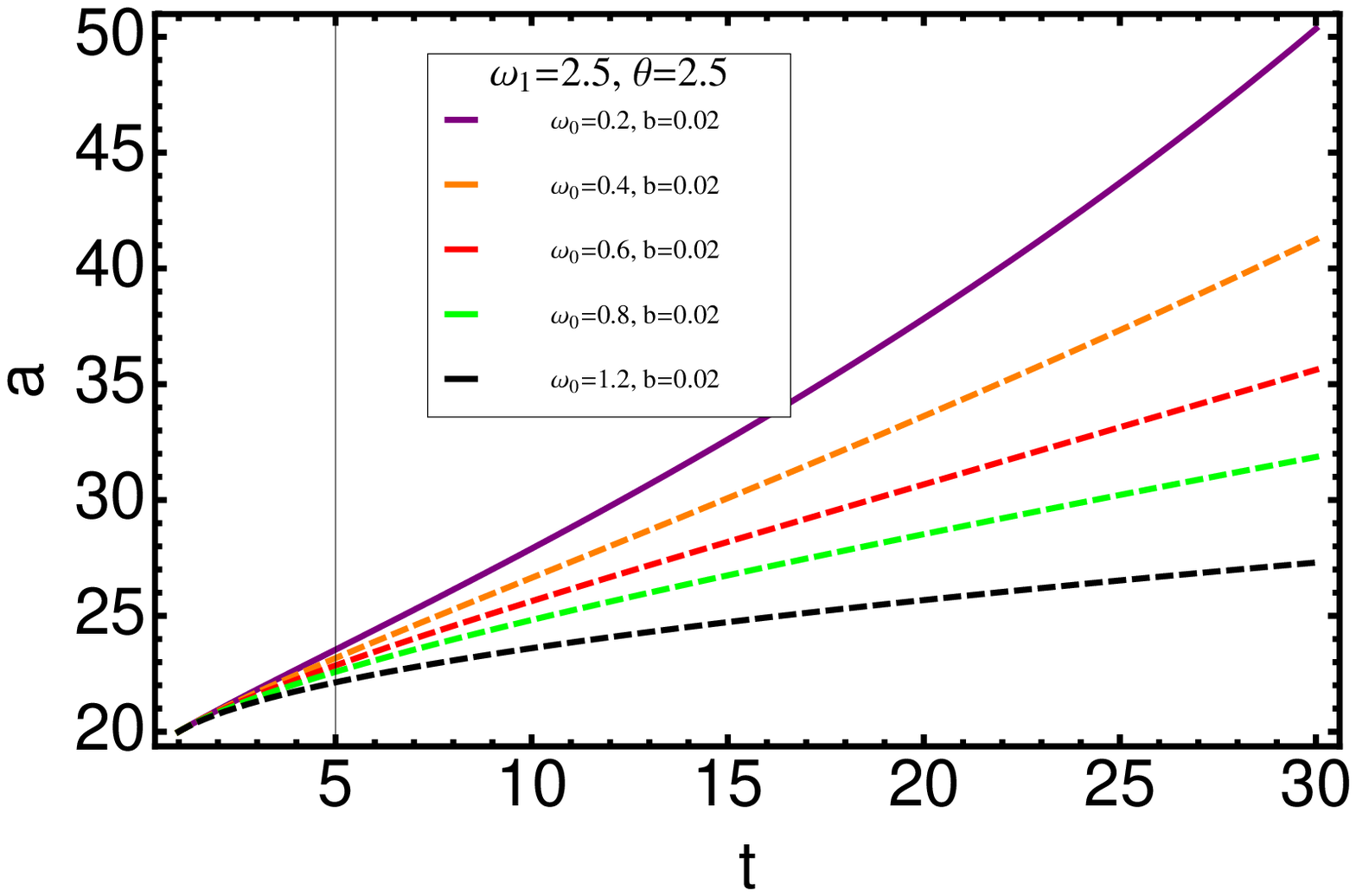} &
\includegraphics[width=50 mm]{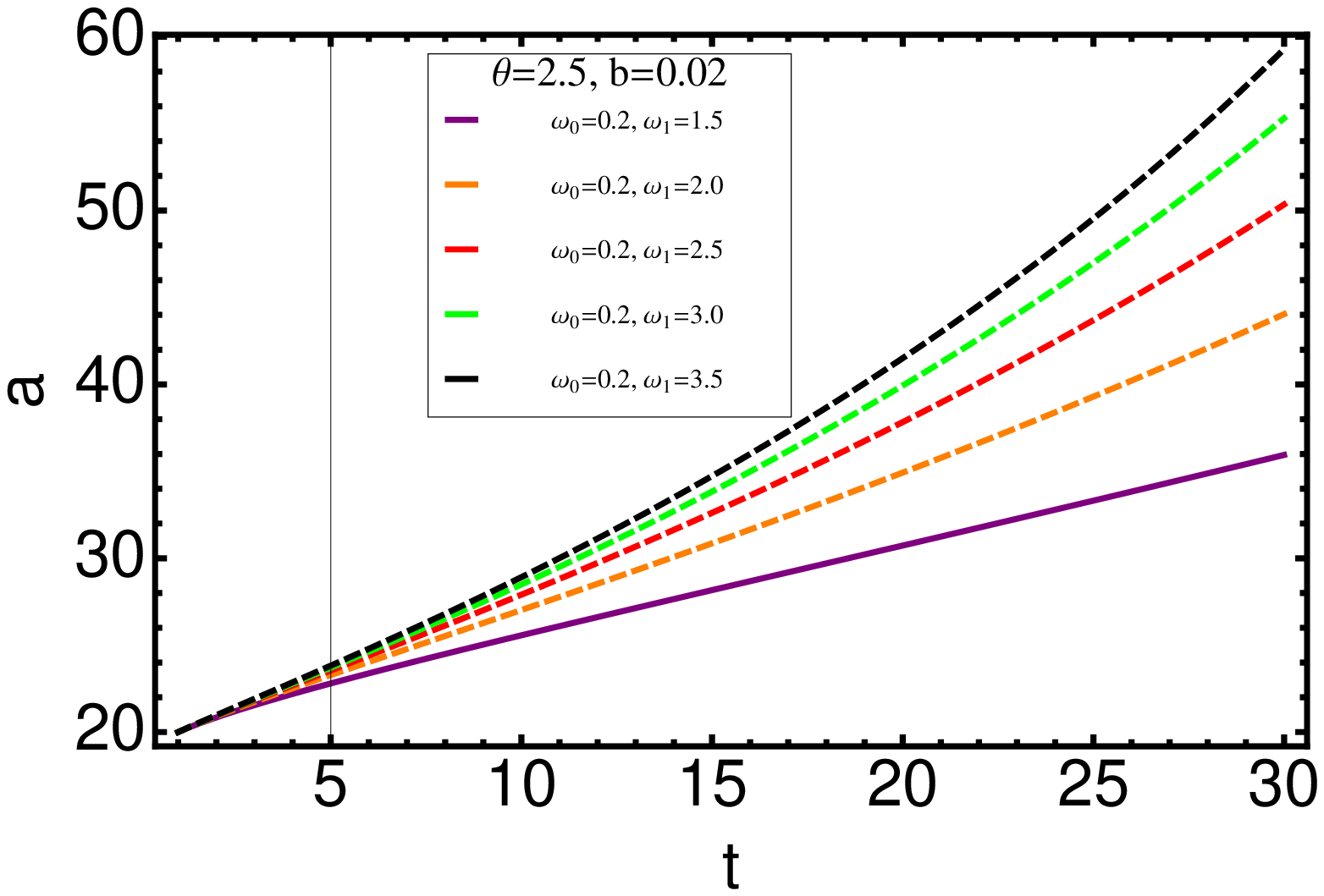}\\
\includegraphics[width=50 mm]{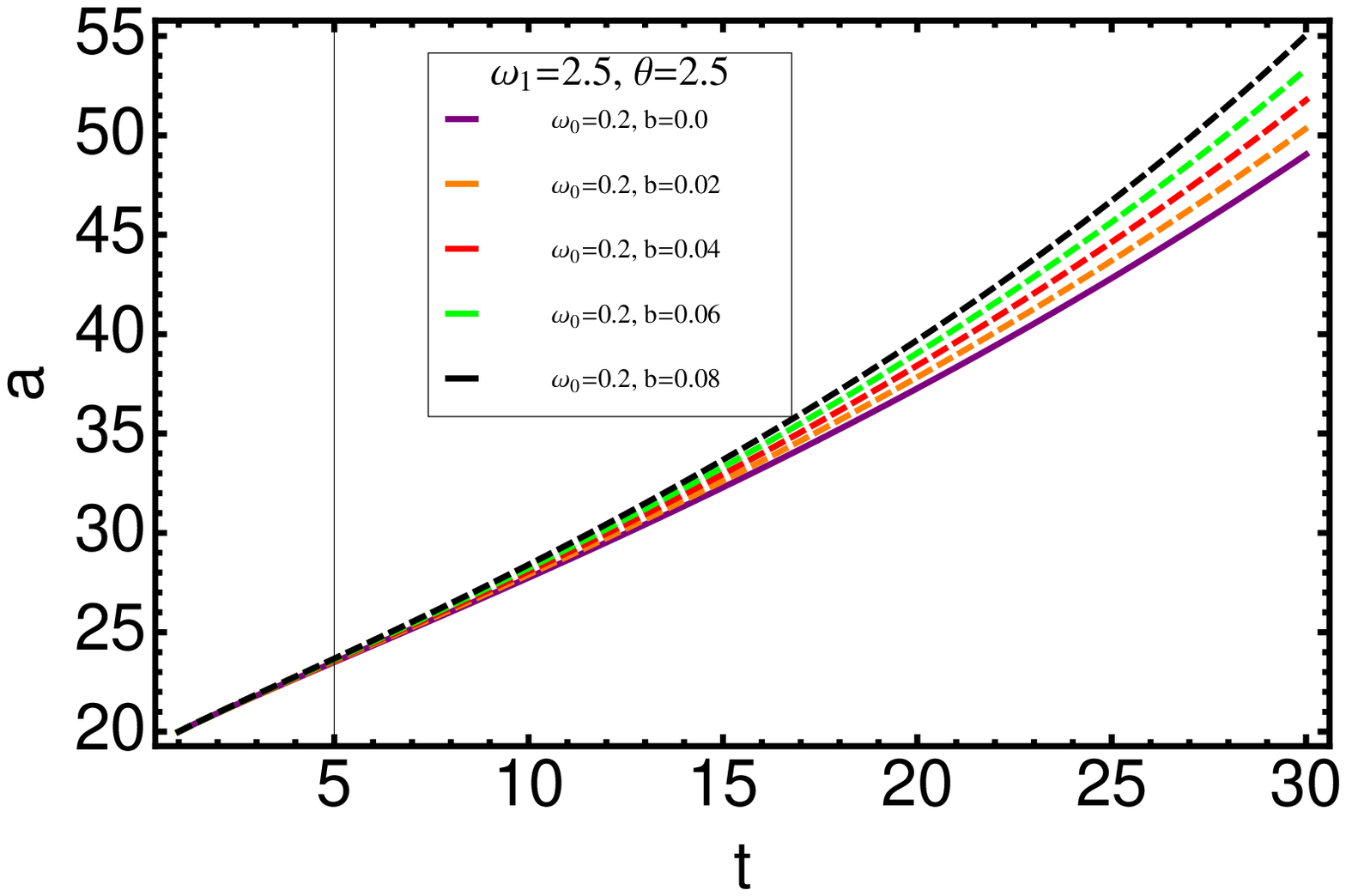} &
\includegraphics[width=50 mm]{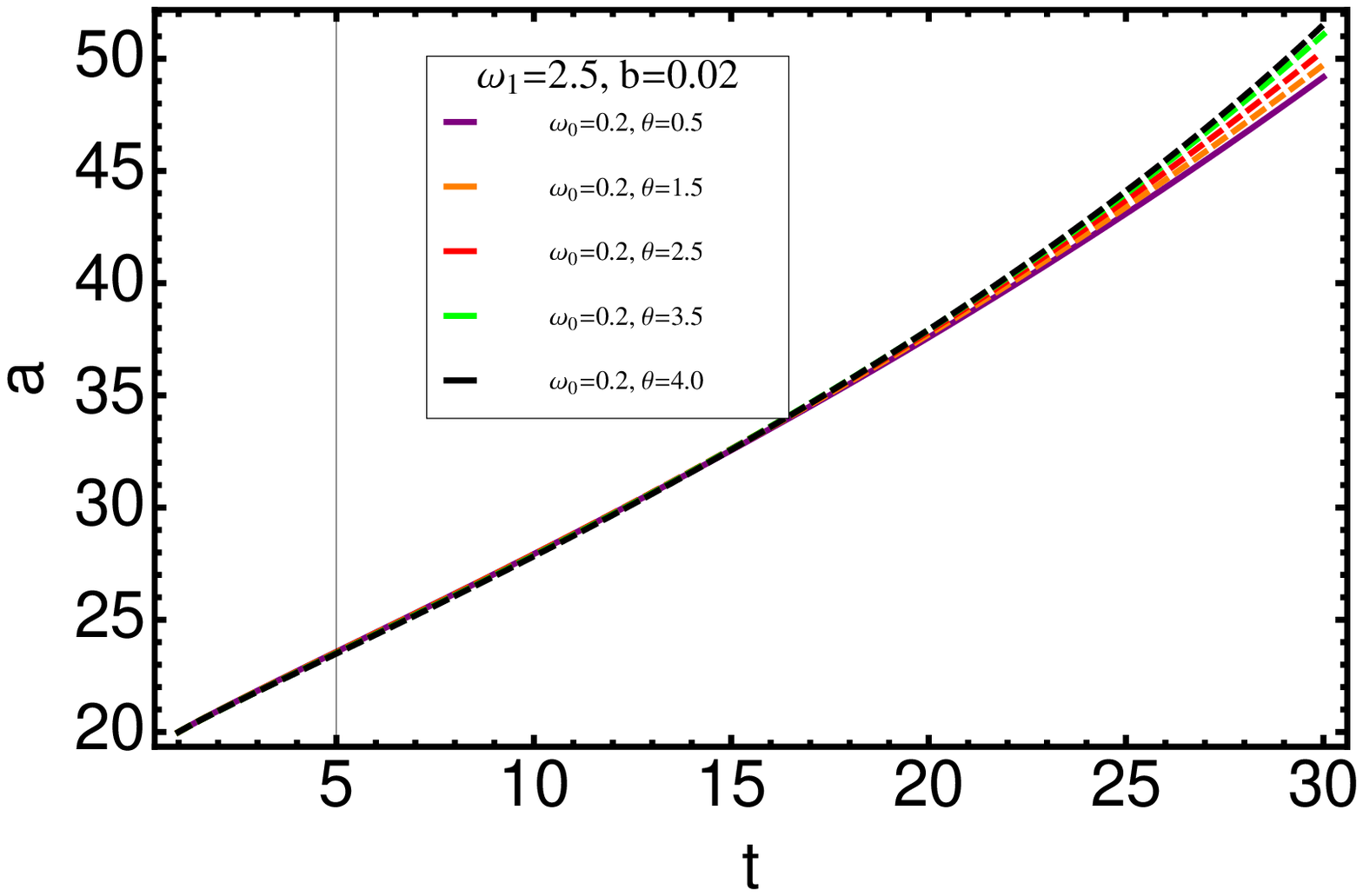}
 \end{array}$
 \end{center}
\caption{Model 2}
 \label{fig:16}
\end{figure}

\begin{figure}[h]
 \begin{center}$
 \begin{array}{cccc}
\includegraphics[width=48 mm]{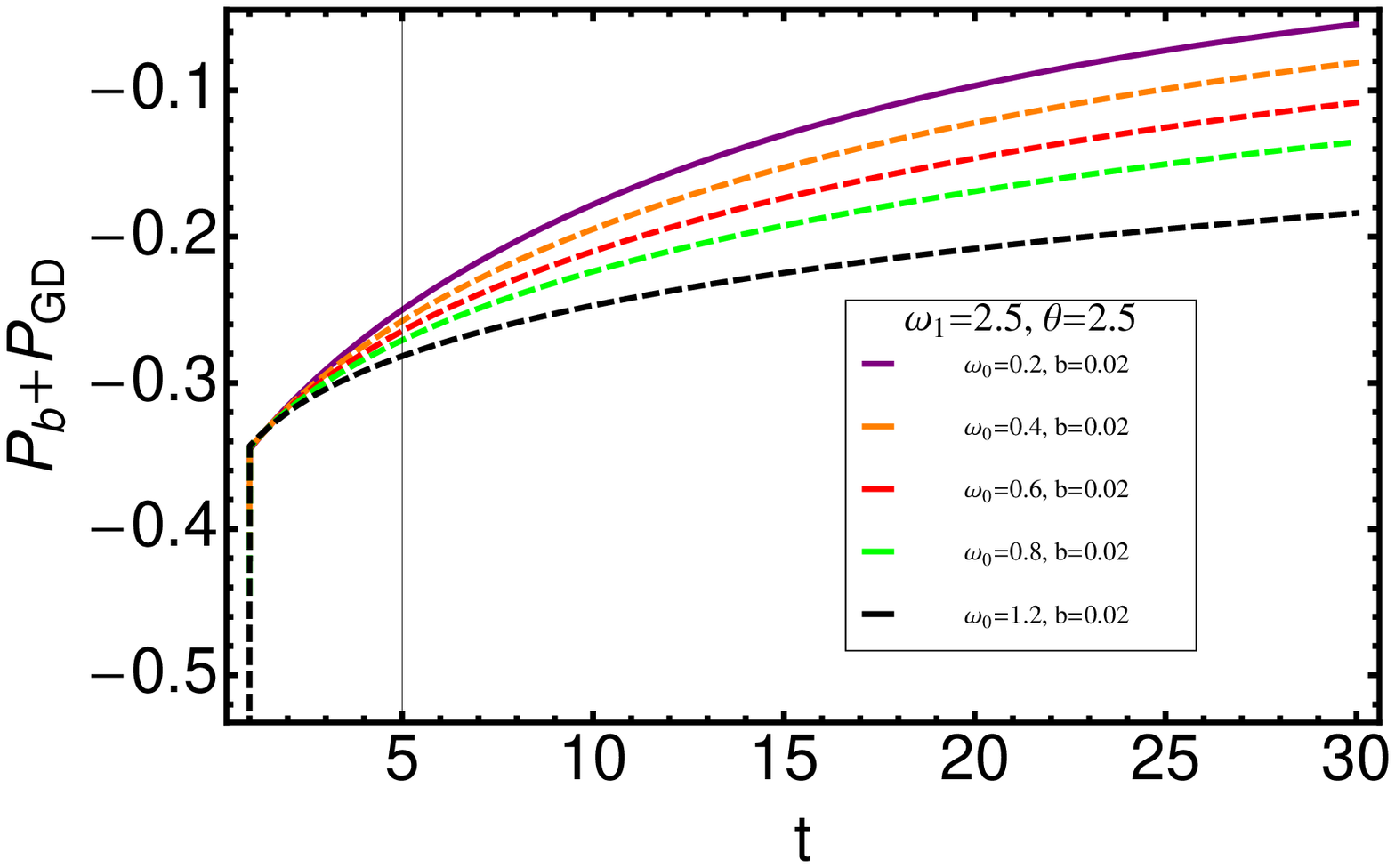} &
\includegraphics[width=48 mm]{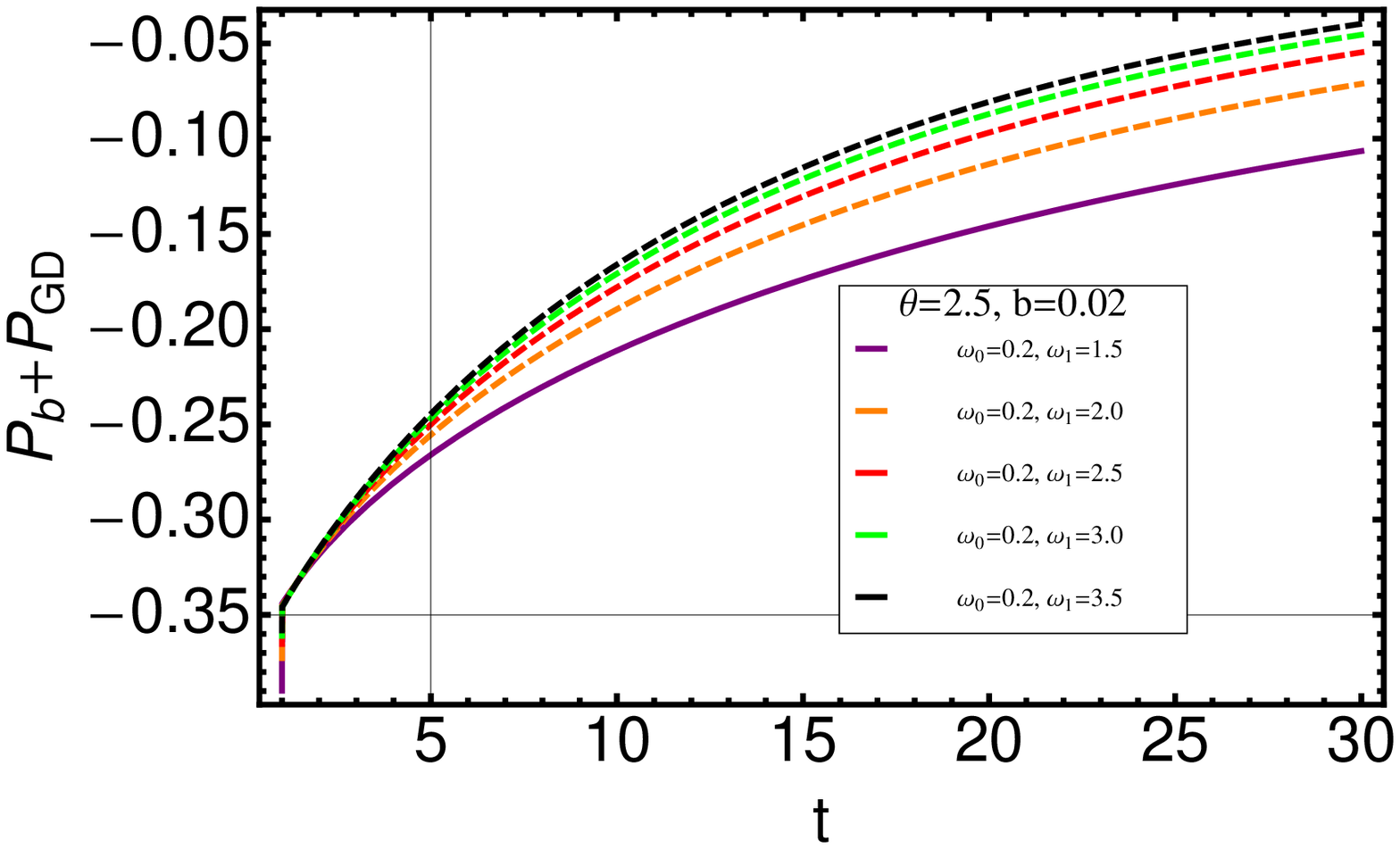}\\
\includegraphics[width=48 mm]{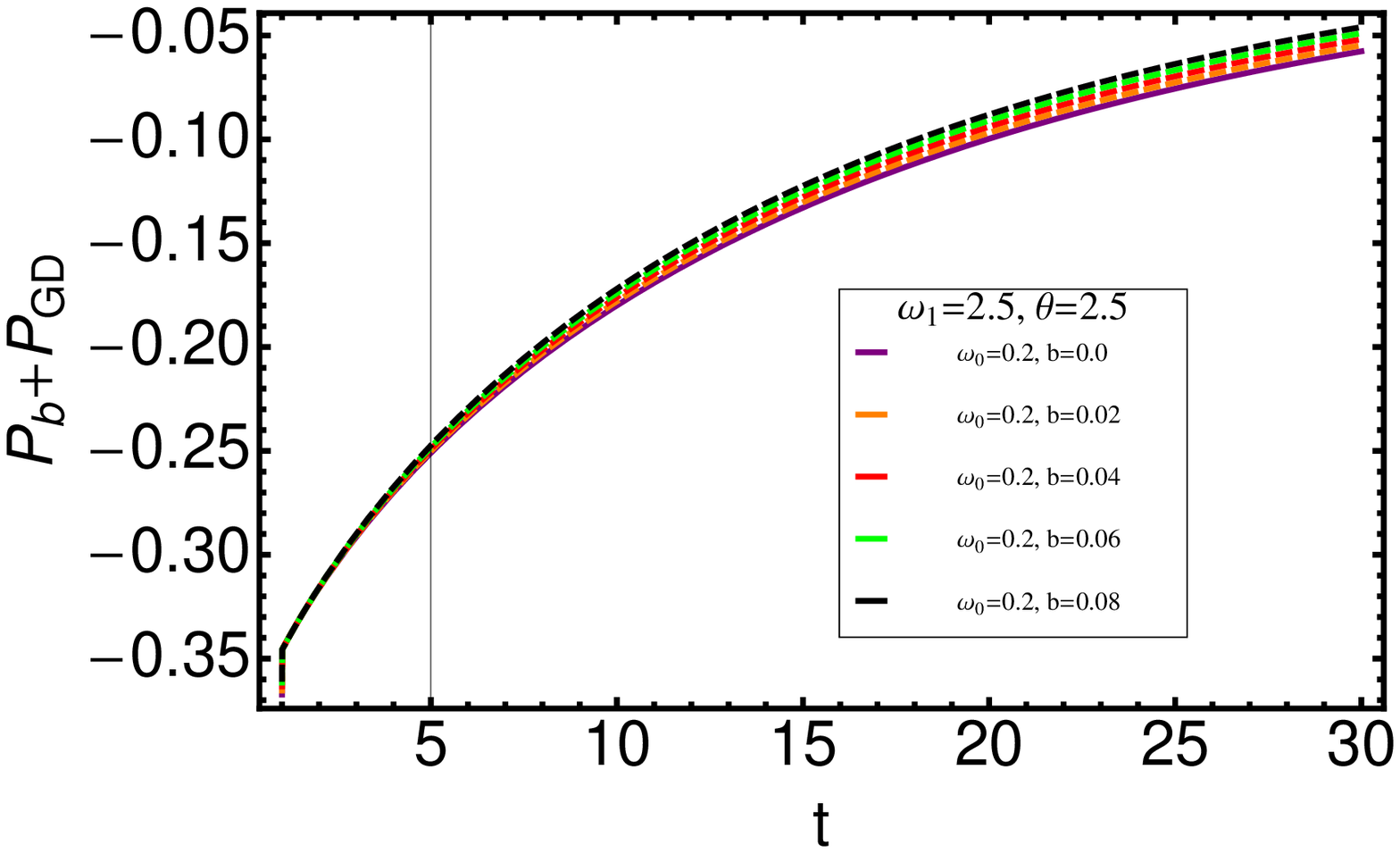} &
\includegraphics[width=48 mm]{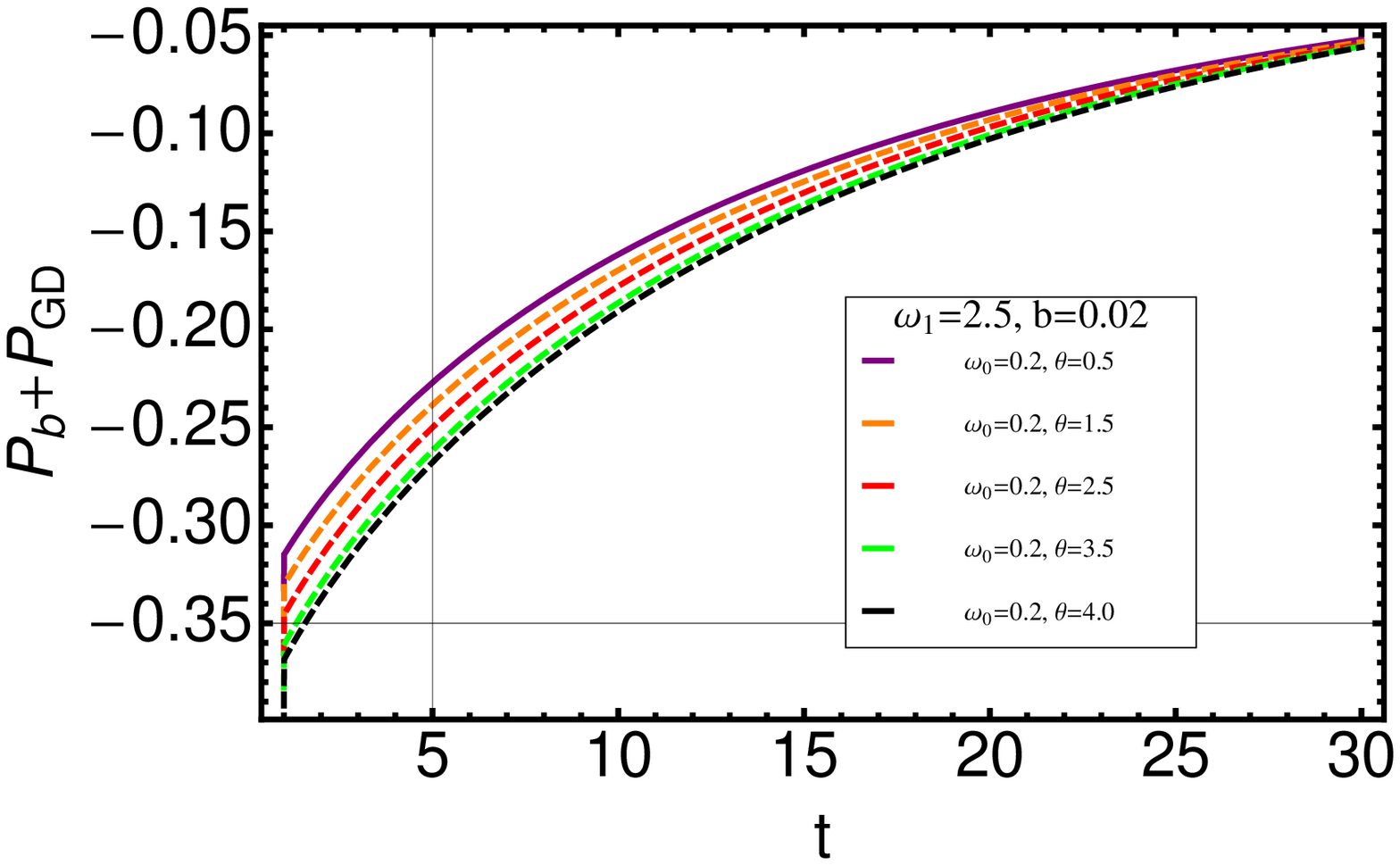}
 \end{array}$
 \end{center}
\caption{Model 2}
 \label{fig:17}
\end{figure}

\begin{figure}[h]
 \begin{center}$
 \begin{array}{cccc}
\includegraphics[width=48 mm]{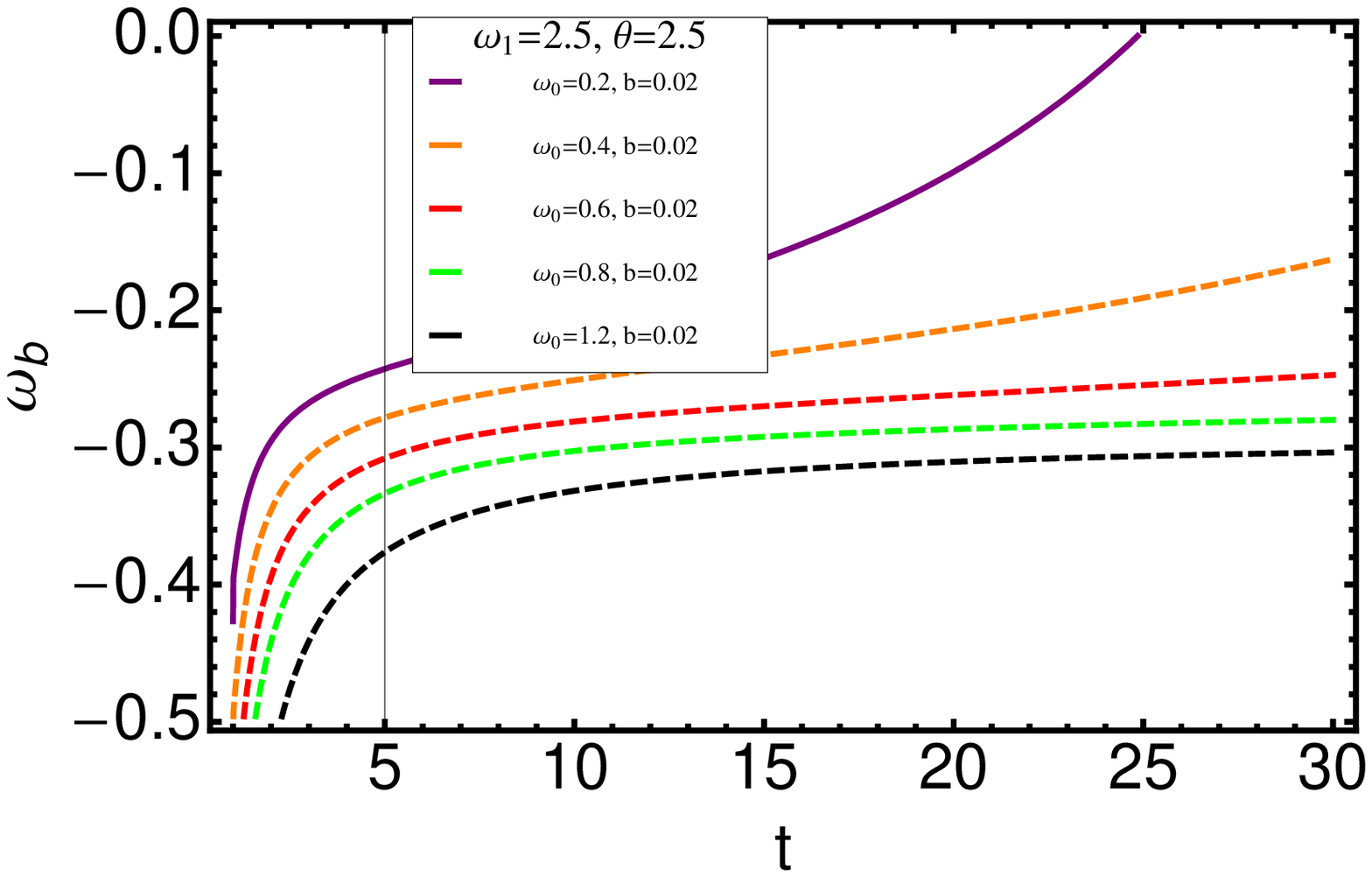} &
\includegraphics[width=48 mm]{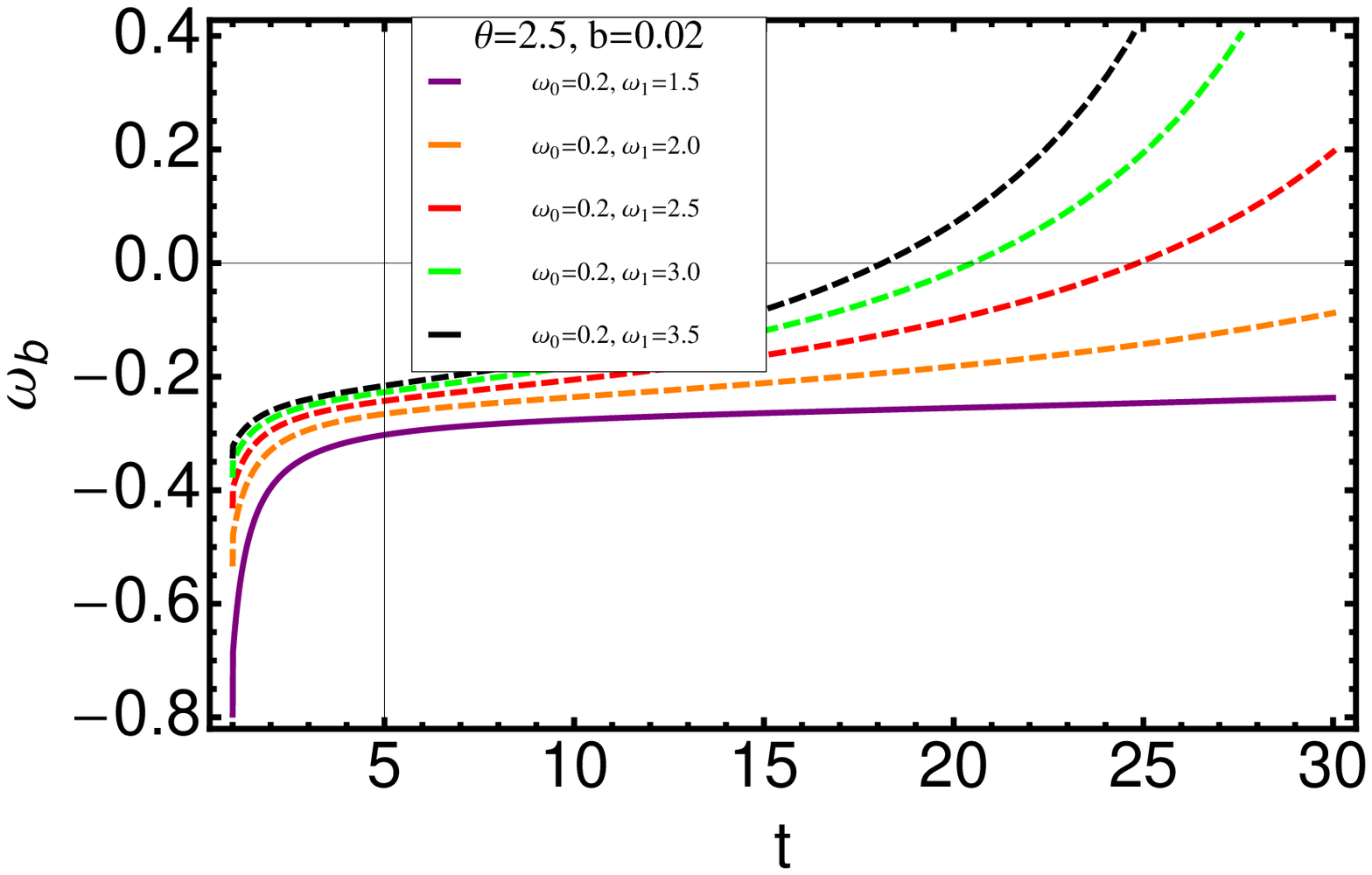}\\
\includegraphics[width=48 mm]{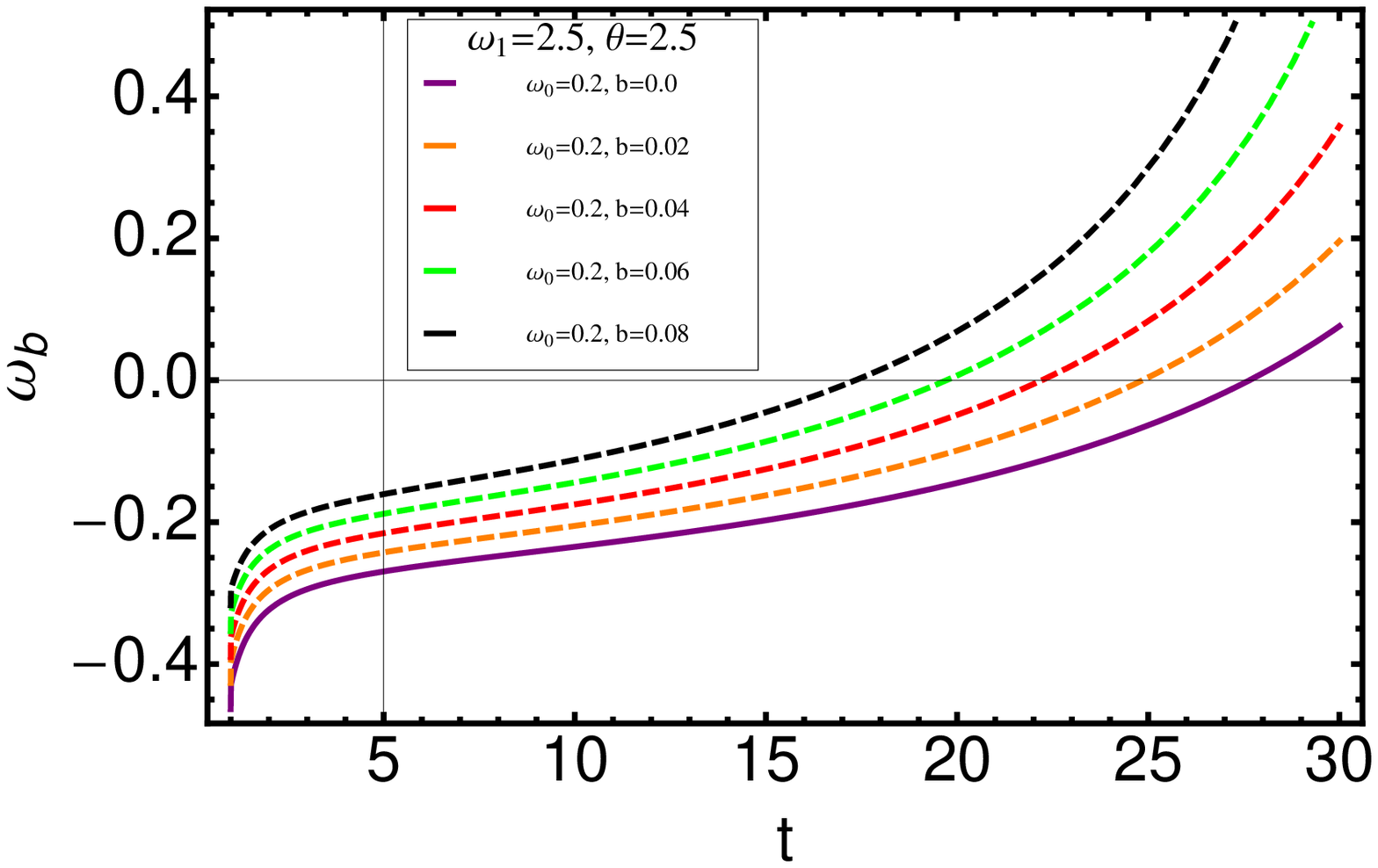} &
\includegraphics[width=48 mm]{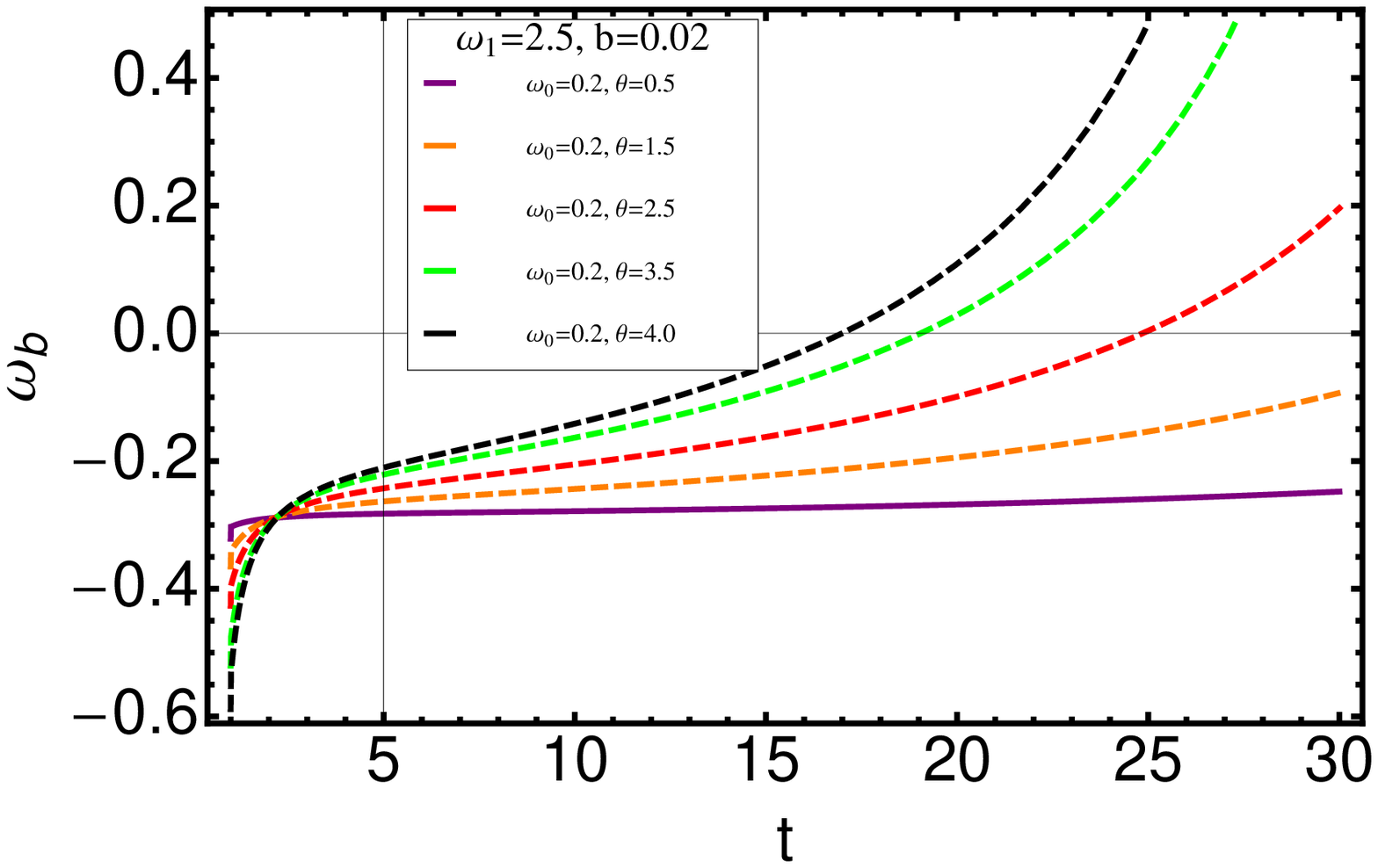}
 \end{array}$
 \end{center}
\caption{Model 2}
 \label{fig:18}
\end{figure}

\begin{figure}[h]
 \begin{center}$
 \begin{array}{cccc}
\includegraphics[width=50 mm]{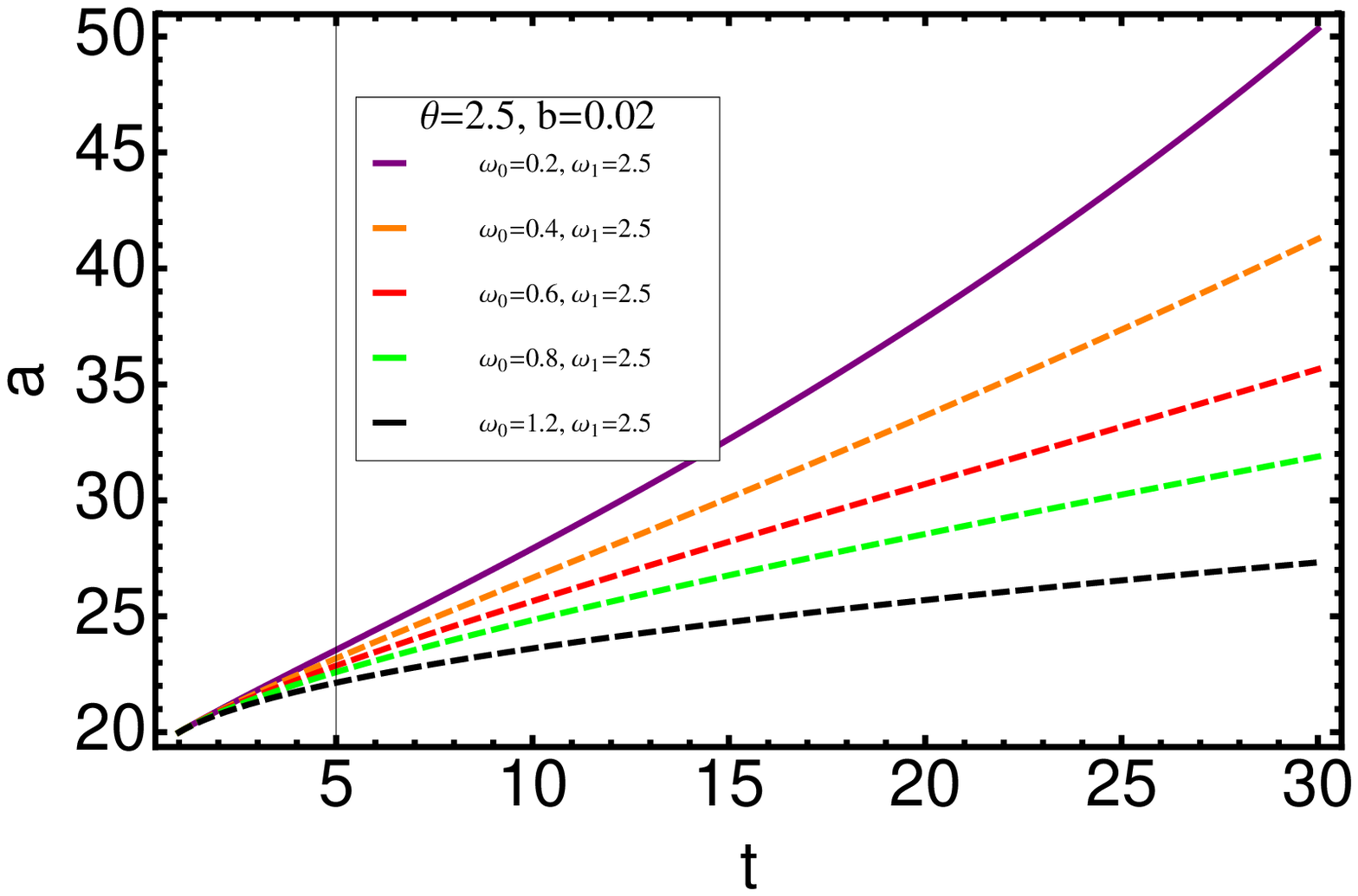} &
\includegraphics[width=50 mm]{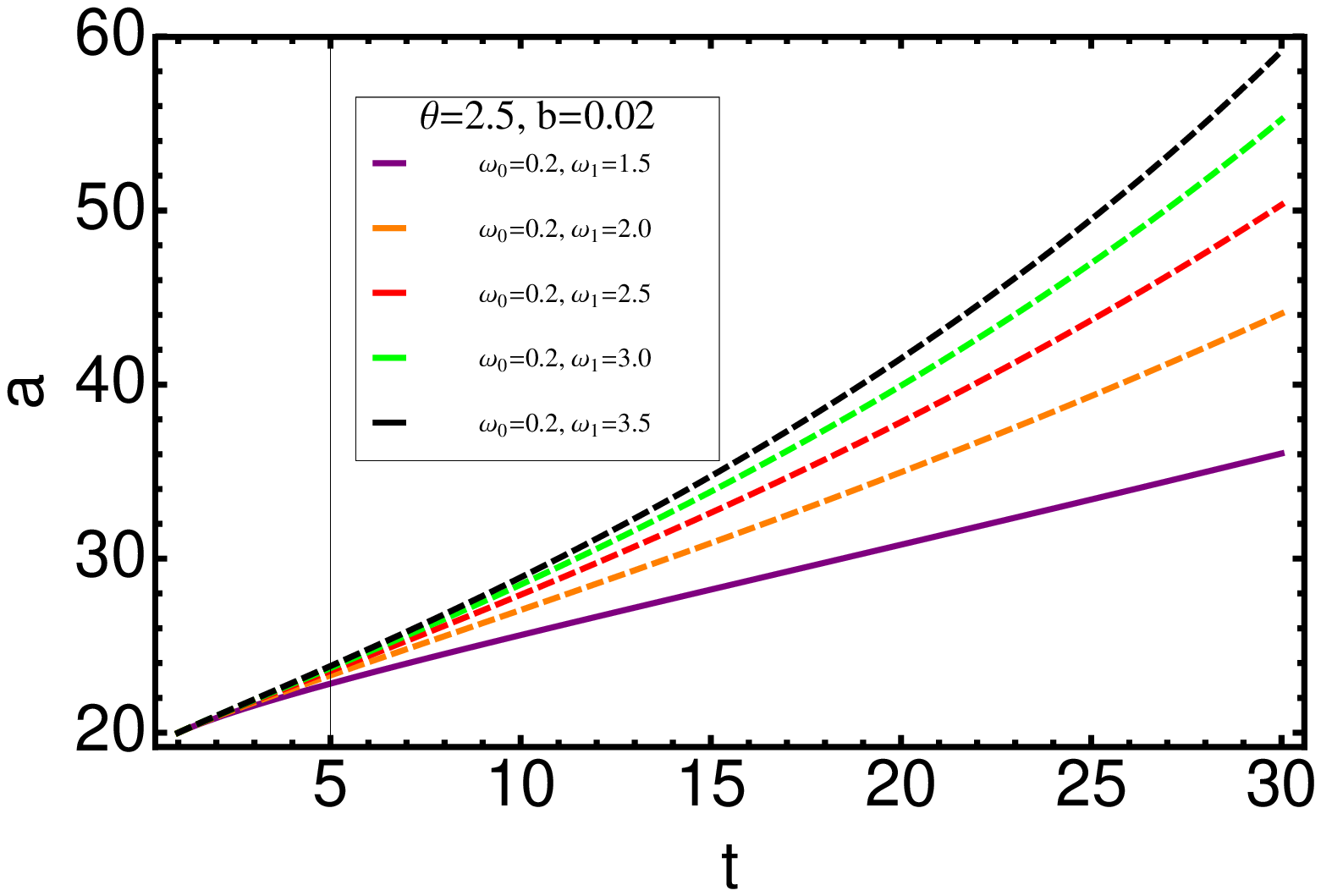}\\
\includegraphics[width=50 mm]{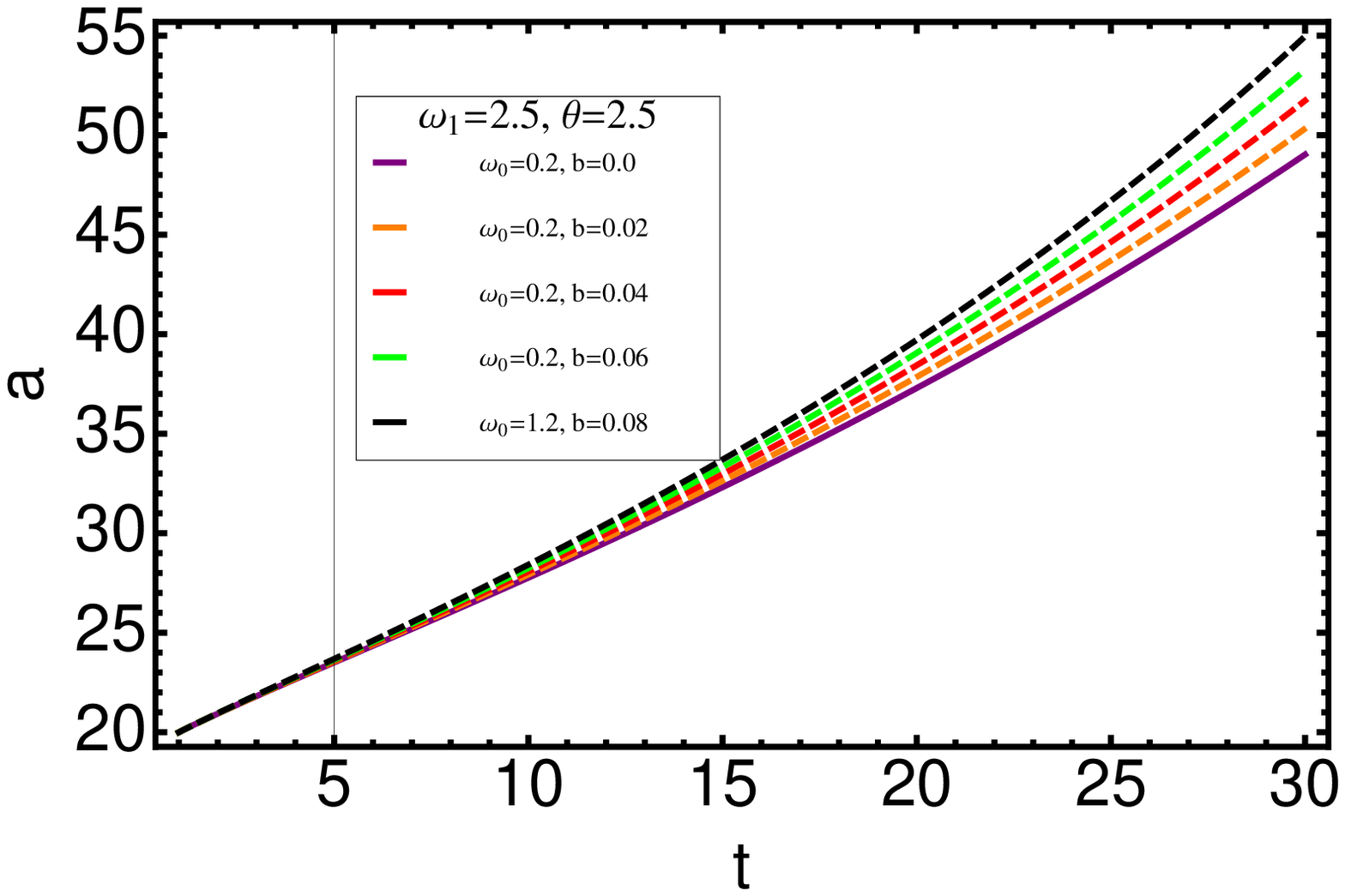} &
\includegraphics[width=50 mm]{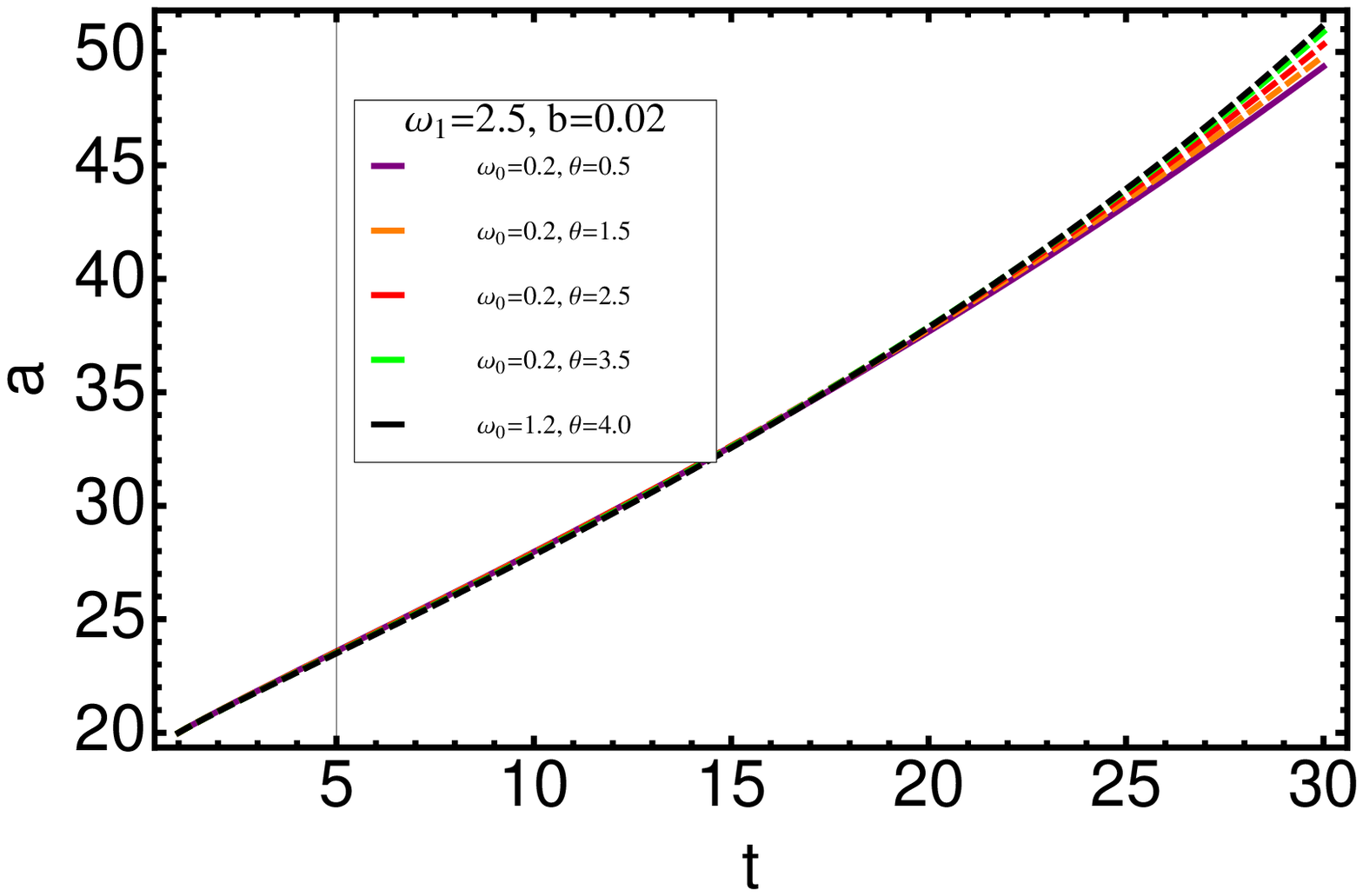}
 \end{array}$
 \end{center}
\caption{Model 3}
 \label{fig:19}
\end{figure}

\begin{figure}[h]
 \begin{center}$
 \begin{array}{cccc}
\includegraphics[width=50 mm]{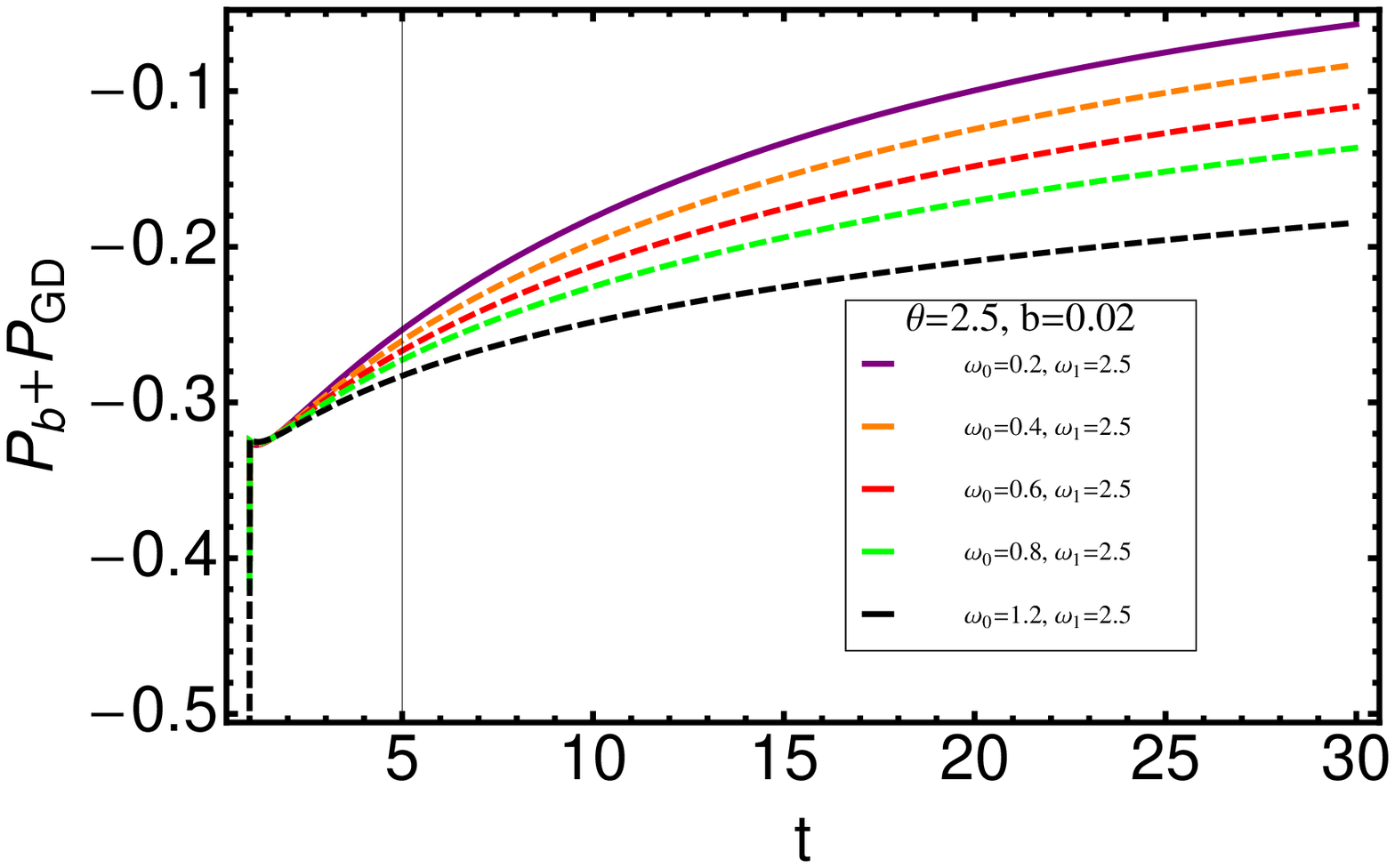} &
\includegraphics[width=50 mm]{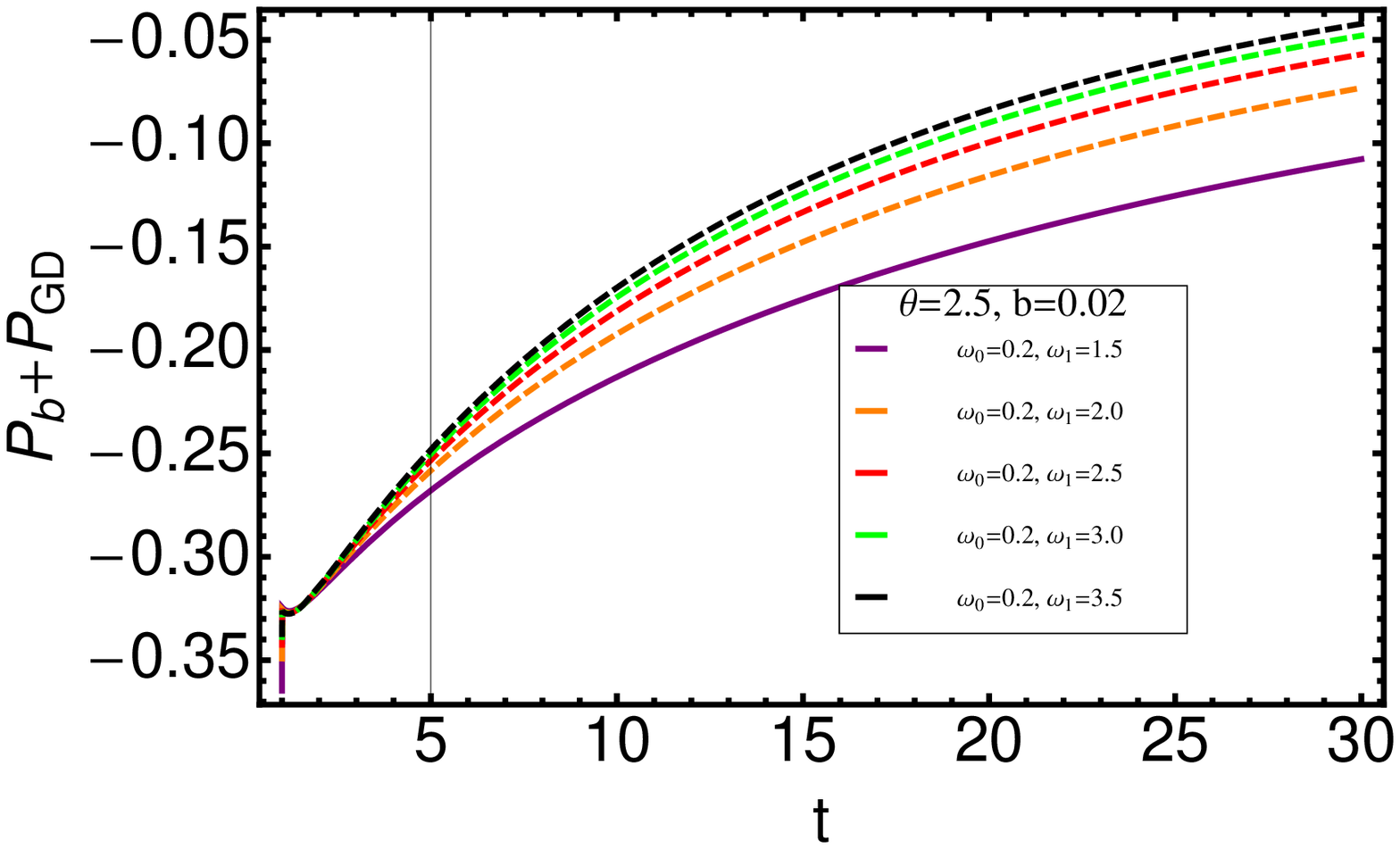}\\
\includegraphics[width=50 mm]{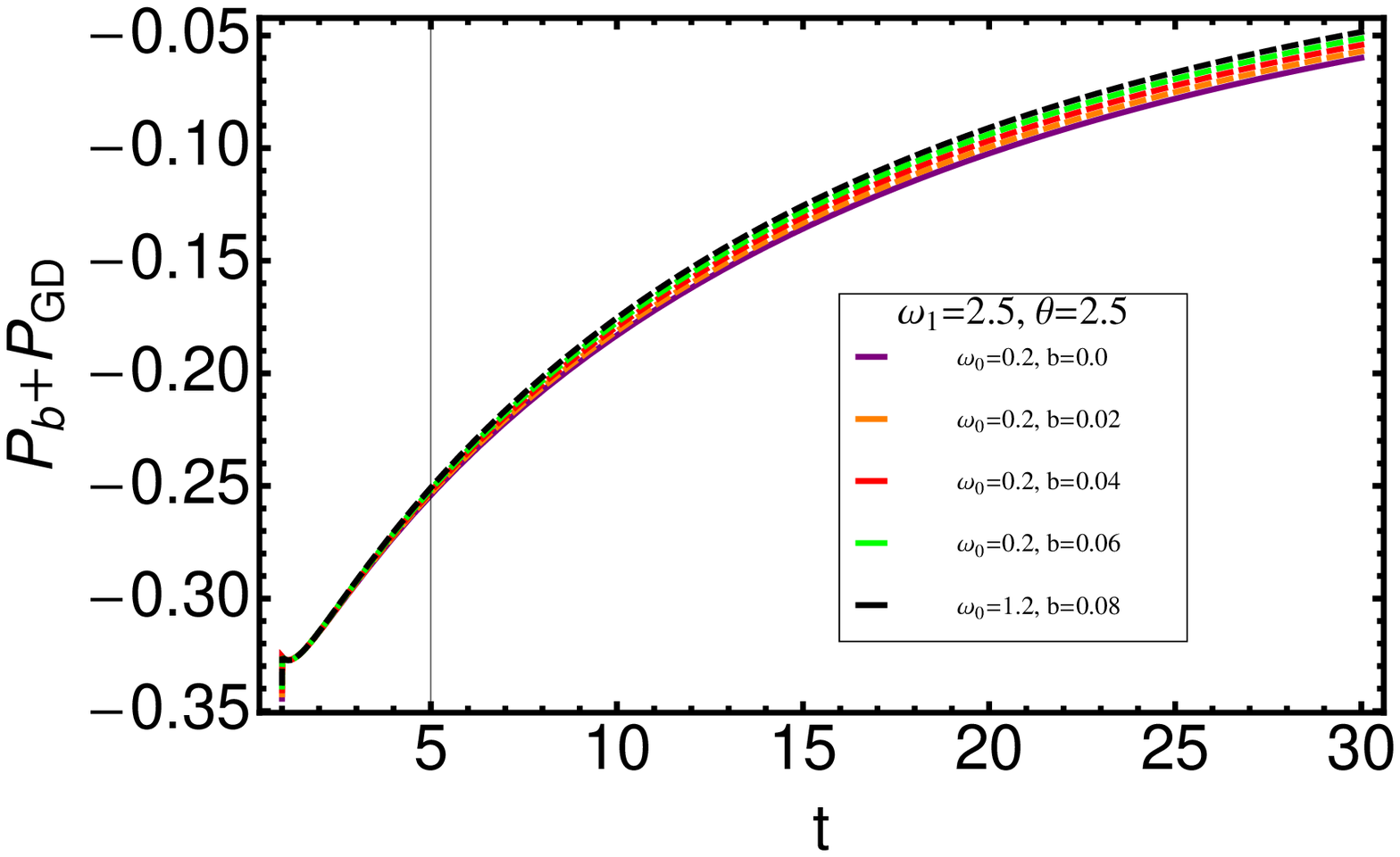} &
\includegraphics[width=50 mm]{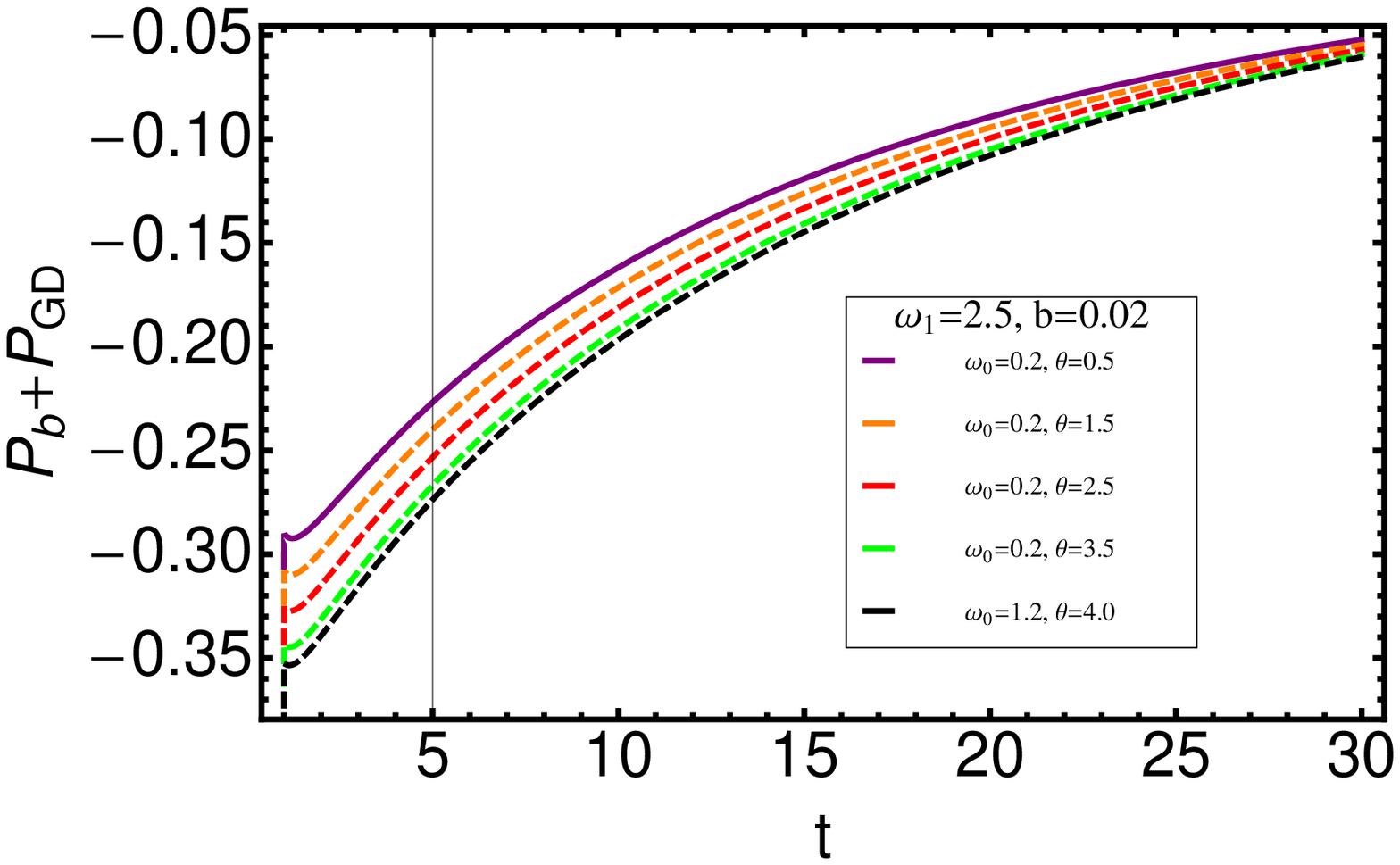}
 \end{array}$
 \end{center}
\caption{Model 3}
 \label{fig:20}
\end{figure}

\begin{figure}[h]
 \begin{center}$
 \begin{array}{cccc}
\includegraphics[width=50 mm]{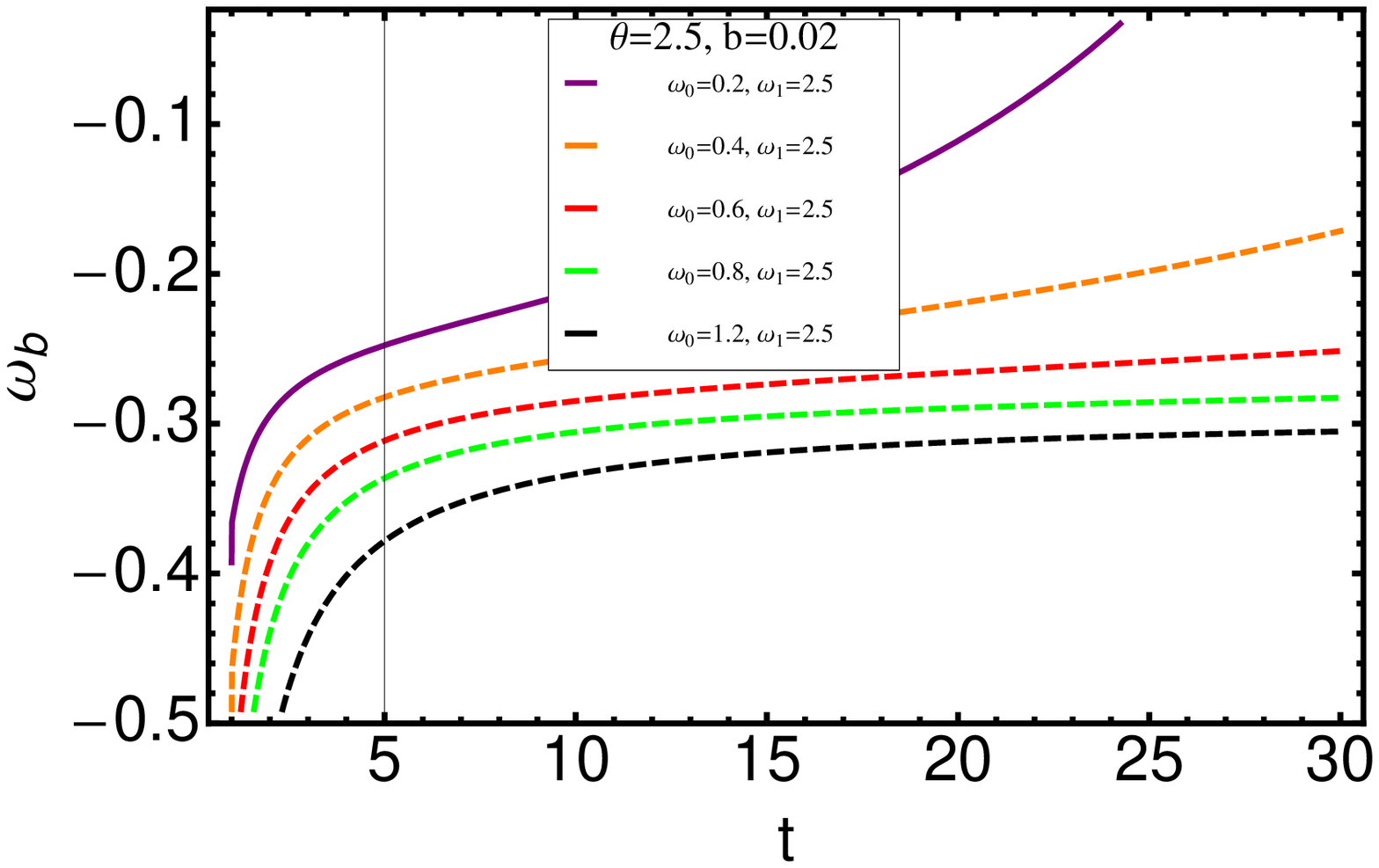} &
\includegraphics[width=50 mm]{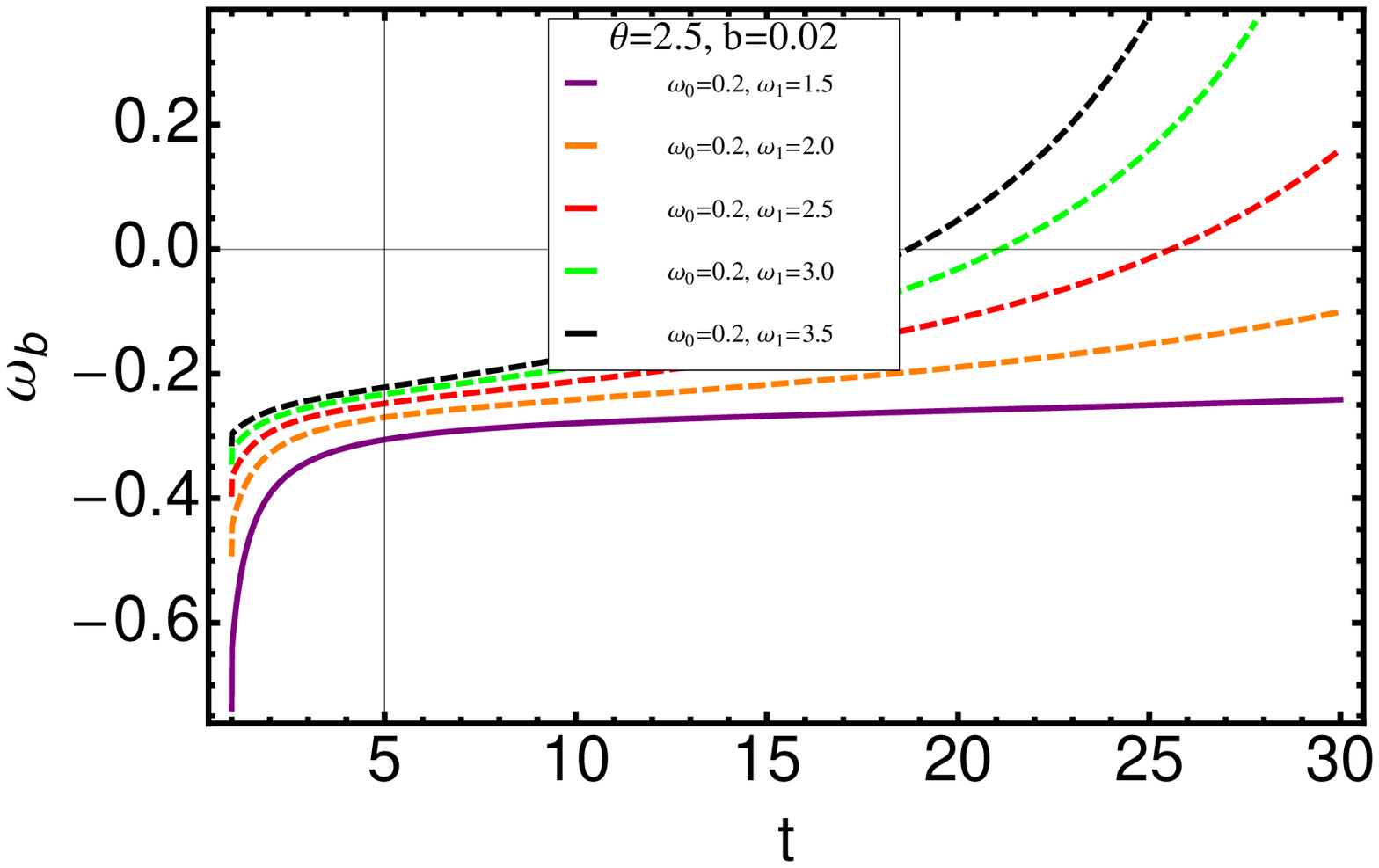}\\
\includegraphics[width=50 mm]{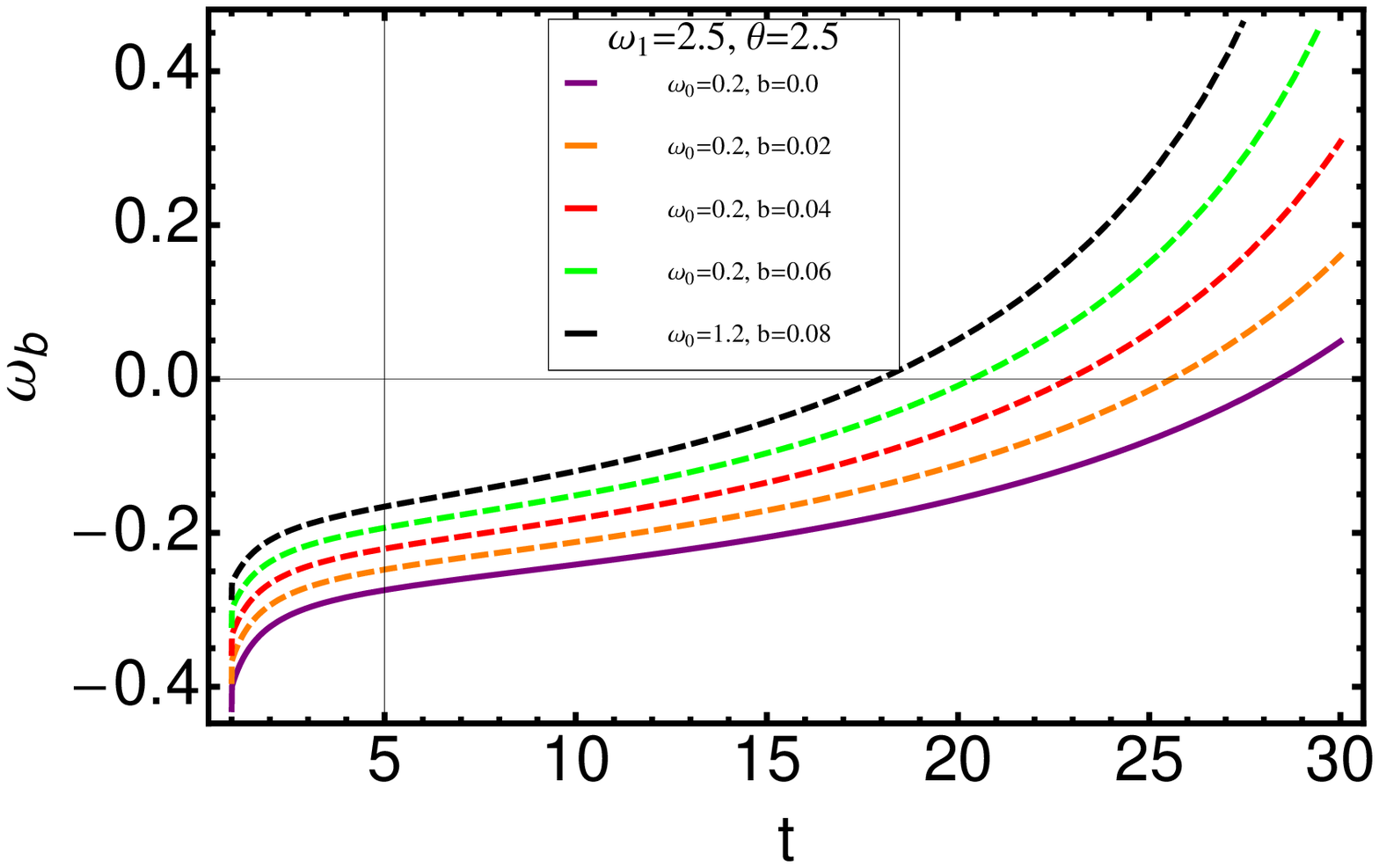} &
\includegraphics[width=50 mm]{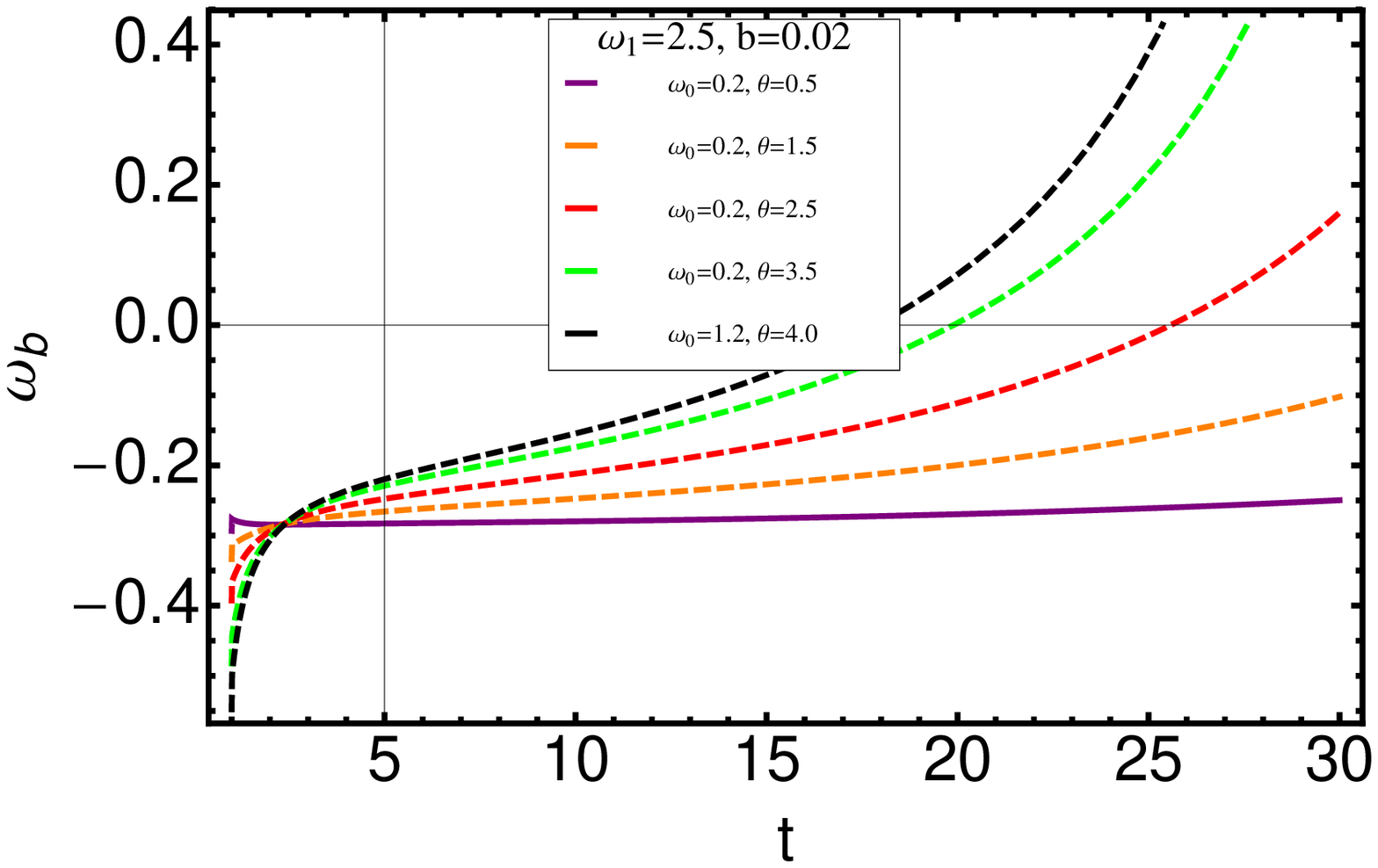}
 \end{array}$
 \end{center}
\caption{Model 3}
 \label{fig:21}
\end{figure}


\begin{thebibliography}{1}
\bibitem{Riess1}
A.G. Riess et al. [Supernova Search Team Colloboration], Astron. J.
116 (1998) 1009
\bibitem{Perlmutter2}
S. Perlmutter et al. [Supernova Cosmology Project Collaboration],
Astrophys. J. 517 (1999) 565
\bibitem{Amanullah3}
R. Amanullah et al., Astrophys. J. 716 (2010) 712
\bibitem{Pope4}
A.C. Pope et al. Astrophys. J. 607 (2004) 655,
[arXiv:astro-ph/0401249]
\bibitem{Spergel5}
D.N. Spergel et al. Astrophys. J. Supp. 148 (2003) 175,
[arXiv:astro-ph/0302209]
\bibitem{Steinhardt6}
P.J. Steinhardt, "Critical Problems in Physics" (1997), Prinston
University Press
\bibitem{Sola7}
J. Sola and H. Stefancic, Phys. Lett. B 624 (2005) 147
\bibitem{Sola8}
I.L. Shapiro and J. Sola, Phys. Lett. B 682 (2009) 105
\bibitem{Ratra9}
B. Ratra and P. J. E. Peebles, Phys. Rev. D 37 (1988) 3406
\bibitem{Wetterich10}
C. Wetterich, Nucl. Phys. B 302 (1988) 668
\bibitem{Zlatev11}
I. Zlatev, L. M. Wang and P. J. Steinhardt, Phys. Rev. Lett. 82
(1999) 896
\bibitem{Guo12}
Z. K. Guo, N. Ohta and Y. Z. Zhang, Mod. Phys. Lett. A 22 (2007) 883
\bibitem{Dutta13}
S. Dutta, E. N. Saridakis and R. J. Scherrer, Phys. Rev. D 79 (2009)
103005
\bibitem{Saridakis14}
E.N. Saridakis and S. V. Sushkov, Phys. Rev. D 81 (2010) 083510
\bibitem{Caldwell15}
R.R. Caldwell, M. Kamionkowski and N.N. Weinberg, Phys. Rev. Lett.
91 (2003) 071301
\bibitem{Caldwell16}
R.R. Caldwell, Phys. Lett. B 545 (2002) 23
\bibitem{Nojiri17}
S. Nojiri and S. D. Odintsov, Phys. Lett. B 562 (2003) 147
\bibitem{Singh18}
P. Singh, M. Sami and N. Dadhich, Phys. Rev. D 68 (2003) 023522
\bibitem{Cline19}
J.M. Cline, S. Jeon and G.D. Moore, Phys. Rev. D 70 (2004) 043543
\bibitem{Onemli20}
V.K. Onemli and R.P. Woodard, Phys. Rev. D 70 (2004) 107301
\bibitem{Hu21}
W. Hu, Phys. Rev. D 71 (2005) 047301
\bibitem{Setare22}
M.R. Setare and E. N. Saridakis, JCAP 0903 (2009) 002
\bibitem{Saridakis23}
E.N. Saridakis, Nucl. Phys. B 819 (2009) 116
\bibitem{Feng24}
B. Feng, X. L. Wang and X. M. Zhang, Phys. Lett. B 607 (2005) 35
\bibitem{Elizalde25}
E. Elizalde, S. Nojiri and S.D. Odintsov, Phys. Rev. D 70 (2004)
043539
\bibitem{Guo26}
Z.K. Guo, et al., Phys. Lett. B 608 (2005) 177
\bibitem{Li27}
M.-Z Li, B. Feng, X.-M Zhang, JCAP, 0512 (2005) 002
\bibitem{Feng28}
B. Feng, M. Li, Y.-S. Piao and X. Zhang, Phys. Lett. B 634 (2006)
101
\bibitem{Capozziello29}
S. Capozziello, S. Nojiri and S.D. Odintsov, Phys. Lett. B 632
(2006) 597
\bibitem{Zhao30}
W. Zhao and Y. Zhang, Phys. Rev. D 73 (2006) 123509
\bibitem{Cai31}
Y.F. Cai, T. Qiu, Y.S. Piao, M. Li and X. Zhang, JHEP 0710 (2007)
071
\bibitem{Saridakis32}
E.N. Saridakis and J.M. Weller, Phys. Rev. D 81 (2010) 123523
\bibitem{Cai33}
Y.F. Cai, T. Qiu, R. Brandenberger, Y.S. Piao and X. Zhang, JCAP
0803 (2008) 013
\bibitem{Setare34}
M.R. Setare and E.N. Saridakis, Phys. Lett. B 668 (2008) 177
\bibitem{Setare35}
M.R. Setare and E.N. Saridakis, Int. J. Mod. Phys. D 18 (2009) 549
\bibitem{Cai36}
Y. F. Cai, E. N. Saridakis, M. R. Setare and J. Q. Xia, Phys. Rept.
493 (2010) 1
\bibitem{Cai37}
T. Qiu, Mod. Phys. Lett. A 25 (2010) 909
\bibitem{Hsu38}
S.D.H. Hsu, Phys. Lett. B 594 (2004) 13
\bibitem{Li39}
M. Li, Phys. Lett. B 603 (2004) 1
\bibitem{Huang40}
Q.G. Huang and M. Li, JCAP 0408 (2004) 013
\bibitem{Ito41}
M. Ito, Europhys. Lett. 71 (2005) 712
\bibitem{Zhang42}
X. Zhang and F.Q. Wu, Phys. Rev. D 72 (2005) 043524
\bibitem{Pavon43}
D. Pavon and W. Zimdahl, Phys. Lett. B 628 (2005) 206
\bibitem{Nojiri44}
S. Nojiri and S.D. Odintsov, Gen. Rel. Grav. 38 (2006) 1285
\bibitem{Elizalde45}
E. Elizalde, S. Nojiri, S.D. Odintsov and P. Wang, Phys. Rev. D 71
(2005) 103504
\bibitem{Li46}
H. Li, Z.K. Guo and Y.Z. Zhang, Int. J. Mod. Phys. D 15 (2006) 869
\bibitem{Saridakis47}
E.N. Saridakis, Phys. Lett. B 660 (2008) 138
\bibitem{Saridakis48}
E.N. Saridakis, JCAP 0804 (2008) 020
\bibitem{Saridakis49}
E.N. Saridakis, Phys. Lett. B 661 (2008) 335
\bibitem{Cai50}
R.G. Cai, Phys. Lett. B 657 (2007) 228
\bibitem{Wei51}
H. Wei and R.G. Cai, Phys. Lett. B 660 (2008) 113
\bibitem{Wei52}
H. Wei and R.G. Cai, Eur. Phys. J. C 59 (2009) 99
\bibitem{Martiros53}
M. Khurshudyan, "A Matter with an effective EoS interacting with a
tachynic field in an accelerating Universe",
[arXiv:1301.0005[physics.gen-ph]]
\bibitem{P54}
A.R. Amani and  B. Pourhassan, "Viscous Generalized Chaplygin gas
with Arbitrary $\alpha$", Int. J. Theor. Phys. 52 (2013) 1309
\bibitem{P55}
H. Saadat and  B. Pourhassan, "Viscous Varying Generalized Chaplygin
Gas with Cosmological Constant and Space Curvature", Int. J. Theor.
Phys. DOI: 10.1007/s10773-013-1676-2
\bibitem{P56}
H. Saadat and  B. Pourhassan, "FRW Bulk Viscous Cosmology with
Modified Chaplygin Gas in Flat Space", Astrophysics and Space
Science 343 (2013) 783
\bibitem{P57}
H. Saadat and  B. Pourhassan, "FRW bulk viscous cosmology with
modified cosmic Chaplygin gas", Astrophysics and Space Science 344
(2013) 237
\bibitem{P58}
B. Pourhassan, "Viscous Modified Cosmic Chaplygin Gas Cosmology"
International Journal of Modern Physics D 22 (9) (2013) 1350061
[arXiv:1305.6054 [gr-qc]]
\bibitem{P59}
J. Sadeghi, M. Khurshudyan, H. Farahani, "Phenomenological Varying
Modified Chaplygin Gas with Variable $G$ and $\Lambda$: Toy Models
for Our Universe", [arXiv:1308.1819 [gr-qc]]
\bibitem{Martiros60}
M. Khurshudyan, "Interaction between Generalized Varying Chaplygin
gas and Tachyonic Fluid", [arXiv:1301.1021 [gr-qc]]
\bibitem{Zhang61}
X. Zhang et al., JCAP 0601 (2006) 003
\bibitem{Chattopadhyay62}
S. Chattopadhyay, U. Debnath, Grav. Cosmol. 14 (2008) 341
\bibitem{Jamil63}
M. Jamil, "Interacting New Generalized Chaplygin Gas", Int. J.
Theor. Phys. 49 (2010) 62
\bibitem{Hao64}
W. Hao, Cosmological Constraints on the Sign-Changeable
Interactions, Common. Theory. Phys. 56 (2011) 972
\bibitem{Freese65}
K. Freese, M. Lewis, "Cardassian Expansion: a Model in which the
Universe is Flat, Matter Dominated, and Accelerating", Phys. Lett.
B540 (2002) 1
\bibitem{Freese66}
K. Freese, "Generalized Cardassian Expansion: Models in Which the
Universe is Flat, Matter Dominated, and Accelerating", Nucl. Phys.
Proc. Suppl. 124 (2003) 50
\bibitem{Liu67}
Dao-jun Liu, Chang-bo Sun, Xin-zhou Li, "Exponential Cardassian
Universe",  Phys. Lett. B634 (2006) 442
\bibitem{Yi68}
Ze-Long Yi, Tong-Jie Zhang, "Statefinder diagnostic for the modified
polytropic Cardassian universe", Phys. Rev. D75 (2007) 083515
\bibitem{Chao69}
Chao-Jun Feng, Xin-Zhou Li, Xian-Yong Shen, "Thermodynamic Origin of
the Cardassian Universe",  Phys. Rev. D83 (2011) 123527
\bibitem{UjjalD70}
P.B. Khatua, U. Debnath, "Natures of Statefinder Parameters and Om
Diagnostic for Cardassian Universe in Horava-Lifshitz Gravity",
[arXiv:1108.1186 [physics.gen-ph]]
\bibitem{Benvenuto71}
O.G. Benvenuto et al. Phys. Rev. D 69 (2004) 082002
\bibitem{Saibal72}
Saibal Ray, Utpal Mukhopadhyay, Partha Pratim Ghosh, "Large Number
Hypothesis: A Review", [arXiv:0705.1836 [gr-qc]]
\bibitem{Abdussattar73}
Abdussattar and R.G. Vishwakarma, Some FRW models with variable $G$
and $\Lambda$, Class. Quantum Grav. 14 (1997) 945
\bibitem{74}
F.R. Urban, A.R. Zhitnitsky, Phys. Rev. D 80 (2009) 063001
\bibitem{75}
F.R. Urban, A.R. Zhitnitsky, JCAP 09 (2009) 018
\bibitem{76}
F.R. Urban, A.R. Zhitnitsky, Phys. Lett. B 688 (2010) 9
\bibitem{77}
F.R. Urban, A.R. Zhitnitsky, Nucl. Phys. B 835 (2010) 135
\bibitem{78}
N. Ohta, Phys. Lett. B 695 (2011)41
\bibitem{79}
R.G. Cai, Z.L. Tuo, H.B. Zhang, [arXiv:1011.3212 [astro-ph.CO]]
\bibitem{80}
A. Sheykhi, A. Bagheri, Europhys. Lett. 95 (2011) 39001
\bibitem{81}
E. Ebrahimi, A. Sheykhi, Phys. Lett. B 705 (2011) 19
\bibitem{82}
E. Ebrahimi, A. Sheykhi, Int. J. Mod. Phys. D 20 (2011) 2369
\bibitem{83}
A. Sheykhi, M.S Movahed, Gen. Relativ. Gravit.
[DOI:10.1007/s10714-011- 1286-3]
\bibitem{Chao-Jun84}
Chao-Jun Feng, Xin-Zhou Li, Ping Xi, "Global behavior of
cosmological dynamics with interacting Veneziano ghost", JHEP 1205
(2012) 046
\bibitem{85}
Chao-Jun Feng, Xin-Zhou Li, Xian-Yong Shen, "Latest Observational
Constraints to the Ghost Dark Energy Model by Using Markov Chain
Monte Carlo Approach", Phys.Rev. D87 (2013) 023006
\bibitem{86}
Chao-Jun Feng, Xin-Zhou Li, Xian-Yong Shen, "Thermodynamic of the
QCD Ghost Dark Energy Universe", Mod. Phys. Lett. A27 (2012) 1250182
\bibitem{Cai87}
R.G. Cai, Z.L. Tuo, Y.B. Wu, Y.Y. Zhao, Phys Rev. D 86 (2012) 023511
\bibitem{AmalyaMartiros88}
M. Khurshudyan, A. Khurshudyan, "A model of a varying Ghost Dark
energy", [arXiv:1307.7859[gr-qc]]
\bibitem{Hao89}
H. Wei, Nucl. Phys. B 845 (2011) 381.
\end{thebibliography}
\end{document}